\documentclass[reprint,superscriptaddress,amsmath,amssymb,aps,pra,floatfix]{revtex4-2}

\usepackage{graphicx}
\usepackage{physics}
\usepackage{amsmath, amssymb}

\usepackage{mathtools}
\usepackage{amsthm}
\usepackage{dsfont}
\usepackage{subfig}
\usepackage{tabularx}
\usepackage{units}
\usepackage{enumerate}

\usepackage{thmtools} 
\usepackage{thm-restate}
\usepackage{mathtools} 
\usepackage[dvipsnames]{xcolor}
\usepackage[colorlinks=true,citecolor=NavyBlue,linkcolor=NavyBlue,urlcolor=NavyBlue]{hyperref}
\usepackage[english,capitalise]{cleveref} 

\newcommand{\R}{\mathbb{R}}

\newcommand{\N}{\mathbb{N}}
\newcommand{\Hil}{\mathcal{H}}

\newcommand{\1}{\mathds{1}}
\newcommand{\Hmin}{H_{\mathrm{min}}}
\newcommand{\Hmine}[1][\varepsilon]{H_{\mathrm{min}}^{#1}}
\DeclareMathOperator{\diag}{diag}

\DeclareMathOperator{\id}{\mathord{\rm id}}
\DeclareMathOperator*{\argmin}{arg\,min}

\newcommand{\leak}{\lambda_{\text{EC}}}
\newcommand{\Sigmat}{\Sigma_{\text{test}}}

\newcommand*\dif{\mathop{}\!\mathrm{d}}

\newcommand{\kappal}[1]{\widetilde{\kappa}^{L}_{#1}}
\newcommand{\kappau}[1]{\widetilde{\kappa}^{U}_{#1}}

\newcommand{\Fobs}{F^\text{obs}}
\newcommand{\bstat}{b_\text{stat}}

\newcommand{\epsAT}{\varepsilon_{\text{AT}}}
\newcommand{\epsPA}{\varepsilon_{\text{PA}}}
\newcommand{\epsEV}{\varepsilon_{\text{EV}}}

\newcommand{\PAcost}{\theta(\epsPA,\epsEV)}
\newcommand{\Nsift}{N_\text{sift}^\text{obs}}
\newcommand{\nsift}{m}

\newcommand{\accsetadapt}[2]{\tilde{Q}_{#1,#2}}

\newcommand{\CEV}{C_\text{E,V}}

\theoremstyle{definition}
\newtheorem{defn}{Definition}

\theoremstyle{plain}
\newtheorem{thrm}[defn]{Theorem}

\theoremstyle{plain}
\newtheorem{lem}[defn]{Lemma}

\theoremstyle{plain}
\newtheorem{cor}[defn]{Corollary}

\theoremstyle{definition}
\newtheorem{prot}{Protocol}

\theoremstyle{remark}
\newtheorem{rem}{Remark}

\begin{document}
	
	\title{Improved finite-size effects in QKD protocols with applications to decoy-state QKD}
	
	\author{Lars Kamin}
	\email{lars.kamin@outlook.com}
	\affiliation{Institute for Quantum Computing and Department of Physics and Astronomy, University of Waterloo, Waterloo, Ontario N2L 3G1, Canada}
	
	\author{Devashish Tupkary}
	\email{tupkary.devashish@gmail.com}
	\affiliation{Institute for Quantum Computing and Department of Physics and Astronomy, University of Waterloo, Waterloo, Ontario N2L 3G1, Canada}
		
	\author{Norbert L{\"u}tkenhaus}
	\email{lutkenhaus.office@uwaterloo.ca}
	\affiliation{Institute for Quantum Computing and Department of Physics and Astronomy, University of Waterloo, Waterloo, Ontario N2L 3G1, Canada}
	
	\date{\today}
	
	\begin{abstract}
    We present a finite-size security proof for generic quantum key distribution protocols against independent and identically distributed collective attacks and extend it to coherent attacks using the postselection technique. This work introduces two significant improvements over previous results. First, we achieve tighter finite-size key rates by employing refined concentration inequalities in the acceptance testing phase. Second, we improve second-order correction terms in the key rate expression, by reducing them to scale with the number of sifted rounds rather than the total number of protocol rounds. We apply these advancements to compute finite-size key rates for a qubit and decoy-state BB84 protocol, accommodating arbitrary protocol parameters. Finally, we extend our finite-size security proof to coherent attacks and variable-length protocols and present our results for the decoy-state 4-6 protocol incorporating imperfections such as unequal intensity settings.
	\end{abstract}
	
	\maketitle

\section{Introduction}

Decoy-state methods using weak coherent pulse (WCP) sources \cite{Ma2005Phys.Rev.A,Lo2004Phys.Rev.Lett.,Hwang2002Phys.Rev.Lett.,Wang2005Phys.Rev.Lett.} have emerged as a key step toward realizing optical Quantum Key Distribution (QKD) protocols over long distances due to their practicality and feasibility. At the same time, real-world implementations of QKD require security proofs that address all aspects of executing a QKD protocol with a finite number of signals (finite-size effects).

Decoy-state protocols aim to estimate the adversary's (Eve) influence on the single-photon signals by collecting additional data through transmitting signals with different intensities. Due to the structure of states emitted by fully phase-randomized WCP sources, Eve cannot determine the signal's intensity solely from its photon number. Thus, the additional data can be used to perform a partial channel tomography and bound Eve's influence on single-photon signals.

Previous results for finite-size key rates of decoy-state protocols relied on the entropic uncertainty relation (EUR) \cite{Tomamichel2011Phys.Rev.Lett.,Lim2014Phys.Rev.A,Rusca2018Appl.Phys.Lett.} or the phase error correction approach \cite{Koashi2009,Hayashi2014}. These approaches require the assumption that the probability of a detection in Bob's setup is independent of the receiver's basis choice \cite[Sec 7.1]{Tomamichel2017Quantum}. If the assumptions are satisfied, these approaches are expected to lead to tighter finite-size key rates than the ones presented in this work. Although the basis-independent loss assumption has been partly removed by the recent work of Ref.~\cite{Tupkary2024}, the analysis there currently only applies to the BB84 protocol. In contrast, the approach presented in our work can be applied to generic QKD protocols with active basis choices. 

In this work, we provide an improved finite-size security proof that applies more broadly to unstructured QKD protocols, not only the WCP BB84 decoy. While the core of our work addresses the security proof against independent and identically distributed (IID) collective attacks, we extend it to coherent attacks using recent advancements \cite[Corollary 3.1]{nahar2024postselection} in the postselection technique \cite{Christandl2009Phys.Rev.Lett.}.

We improve upon the underlying IID security proof in two main ways. First, we derived methods which reduce the second-order corrections of the final key length expression such that they scale with the number of sifted signals, rather than the total number of signals sent. Second, we utilize an improved statistical analysis using individual (or entry-wise) constraints on each observation rather than relying on the full frequency vector of observations (as is done in \cite{George2020Phys.Rev.Res.}). The choice of entry-wise constraints was additionally motivated by the requirements of typical decoy-state methods. Both contributions lead to significant improvements to the key rates obtained.

We apply this refined IID proof to decoy-state protocols, which traditionally find upper and lower bounds on individual observations, such as error rates. Therefore, a finite-size analysis based on entry-wise constraints is more natural, and as we will show throughout this work, leads to higher secret key rates.

The structure of the paper is as follows: In \cref{sec:Security definitions and Notation}, we define key security concepts and introduce the notation used throughout the paper. \cref{sec:Improved Prepare-and-Measure Finite-size} develops a finite-size security proof for generic prepare-and-measure (P\&M) protocols, with several improvements on previous works. Afterwards, in \cref{sec:Decoy-State Methods in Finite-size regime}, we apply these findings to decoy-state protocols. Then, \cref{sec:Optimizing Security parameters} explores how to heuristically optimize the security parameters, ensuring a more robust and efficient implementation. In
\cref{sec:Optimising expected Key Rates} we calculate expected key rates \cite{Kanitschar2023PRXQuantum,Tupkary2024Phys.Rev.Res.}, which is crucial for real-world implementations. By optimizing the acceptance set based on the expected key rate, one can fine-tune the acceptance criteria for the honest implementations, thus maximizing the efficiency of the protocol.
Later, in \cref{sec:Examples}, we present two practical examples under a collective IID attack: the BB84 protocol with a perfect qubit source and its corresponding decoy-state protocol using a WCP source.
Finally, in \cref{sec:variable} and \cref{sec:Coherent Attacks} we extend these results to variable key lengths and coherent attacks. Additionally, we provide a decoy-state example of the reference-frame-independent (RFI) 4-6 protocol \cite{Laing2010Phys.Rev.A}, which incorporates both decoy states and a WCP source, further demonstrating the flexibility of our methods across different protocols.

\section{Security definitions \& Notation}\label{sec:Security definitions and Notation}
Let \(\sigma_{K_AK_BE^N}\) denote the output state after \(N\) rounds of a QKD protocol, where $K_A,K_B$ denotes Alice and Bob's final key registers, and $E^N$ denotes Eve's side information at the end of the QKD protocol, and includes all classical announcements. We make use of the following definitions of security as in \cite[Sec. III.B]{Portmann2022Rev.Mod.Phys.}.

\begin{defn}[$\varepsilon$-security of fixed-length protocols]\label{Def:eps security}
	Let \(\Omega_{\text{acc}}\) be the event of a QKD protocol not aborting. Then, a fixed-length QKD protocol is \emph{\(\varepsilon_{\text{sec}}\)-secure} if for all input states $\sigma_{A^NB^N}$, the resulting output state \(\sigma_{K_AK_BE^N}\) satisfies
	\begin{equation} \label{eq:security}
        \begin{split}
             \frac{1}{2} \Pr[\Omega_{\text{acc}}] \Bigg\lVert &\sigma_{K_AK_BE^N {|{\Omega_{\text{acc}}}}} \\ &- \sum_{k \in \{0,1\}^l } \frac{\dyad{kk}}{2^l} \otimes \sigma_{E^N {|{\Omega_{\text{acc}}}}} \Bigg\rVert_1 \leq \varepsilon_{\text{sec}}.
         \end{split}
	\end{equation}
   Furthermore, a fixed-length QKD protocol is \emph{\(\varepsilon_{\text{sct}}\)-secret} if
	\begin{equation}
		\begin{split}
			\frac{1}{2} \Pr[\Omega_{\text{acc}}] \Bigg\lVert &\sigma_{K_AE^N {|{\Omega_{\text{acc}}}}} \\ &- \sum_{k \in \{0,1\}^l} \frac{\dyad{k}}{2^l} \otimes \sigma_{E^N {|{\Omega_{\text{acc}}}}} \Bigg\rVert_1 \leq \varepsilon_{\text{sct}},
		\end{split}
	\end{equation}
	and \emph{\(\varepsilon_{\text{cor}}\)-correct} if
	\begin{equation}
		\Pr[K_A \neq K_B \wedge \Omega_{\text{acc}}] \leq \varepsilon_{\text{cor}}.
	\end{equation}
    Moreover, a QKD protocol that is \(\varepsilon_{\text{cor}}\)-correct and \(\varepsilon_{\text{sct}}\)-secret, is \(\varepsilon_{\text{sec}} = \varepsilon_{\text{sct}} + \varepsilon_{\text{cor}} \) secure \cite{Portmann2022Rev.Mod.Phys.}.
\end{defn}

For P\&M protocols, we restrict the input states \(\sigma_{A^NB^N}\) such that
\begin{equation}\label{eq:marginal state constraint sec def}
    \Tr_B^N\left[\sigma_{A^NB^N}\right] = \tau_A^{\otimes N},
\end{equation}
for some state \(\tau_A\). We make this restriction because we assume Alice's lab is inaccessible to Eve and thus the final shared state between Alice and Bob must satisfy the marginal state constraint given by Alice's device settings.

In all of this paper, we will use the following protocol parameters.

\textbf{Protocol Parameters (for fixed length):} \\
\begin{tabularx}{0.9\linewidth}{r c X}\label{Tab:Protocol parameters}
	\(N \in \N_0\) 						&:& 	Total number of signals sent\\
	\(l \in \N_0\)						&:&		length of final key\\
	\(\{\ket{s_i}\}_{i=1 \dots d_A}\) 	&:& 	States sent by Alice \\
    \(d_A\)                             &:&     Number of Alice's signal choices\\
	\(\leak\)							&:& 	length of bits exchanged during error correction  \\
	\(\varepsilon_{\text{AT}}\)			&:&		tolerated error during acceptance test \\
	\(\varepsilon_{\text{PA}}\)			&:&		tolerated error during privacy amplification \\
	\(\varepsilon_{\text{EV}}\)			&:&		tolerated error during error verification \\
	\(\bar{\varepsilon}\)				&:&		smoothing parameter \\
	\(\mathcal{Q}\)						&:& 	Acceptance set of accepted frequencies \\
	\(S_{\vec{\nu}}\)					&:& 	Feasible set \\
	\(\vec{\nu}\)						&:& 	variational bounds \\
	\(\vec{t}\)							&:&		tolerated fluctuation from expected frequencies \\
	\(\bar{F}\)							&:&		expected frequencies for test and generation rounds \\
	\(F^{\text{obs}}\)					&:&		observed frequencies from test and generation rounds \\
	\(\Sigmat\)							&:&		alphabet labeling the observations used for testing \\
	\(\Omega_{\text{AT}}\)				&:&		Event of passing the acceptance test \\
	\(\Omega_{\text{EV}}\)				&:&		Event of the error verification succeeding \\
	\(\Omega_{\text{acc}} = \Omega_{\text{AT}} \wedge \Omega_{\text{EV}}\)				&:&		Event of the protocol not aborting \\
\end{tabularx}

Note that in the literature, for example \cite{Renner2006,George2020Phys.Rev.Res.} \(\mu\) instead of \(\nu\) is often used for the variational bound, but to avoid confusion with the intensity settings \(\mu_i\) we chose a different notation.

\section{Improved Prepare-and-Measure Finite-size}\label{sec:Improved Prepare-and-Measure Finite-size}
In this section, we first improve finite-size key rates for generic P\&M protocol and later apply those results to decoy-state protocols in \cref{sec:Decoy-State Methods in Finite-size regime}. For completeness, we describe the steps of a generic P\&M protocol in \cref{Prot:PM Protocol}.

\subsection{Protocol Description}\label{subsec:Protocol Description}
\begin{prot}[Generic Prepare-and-Measure Protocol]\label{Prot:PM Protocol}
	We assume an authenticated classical channel as a resource for the following protocol. \\
	\textbf{Protocol steps:}
	\begin{enumerate}
		\item \textbf{State preparation and transmission:} In each round Alice decides with probability \(p(\text{gen})\) whether she sends a test or generation round. Then, Alice prepares one of \(d_A\) states \(\{\ket{s_i}\}_{i=1\dots d_A}\), which in general can depend on the choice of a test or generation round and she stores a label for her signal state in register \(X_i\) with alphabet \(\mathcal{X}\). Depending on her choices Alice computes her announcement (partition of all local available information) and stores it in register \(C_A\). Finally, Alice sends the signal state to Bob via a quantum channel. At this point under the IID assumption Alice, Bob and Eve share the state \(\rho_{\vec{X}\vec{B}\vec{E}} = \otimes_{i=1}^{N} \sigma_{ZYE}\).
		
		\item \textbf{Measurements:} Bob measures his received states, which can be phrased as applying one of his POVM elements \(\{M_k^B\}_{k=1 \dots d_B}\), and stores his results in a register \(Y_i\) with alphabet \(\mathcal{Y}\). Furthermore, he partitions his local available information and stores his public data containing the partition in \(C_B\).
	\end{enumerate}
	
	Alice and Bob repeat steps 1. and 2. \(N\) times
	\begin{enumerate}
		\setcounter{enumi}{2}
		\item \textbf{Public announcement and sifting:} Alice and Bob have a public discussion  and announce their public data contained in \(C\). (Note that one could also consider multi-round announcements, where \(C\) is a more complicated function of Alice and Bob's choices and measurement results.)
  
		Then, they sift their secret data based on their public communication. Including the public information the state shared has now the form \(\rho_{\vec{X}\vec{Y}\vec{E}\vec{C}}\).
        
		\item \textbf{Acceptance test (parameter estimation):} Based on their public data, Alice and Bob calculate the observed frequency of outcomes \(\Fobs\) in the test and generation rounds. Then, they test if \(F^{\text{obs}} \in \mathcal{Q} \), where \(\mathcal{Q}\) is the predetermined acceptance set. If \(F^{\text{obs}} \in \mathcal{Q} \) they proceed, otherwise, they abort the protocol. We will call the event of passing this step \(\Omega_{\text{AT}}\). (In this work we will consider acceptance sets which allow for a tolerance around some fixed value \(\bar{F}\), but this could be chosen differently.)
        
		\item \textbf{Key map:} Alice (or Bob) performs the key map \(g:\mathcal{XC} \rightarrow \mathcal{Z}\) and stores their raw key in register \(Z\) with alphabet \(\mathcal{Z}\). If Alice (or Bob) discard a round, due to e.g. it being a test round they map it to \(\perp\). The state after this step is \(\rho_{\vec{Z}\vec{Y}\vec{E}\vec{C} \vec{E} {|{\Omega_{\text{AT}}}}}\). \footnote{This is done to maintain a fixed length string going into privacy amplification for technical reasons. In an experimental implementation this step is not required if the techniques from \cite{Tupkary2024Phys.Rev.Res.} are used. In particular, Alice and Bob may \textit{discard} rounds as long as the positions of the discarded rounds can be computed from public announcements.}
        
 		\item \textbf{Error correction and verification:} Using the classical channel, Alice and Bob communicate to correct errors in their raw data. In the process, they send error correction information of length \(\leak\). After the error correction step, Alice sends hash of her raw key \(\vec{Z}\) of length \(\log(2/\varepsilon_{\text{EV}})\) to Bob. Bob then compares it to a hash of his own guess of \(\vec{Z}\) and aborts if the hashes do not match.
        In total they reveal \(\leak + \log(2/\epsEV) \) bits and the resulting state is \(\rho_{\vec{Z}\vec{Y}\vec{C} \CEV \vec{E}{|{\Omega_{\text{acc}}}} }\), which is now also conditioned on the event of both passing the error correction and verification step \(\Omega_{\text{EV}}\). Thus, $\Omega_\text{acc} = \Omega_\text{AT} \wedge \Omega_\text{EV}$.
        
		\item \textbf{Privacy Amplification:} Alice and Bob randomly choose a two-universal hash function mapping the length of the key bit register to \(l\) bits, and record their choice of hash function in register \(C_P\). Then, they apply it to their raw keys, producing their final secure keys of \emph{fixed} length \(l\). We denote the final state of the protocol as \(\rho_{K_A K_B\vec{C} C_P \CEV \vec{E}{|{\Omega_{\text{acc}}}} }\), where \(K_A\) and \(K_B\) are Alice's and Bob's final keys.
	\end{enumerate}
\end{prot}

\subsection{Acceptance Test and Resulting Entropy Bounds}\label{subsec:Acceptance Test}

After setting up the protocol, we will work towards the security proof of \cref{Thrm:PM QKD security} of a generic P\&M protocol, which is proven in \cref{subsec:Security Proof against IID Attacks}. However, first, we show the two improvements of the upcoming \cref{Lem:Bound Smooth Min-Entropy sifted signals} and \cref{Thrm:Secrecy outside Feasible Set}, which are a direct consequence of the acceptance test and its underlying acceptance and feasible sets, defined in this section. 

For this purpose, let us briefly introduce some common notation used throughout the paper, starting with the description of Alice's source. In the source replacement scheme \cite{Bennett1992Phys.Rev.Lett., Ferenczi2012Phys.Rev.A}, the states Alice sends can be rewritten in the following form
\begin{equation}
	\ket{\psi}_{AA'} = \sum_{a=1}^{d_A}  \sqrt{p(a)} \ket{a}_A \ket{s_a}_{A'},
\end{equation}
where \(p(a) = p(\text{gen}) p(a|\text{gen}) + p(\text{test}) p(a|\text{test})\) is the total probability of Alice sending state \(\ket{s_a}\). In this equivalent picture, Alice now acts with the POVM elements \(\{\ketbra{a}\}\) on her system \(A\).

Furthermore, the single round state after Eve's channel has the form
\begin{equation}
	\sigma_{ABE} = \left( \id_A \circ \:\mathcal{E}_{A' \rightarrow B} \right) \left[\dyad{\psi}_{AA'} \right].
\end{equation}
Under the assumption of an IID collective attack, the final \(N\)-round state can thus be written as
\begin{equation}
	\rho_{\vec{A}\vec{B}\vec{E}} = \bigotimes_{i=1}^{N} \sigma_{ABE},
\end{equation}
where the vector is used to indicate that the register contains multiple rounds. We will use this notation extensively in the following. Due to the source replacement scheme, we only have to consider states which satisfy 
\begin{equation} \label{eq:marginal}
    \Tr_{A'}[\ketbra{\psi}] = \sigma_{A},
\end{equation}
because Eve has no access to Alice's lab, as previously described below the security definition \cref{eq:marginal state constraint sec def}.

Now, we are in a position to discuss some important constructions such as the acceptance and feasible set, necessary for the security proof. Before we formally introduce them in \cref{Thrm:Secrecy outside Feasible Set}, let us start by discussing their role in the security proof to gain some intuition. The acceptance set is used in the acceptance test of \cref{Prot:PM Protocol} to make the decision whether we accept the observations or abort the protocol. The acceptance sets we will consider in this work are always characterized by the expected frequency \(\bar{F}\) and an allowed tolerance \(\vec{t}\), i.e. we accept if the observed frequencies lie in some tolerance around the expected frequencies.

The feasible set \(S_{\vec{\nu}}\) splits the set of all density matrices (with the marginal constraint originating from the source replacement scheme, \cref{eq:marginal}) into the states in \(S_{\vec{\nu}}\) and outside of \(S_{\vec{\nu}}\). For the states outside of \(S_{\vec{\nu}}\) we would want to show that they will be accepted with low probability, i.e. \( \Pr[\Omega_{\text{acc}}] \leq \varepsilon_{\text{AT}} \) and the protocol will abort. In \cref{Thrm:Secrecy outside Feasible Set}, we will construct the acceptance set such that the probability of the protocol accepting any state outside of the feasible set \(S_{\vec{\nu}}\) is small. 

\begin{figure}[ht]
    \centering
    \includegraphics[width=1.2\linewidth]{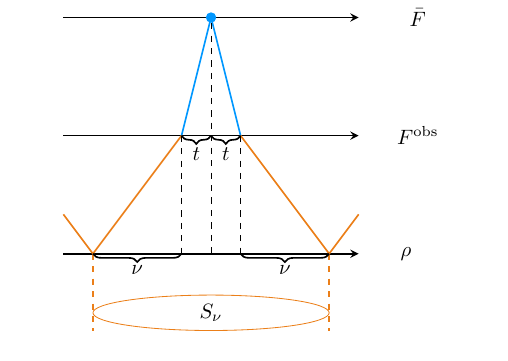}
    \caption{One dimensional representation of the construction of the feasible set from the expected frequency \(\bar{F}\) with equal upper and lower \(\nu\).}
    \label{fig:feasibleset}
\end{figure}

A one dimensional representation of the construction of the feasible set is depicted in \cref{fig:feasibleset}. It is determined by the expected frequency \(\bar{F}\), the tolerated fluctuation \(t\) and the parameter \(\nu\) characterizing the uncertainty due to finite-size effects. As shown in \cref{fig:feasibleset} in \(S_{\vec{\nu}}\) we need to consider all density matrices for which there exists an observed frequency which is at most \(\nu\) away from the true, but unknown value \(\Tr[\Gamma \rho]\), and still within the tolerated fluctuation from \(\bar{F}\). All those density matrices make up the feasible set \(S_{\vec{\nu}}\).

For the states in \(S_{\vec{\nu}}\) we will later derive the secure key length in \cref{Thrm:PM QKD security}. Hence, the feasible set allows us to split the set of density matrices and thus, split the security proof by densities matrices in \(S_{\vec{\nu}}\) and density matrices outside of \(S_{\vec{\nu}}\). 

Now, we will start with the security proof by proving secrecy for states outside the feasible set \(S_{\vec{\nu}}\) in \cref{Thrm:Secrecy outside Feasible Set}.

	\begin{thrm}[Accept Probability outside Feasible Set]\label{Thrm:Secrecy outside Feasible Set}
		Let \(\Sigma = \Sigmat \cup \{\text{sift} \; , \perp \}\) be the alphabet of observations from test and generation rounds and \(\{\Gamma_{k}\}_{k \in \Sigma}\) the corresponding POVM elements. Furthermore, let the acceptance set \(\mathcal{Q}\) and the feasible set \(S_{\vec{\nu}}\) of \cref{Prot:PM Protocol} be defined as
		\begin{align} \label{eq:QSsets}
			\mathcal{Q} &:= \mathcal{Q}_1 \cap \mathcal{Q}_2, \\
			\mathcal{Q}_1 &:= \{F^{\text{obs}} \in \mathcal{P}(\Sigma) \; |  \abs{F^{\text{obs}}_k - \bar{F}_k } \leq t_k \; \forall k \in \Sigmat \} \\
			\mathcal{Q}_2 &:= \{ F^{\text{obs}} \in \mathcal{P}(\Sigma) \; | \abs{ F^{\text{obs}}_{\text{sift}} - \frac{n_{\text{sift}}}{N} } \leq  t_{\text{sift}} \}, \label{eq:Q2} \\
			S_{\vec{\nu}} &:= \{\rho \in S_{\circ}\left(\Hil_A \otimes \Hil_B \right)| \exists F^{\text{obs}} \in \mathcal{Q}, \text{s.t. } \forall k \in \Sigma \nonumber \\ 
			&\hspace{20pt} F^{\text{obs}}_k - \nu_k^L \leq \Tr[ \Gamma_k \rho ] \leq F^{\text{obs}}_k  + \nu_k^U \: , \\
			&\hspace{20pt}\rho_A = \Tr_{A'}[\ketbra{\psi}]\} \nonumber .
		\end{align}
		Furthermore, define \(n_{\text{sift}}\) in \(\mathcal{Q}_2\) as the total expected click count in the generation rounds surviving sifting. Then, the variational bounds \(\nu_k^{L/U}\) can be chosen as
        \begin{equation}
			\nu_k^{U/L} = \min_{\nu'} \{\nu' \in [0,1] \; | \; C_k^{U/L}(\nu') = \varepsilon_{AT} \},
		\end{equation}
		such that for all \(\rho_{\vec{A}\vec{B}\vec{E}} = \bigotimes_{i=1}^{N} \sigma_{ABE}\) and \(\sigma \notin S_{\vec{\nu}}\),
		\begin{equation}
			\Pr[\Omega_\text{acc}] \leq \varepsilon_{\text{AT}}.
		\end{equation}
		Furthermore, each \(C_k^{L} \) and \(C_k^{U} \) is defined by 
		\begin{align}
			C_k^{L}(\nu) &:= 1- I_{1-\left(F^L_k - \nu \right)} \left( N + 1 - \lceil N F^{L}_k \rceil , \lceil N F^{L}_k  \rceil  \right) , \\
			C_k^{U}(\nu) &:= I_{1-\left(F^U_k +\nu \right)} \left( N - \lfloor N F^U_k  \rfloor , \lfloor N F^U_k \rfloor +1 \right) ,
		\end{align}
		where \(F^{L/U}_k = \bar{F}_k \mp t_k\) and \(I_x[a,b]\) is the incomplete beta function.
	\end{thrm}
Although it is not completely obvious at this point our acceptance set \(\mathcal{Q}\) uses observations from both test and generation rounds. The subset \(\mathcal{Q}_1\) intuitively serves the purpose of constraining Eve's attack by observing the outcomes of the test rounds. Even simpler, this could be viewed as if one had observed the error rate only and had made an accept decision based on that. On the other hand, the set \(\mathcal{Q}_2\) serves the main purpose of reducing the finite-size corrections.
This can be understood as follows, if in key generation rounds Alice and Bob announce which signals survived the sifting stage, one can use this additional information as well to achieve tighter finite-size bounds. Therefore, we added the events \(\{\text{sift} \; , \perp \}\) to the alphabet of test rounds \(\Sigmat\) to indicate that a signal in a key generation survived sifting or not, respectively. Later, in \cref{Lem:Bound Smooth Min-Entropy sifted signals} we will utilize this additional information contained in the set \(\mathcal{Q}_2\) for a better bound on the smooth min-entropy by reducing the finite-size corrections.

Finally, the appearance of the incomplete beta function might seem surprising, but under a collective IID attack, we know that each individual observation is binomial distributed and the incomplete beta function is the cumulative distribution function of the binomial distribution.

\begin{proof}
    For simplicity define \(p_k := \Tr[\Gamma_k \sigma]\) and let \(X_1, \dots ,X_{\abs{\Sigma}}\) be the random variables corresponding to the click statistics for each POVM element \(\Gamma_k\), i.e. \(\frac{X_k}{N} = F^{\text{obs}}_k\). 
    
    We need to show that for all \(\sigma \notin S_{\vec{\nu}} \), \(\Pr[\Omega_{\text{acc}}] \leq \varepsilon_{\text{AT}}\). Since \(\Omega_{\text{acc}} = \Omega_{\text{AT}} \wedge \Omega_{\text{EV}}\), we find an upper bound as
    \begin{align}
        &\Pr[\Omega_{\text{acc}}] \leq \Pr[\Omega_{\text{AT} }] = \Pr[F^{\text{obs}} \in \mathcal{Q} ] \\
        &\leq \Pr[\abs{F^{\text{obs}}_k - \bar{F}_k} \leq t_k] \: \forall k \in \Sigma,
    \end{align}
    which holds due to the union bound. For each \(k\), this can be split in two cases as
    \begin{equation}
        \begin{split}
            &  \Pr[\abs{F^{\text{obs}}_k - \bar{F}_k} \leq t_k] \\ 
            \leq &\max\left\{ \Pr[\abs{F^{\text{obs}}_k - \bar{F}_k} \leq t_k]\Big|_{p_k > \bar{F}_k +t_k +\nu_k^{U}} \right.,  \\  &\hspace{27pt} \left.\Pr[\abs{F^{\text{obs}}_k - \bar{F}_k} \leq t_k]\Big|_{p_k < \bar{F}_k -t_k -\nu_k^{L}} \right\}.
        \end{split}
    \end{equation}
    As can be seen in \cref{fig:feasibleset} only these two cases are possible for \(\sigma \notin S_{\vec{\nu}}\), as otherwise the true value \(p_k\) could be generated from states inside the feasible set. Furthermore, since \(\sigma \notin S_{\vec{\nu}} \), there must exist at least one \(k\) such that either \(p_k > \bar{F}_k +t_k +\nu_k^{U}\) or \(p_k < \bar{F}_k -t_k -\nu_k^{L}\).
    
    First, consider \(p_k > \bar{F}_k +t_k +\nu\), for some \(\nu\) which we aim to find and will be denoted by \(\nu_k^{U}\). For a single index \(k\) we find
    \begin{align}
        &\Pr[\abs{F^{\text{obs}}_k - \bar{F}_k} \leq t_k]\Big|_{p_k > \bar{F}_k +t_k+\nu} \\
        &= \Pr[F^{\text{obs}}_k \leq \bar{F}_k + t_k]\Big|_{p_k > \bar{F}_k +t_k +\nu} \\
        &\hspace{10pt} - \Pr[F^{\text{obs}}_k < \bar{F}_k - t_k]\Big|_{p_k > \bar{F}_k +t_k +\nu} \\
        &\leq \Pr[F^{\text{obs}}_k \leq \bar{F}_k + t_k]\Big|_{p_k > \bar{F}_k +t_k +\nu}
    \end{align}
    Under a collective IID attack, we know that each \(X_k \sim \mathrm{Bin}_N(p_k)\). Then, identifying \(F^{\text{obs}}_k = \frac{X_k}{N} \), lets us write
    \begin{equation}
        F^{\text{obs}}_k \leq \bar{F}_k + t \Longleftrightarrow X_k \leq \lfloor N \bar{F}_k + N t_k\rfloor .
    \end{equation}
    We can use the fact that \(X_k \sim \mathrm{Bin}_N(p_k)\) because observing outcome \(k\) alone is a \(N\)-times repeated Bernoulli trial. Then, note that the incomplete beta function \(I_x(a,b)\) is the cummulative distribution function of the binomial distribution, which allows us to write \(\Pr[X_k \leq m ] = I_{1 - p_k}(N-m,m+1) \). Thus,
    \begin{align}
        &\Pr[\abs{F^{\text{obs}}_k - \bar{F}_k} \leq t_k]\Big|_{p_k > \bar{F}_k +t +\nu} \\
        \leq &I_{1-p_k} \left( N - \lfloor N\left( \bar{F}_k +t_k\right)  \rfloor , \lfloor N\left( \bar{F}_k +t_k\right)  \rfloor +1  \right) \\
        \leq &I_{1-\left(F^U_k +\nu \right)} \left( N - \lfloor N F^U_k  \rfloor , \lfloor N F^U_k  \rfloor +1  \right) \\
        = &C_k^{U}(\nu),
    \end{align}
    where for the second inequality, we used that \(p_k > \bar{F}_k +t_k +\nu = F^U_k + \nu \) and that \(I_x(a,b)\) is monotonically increasing in \(x\). A similar consideration can be done for \(p_k < \bar{F}_k -t_k -\nu = F^L_k - \nu\) and in summary one finds
    \begin{align}
      C_k^{U}(\nu) &\geq \Pr[\abs{F^{\text{obs}}_k - \bar{F}_k } \leq t_k]\Big|_{p_k > F_k^U +\nu}, \\
      C_k^{L}(\nu) &\geq \Pr[\abs{F^{\text{obs}}_k - \bar{F}_k} \leq t_k]\Big|_{p_k < F_k^L -\nu}.
    \end{align}
    For each \(k\) one can pick the smallest  \(\nu_k^L\) and \(\nu_k^U\) such that \(C_k^{L}(\nu_k^L)= \varepsilon_{AT}\) and \(C_k^{U}(\nu_k^U) = \varepsilon_{AT}\), and it holds \( \Pr[\abs{F^{\text{obs}}_k - \bar{F}_k} \leq t_k] \leq \varepsilon_{AT}\). Thus, we define each \(\nu_k^{U/L}\) as
    \begin{equation}
        \nu_k^{U/L} = \min_{\nu'} \{\nu' \in [0,1] |  C_k^{U/L}(\nu') = \varepsilon_{AT} \},
    \end{equation}
as in the theorem statement. Therefore, we find for \(\sigma \notin S_{\vec{\nu}}\),
    \begin{equation} \label{eq:concentration}
        \begin{split}
            \Pr[\Omega_{\text{acc}}] &\leq \max_k \Pr[\abs{F^{\text{obs}}_{k} - \bar{F}_{k}} \leq t_k]  \\
            &\leq \max_k\{C_k^{L}(\nu_k^L) , C_k^{U}(\nu_k^U) \} \leq \varepsilon_{AT},
        \end{split}
    \end{equation}
due to the definition of each \(\nu_k^{U/L}\).
\end{proof}

As stated previously, our acceptance set \(\mathcal{Q}\) uses some information of the generation rounds. 
Now, we will utilize these additional constraints originating from \(\mathcal{Q}_2\) to find a better lower bound on the smooth min-entropy, which will ultimately reduce the finite-size corrections.

In preparation for the next \cref{Lem:Bound Smooth Min-Entropy sifted signals}, let us split off the event \( \Omega_{\text{sift}} = \{ \Fobs \in \mathcal{Q}_2 \} \subseteq \Omega_{\text{AT}} \) corresponding to having sufficient sifted signals during generation rounds. Furthermore, let us introduce a register \(\vec{D}\) storing the information of a detection of a sifted signal. Here, we assign \(D_i = 1\) if the signal \(i\) was part of a generation round, detected and sifted, otherwise we assign \(D_i = 0\).
	
\begin{lem}[Bounds on the Smooth Min-Entropy for States in the Feasible Set]\label{Lem:Bound Smooth Min-Entropy sifted signals}
    Let \(\vec{Z},\vec{Y},\vec{C},\vec{E},\vec{D}\) be the registers corresponding to Alice's and Bob's raw keys, public announcements, Eve's quantum side information and the locations of the sifted detections, respectively. We label \(D_i=1\) if a signal in a generation round passed sifting in round \(i\) and \(D_i=0\) otherwise. Thus, \(D_i\) is a deterministic function of \(C_i\). For \(\rho_{\vec{Z}\vec{Y}\vec{C}\vec{E}\vec{D}}\) such that
    \begin{equation}        \rho_{\vec{Z}\vec{Y}\vec{C}\vec{E}\vec{D}} = \bigotimes_{i=1}^{N} \sigma_{ZYCED},
    \end{equation}
    and \(\sigma \in S_{\vec{\nu}}\), with \(S_{\vec{\nu}}\) and \(\mathcal{Q}\) defined as in \cref{Thrm:Secrecy outside Feasible Set}, it holds
    \begin{equation}
        \Hmine[\bar{\varepsilon}](\vec{Z}|\vec{C}\vec{E}\vec{D})_{\rho_{|{\Omega_{\text{sift}}}}} \geq \Hmine[\bar{\varepsilon}](\vec{Z}_{\text{sift}}| \vec{C}_{\text{sift}} \vec{E}_{\text{sift}} )_{\tau^{\text{sift}}_{|{\vec{D}=\vec{1}}}},
    \end{equation}
    where
    \begin{equation}
        \tau^{\text{sift}}_{\vec{Z}_{\text{sift}}\vec{Y}_{\text{sift}}\vec{C}_{\text{sift}}\vec{E}_{\text{sift}}| \vec{D}=\vec{1}} := \bigotimes_{i=1}^{\lfloor n_{\text{sift}} -Nt_{\text{sift}} \rfloor} \sigma_{ZYCE|D=1}.
    \end{equation}
\end{lem}
\begin{proof}
    The proof proceeds with the following steps, first, we will show that conditioning on a specific number of signals passing sifting maintains the IID structure. Then, we will find a lower bound on the smooth min-entropy by using the constraints from set \(\mathcal{Q}_2\).
    
    Let \(\Omega_{\vec{d}} \subseteq \Omega_{\text{sift}} =\{ \Fobs \in \mathcal{Q}_2 \}\) be the event of a single string \(\vec{d}\), i.e. a particular instance of sifted detection events. By \cref{Lem:Conditioning on Classical Register} in \cref{App:Technical Definitions and Lemmas}, we can lower bound the smooth min-entropy by
    \begin{equation}\label{eq:min over d smooth min entropy}
        \Hmine[\bar{\varepsilon}](\vec{Z}|\vec{E}\vec{D})_{\rho_{|{\Omega_{\text{sift}}}}} \geq \min_{\substack{\Omega_{\vec{d}} \in \Omega_{\text{sift}} \\ \text{s.t. } \Pr[\Omega_{\vec{d}}]\neq 0 }} \Hmine[\bar{\varepsilon}](\vec{Z}|\vec{E}\vec{D})_{\rho_{|{\Omega_{\vec{d}}}}} .
    \end{equation}
    Next, consider the collective IID assumption again, where we have
    \begin{equation}
        \rho_{\vec{Z}\vec{Y}\vec{C}\vec{E}\vec{D}} = \bigotimes_{i=1}^{N} \sigma_{ZYCED}.
    \end{equation}
    Then, for all \(\vec{d}\) and \(\rho_{\vec{Z}\vec{Y}\vec{C}\vec{E}\vec{D}{|{\Omega_{\vec{d}}}}}\) one can write
    \begin{equation}
        \rho_{\vec{Z}\vec{Y}\vec{C}\vec{E}\vec{D}{|{\Omega_{\vec{d}}}}} = \sigma_{ZYCE|D = d_1}^{(1)} \otimes \dots \otimes \sigma_{ZYCE|D = d_N}^{(N)}.
    \end{equation}
    For all \(\vec{d}\), we know that at least \(n_{\text{sift}} -Nt_{\text{sift}}\) signals are detected and sifted. The remaining number of signals might contain detected signals but we need to omit those in order to achieve a fixed length of the raw key register. (Note that this is only required in this proof due to subtleties of handling variable-length registers. In practice, we do not need to omit extra rounds in the raw key register due to \cite[Section VII]{Tupkary2024Phys.Rev.Res.}.) 
    
    If one does not detect a signal or sifts it out, no key is produced from this particular signal, and we assign \(D_i=0\) . Therefore, we split \(\vec{Z}\vec{C}\) and \(\vec{E}\) in \(\vec{Z}=\vec{Z}_{\text{sift}}\vec{Z}_{\text{no}}\), \(\vec{C}=\vec{C}_{\text{sift}}\vec{C}_{\text{no}}\) and \(\vec{E}=\vec{E}_{\text{sift}}\vec{E}_{\text{no}}\), where all strings in \(\vec{Z}_{\text{sift}}\), \(\vec{E}_{\text{sift}}\) have fixed length \(n_{\text{sift}} -Nt_{\text{sift}}\). Thus, we find up to reordering
    \begin{equation}
        \begin{split}                   \rho_{\vec{Z}\vec{Y}\vec{C}\vec{E}\vec{D}{|{\Omega_{\vec{d}}}}} = & \left( \bigotimes_{i=1}^{\lfloor n_{\text{sift}} -Nt_{\text{sift}} \rfloor} \sigma_{ZYCE|D=1} \right) \\ 
            &\quad \otimes \tau^{\text{no}}_{\vec{Z}_{\text{no}}\vec{Y}_{\text{no}}\vec{C}_{\text{no}} \vec{E}_{\text{no}} \vec{D}_{\text{no}}},
        \end{split}
    \end{equation}
    where \(\tau^{\text{no}}\) contains all remaining rounds. All registers in this expression have a fixed length. For simplicity, let us also define the state
    \begin{align}
        \tau^{\text{sift}}_{\vec{Z}_{\text{sift}}\vec{Y}_{\text{sift}}\vec{C}_{\text{sift}}\vec{E}_{\text{sift}}| \vec{D}=\vec{1}} &:= \bigotimes_{i=1}^{\lfloor n_{\text{sift}} -Nt_{\text{sift}} \rfloor} \sigma_{ZYCE|D=1}.
    \end{align}
 Then, we can rewrite \(\rho_{\vec{Z}\vec{Y}\vec{C}\vec{E}\vec{D}{|{\Omega_{\vec{d}}}}}\) as
    \begin{align}
        &\rho_{\vec{Z}\vec{Y}\vec{C}\vec{E}\vec{D}{|{\Omega_{\vec{d}}}}} = \rho_{\vec{Z}_{\text{sift}}\vec{Z}_{\text{no}}\vec{Y}\vec{C}\vec{E}_{\text{sift}}\vec{E}_{\text{no}}\vec{D}{|{\Omega_{\vec{d}}}}} \\
        &=  \tau^{\text{sift}}_{\vec{Z}_{\text{sift}}\vec{Y}_{\text{sift}}\vec{C}_{\text{sift}}\vec{E}_{\text{sift}}|\vec{D}=\vec{1}} \otimes \tau^{\text{no}}_{\vec{Z}_{\text{no}}\vec{Y}_{\text{no}}\vec{C}_{\text{no}}\vec{E}_{\text{no} }\vec{D}_{\text{no}}}.
    \end{align}
    
    Since both \(\vec{Z}_{\text{sift}}\) and \(\vec{Z}_{\text{no}}\) are classical, we can bound the smooth min-entropy regarding \(\vec{Z}_{\text{no}}\) by zero \cite[Lemma 6.7]{Tomamichel2016} and thus,
    \begin{equation}
         \Hmine[\bar{\varepsilon}](\vec{Z}|\vec{C}\vec{E}\vec{D})_{\rho_{|{\Omega_{\vec{d}}}}} \geq \Hmine[\bar{\varepsilon}](\vec{Z}_{\text{sift}}|\vec{C}\vec{E}\vec{D})_{\rho_{|{\Omega_{\vec{d}}}}}.
    \end{equation}
    Furthermore, since \(\vec{C}_{\text{no}}\vec{E}_{\text{no} }\) are uncorrelated with \(\vec{Z}_{\text{sift}}\) one can apply the data processing inequality \cite[Theorem 6.2]{Tomamichel2016} and remove them from the conditioning register. Hence, we find in summary,
    \begin{equation}
        \Hmine[\bar{\varepsilon}](\vec{Z}|\vec{C}\vec{E}\vec{D})_{\rho_{|{\Omega_{\vec{d}}}}} \geq \Hmine[\bar{\varepsilon}](\vec{Z}_{\text{sift}}|\vec{C}_{\text{sift}} \vec{E}_{\text{sift}})_{\tau^{\text{sift}}_{|{\vec{D}=\vec{1}}}},
    \end{equation}
    for all \(\Omega_{\vec{d}} \in \Omega_{\text{sift}}\). Since this holds for all \(\Omega_{\vec{d}}\), it especially holds for the minimum in \cref{eq:min over d smooth min entropy} and we reach the theorem statement
    \begin{align}
        &\Hmine[\bar{\varepsilon}](\vec{Z}|\vec{C}\vec{E}\vec{D})_{\rho_{|{\Omega_{\text{sift}}}}} \\
        &\geq \min_{\substack{\Omega_{\vec{d}} \in \Omega_{\text{sift}} \\ \text{s.t. } \Pr[\Omega_{\vec{d}}]\neq 0 }} \Hmine[\bar{\varepsilon}](\vec{Z}|\vec{E}\vec{D})_{\rho_{|{\Omega_{\vec{d}}}}} \\
        &\geq \Hmine[\bar{\varepsilon}](\vec{Z}_{\text{sift}}|\vec{C}_{\text{sift}} \vec{E}_{\text{sift}})_{\tau^{\text{sift}}_{|{\vec{D}=\vec{1}}}}.
    \end{align}
\end{proof}

To gain some insight into the usefulness of this lemma, consider the following situation. In most QKD protocols, one sifts certain events (and discards the remaining outcomes). Without applying \cref{Lem:Bound Smooth Min-Entropy sifted signals}, one would be required to map these outcomes to e.g. \mbox{\(0\)} to keep the input to the privacy amplification procedure of constant length.

Now, one could apply the applying the asymptotic equipartition property directly \cite[Corollary 4.10]{Dupuis2020Commun.Math.Phys.}, achieving a scaling of the following form 
\begin{equation}
     \Hmine[\bar{\varepsilon}](\vec{Z}|\vec{C}\vec{E})_{\rho_{ZCE}^{\otimes N}} \geq N H(Z|EC)_{\rho} - \sqrt{N}g(\bar{\varepsilon}),
\end{equation}
where \(g\) is a correction term depending on \(\bar{\varepsilon}\) and the dimension of \(Z\) only. However, the von Neumann entropy still scales with the number of sifted signals, whereas the correction term is constant in the total number of signals sent. Hence, at some point the correction term will dominate; leading to zero key rate.

Our \cref{Lem:Bound Smooth Min-Entropy sifted signals} improves on this situation by having a correction factor which also depends on the sifted signals. Thus, key rates to much higher losses are possible.

\subsection{Security Proof against IID Attacks for Generic P\&M Protocols}\label{subsec:Security Proof against IID Attacks}
Next, continuing with our security proof, in \cref{Thrm:PM QKD security} we tie all of our previous theorems and lemmas together and prove the security of the prepare-and-measure \cref{Prot:PM Protocol} against IID attacks. In particular, we will make use the definition of the acceptance set from \cref{Thrm:Secrecy outside Feasible Set}. Furthermore, we will apply \cref{Lem:Bound Smooth Min-Entropy sifted signals} to reduce the influence of second order corrections which achieves e.g. better scaling with loss. Most of our improvements in terms of key rate stem from this refined security analysis for IID attacks, which will later be lifted to coherent attacks (see \cref{sec:Coherent Attacks}).

\begin{thrm}[Secure Key Rate of a Prepare-and-Measure QKD Protocol]\label{Thrm:PM QKD security}
    Fix the parameters of \cref{Prot:PM Protocol} and define \(\mathcal{Q}\) and \(S_{\vec{\nu}}\) as in \cref{Thrm:Secrecy outside Feasible Set}. Assuming a collective IID attack, the QKD protocol is \(\varepsilon_{\text{sec}} = \varepsilon_{\text{EV}} +  \max\{\varepsilon_{\text{AT}}, \varepsilon_{\text{PA}} + 2\bar{\varepsilon} \}\) secure, if the key length \(l\), in the case of not aborting, is chosen to be
\begin{equation} \label{eq:lvalue}
    \begin{split}
        l \leq \; &\lfloor n_{\text{sift}} - Nt_{\text{sift}} \rfloor \min_{\rho \in S_{\vec{\nu}} } \frac{H(Z|EC)_{\rho}}{\Pr(\text{sift})} - \leak - \log(2/\epsEV)\\
        &- \sqrt{\lfloor n_{\text{sift}}-Nt_{\text{sift}} \rfloor} \Delta\left(\bar{\varepsilon} \right) - 2\log(\frac{1}{2\varepsilon_{\text{PA}}}),
    \end{split}
\end{equation}
where
\begin{align}
    \Delta(\bar{\varepsilon}) &= 2\log(1+2\dim(Z)) \sqrt{\log(\frac{2}{\bar{\varepsilon}^2})}.
\end{align}
\end{thrm}
Before we state the proof, let us note a few points. In this theorem, the label ``\(\text{sift}\)" refers to signals passing the sifting step in key generation rounds. 

Additionally, in theoretical analysis the leakage term \(\leak\) is typically bounded as \cite{Portmann2022Rev.Mod.Phys.}
\begin{equation}
    \begin{split}
        \leak = &N \Pr(\text{sift}) f_{EC} H(X| Y; \text{sift}) 
    \end{split}
\end{equation}
which we will also do for our examples later on to estimate the error correction cost. However, in practice it should be chosen corresponding to the maximum data exchange allowed during the error correction step. It is important to note that the security proof still holds if the leakage term is changed.

Finally, the fraction in \cref{Thrm:PM QKD security} in the optimization problem makes the optimization problem non-convex. However, one can utilize the constraints implied by the acceptance set \(\mathcal{Q}_2\), see \cref{eq:Q2}, to achieve a simple lower bound given by 
\begin{equation}\label{eq:Lower bound psift}
    \min_{\rho \in S_{\vec{\nu}} } \frac{H(Z|EC)_{\rho}}{\Pr(\text{sift}) } \geq \frac{\min_{\rho \in S_{\vec{\nu}} } H(Z|EC)_{\rho}}{\bar{F}_\text{sift} + t_{\text{sift}} + \nu^U_{\text{sift}}},
\end{equation}
which we will later use as well in \cref{sec:Decoy-State Methods in Finite-size regime} and \cref{sec:Examples}. 

\begin{proof}
As stated in \cref{sec:Security definitions and Notation}, secrecy and correctness imply security. Therefore, we prove each part separately.

For correctness, let us consider Bob's guess \(\vec{Z}_B\) of Alice's raw key \(\vec{Z}\).
Then, correctness follows immediately as a property of hashing used in the error verification step since
    \begin{equation}
    \begin{split}           
        &\Pr[K_A \neq K_B \wedge \Omega_{\text{acc}}] \leq \Pr[K_A \neq K_B \wedge \Omega_{\text{EV}}] \\ &\leq \Pr[(\vec{Z} \neq \vec{Z}_B) \wedge \Omega_{\text{EV}}] \leq \Pr[\Omega_{\text{EV}}| (\vec{Z} \neq \vec{Z}_B) ] \\
        &\leq \varepsilon_{\text{EV}}.
        \end{split}
    \end{equation}

Turning our attention to secrecy two different cases have to be considered. First note, that under the collective IID assumption \(\rho_{\vec{Z}\vec{Y}\vec{E}\vec{D}}\) can be written as
    \begin{equation}
        \rho_{\vec{Z}\vec{Y}\vec{E}\vec{C} C_{EC}\vec{D}} = \bigotimes_{i=1}^{N} \sigma_{ZYE\vec{C} C_{EC}D}.
    \end{equation}
Thus, we split the proof in two cases \(\sigma \in S_{\vec{\nu}}\) and \(\sigma \notin S_{\vec{\nu}}\).

In the case of \(\sigma \notin S_{\vec{\nu}}\), we can apply \cref{Thrm:Secrecy outside Feasible Set} and it holds
    \begin{align}
        \frac{1}{2} &\Pr[\Omega_{\text{acc}}] \; \norm{\omega_{K_AS^HE_{|_{\Omega_{\text{acc}}}}} - \chi_{K_A} \otimes \omega_{S^HE_{|_{\Omega_{\text{acc}}}}} } \\
        \leq &\Pr[\Omega_{\text{AT}}] \leq \varepsilon_{\text{AT}},
    \end{align}
where we applied the triangle inequality on the \(1\)-norm in the first inequality. 

Next, consider \(\sigma \in S_{\vec{\nu}} \), where we will make use of \cref{Cor:Conditional LHL}. We split \(\Omega_{\text{AT}}\) into two parts \(\Omega_{\text{AT}} = \Omega_{\text{sift}} \wedge \Omega'_{\text{AT}}\), where \(\Omega_{\text{sift}} = \{F_{\text{sift}}^{\text{obs}} \in \mathcal{Q}_2\}\), i.e. the number of detected signals in the generation rounds used for key generation varies at most \(Nt_{\text{sift}}\) from \(n_{\text{sift}}\). The event \(\Omega'_{\text{AT}}\) corresponds to the remaining conditions in \(\Omega_{\text{AT}}\).

Furthermore, let \(\vec{E}'=\vec{E} \vec{C} C_{EC}\vec{D}\) be Eve's side information which contains Eve's quantum side information, public announcements, leakage during error correction and \(\vec{D}\) corresponding to the information of a sifted and detected signals in the key generation rounds.
As in \cref{Lem:Bound Smooth Min-Entropy sifted signals} we assign \(D_i = 1\) if a signal was detected and survived sifting in round \(i\) which was also a generation round, otherwise \(D_i = 0\).
Recall that $\Omega_\text{sift}$ is a necessary condition for the protocol to accept ($\Omega_{\text{acc}}$). Then, by \cref{Cor:Conditional LHL}
\begin{equation}
    \begin{split}
        \frac{1}{2} \Pr[\Omega_{\text{acc}}] \; &\norm{\omega_{KS^H\vec{E}'_{|_{\Omega_{\text{acc}}}}} - \chi_K \otimes \omega_{S^H\vec{E}'_{|_{\Omega_{\text{acc}}}}} } \\ 
        &\leq \frac{1}{2} 2^{-\frac{1}{2} \left(\Hmine[\bar{\varepsilon}](\vec{Z}|\vec{E}')_{\rho_{|_{\Omega_{\text{sift}}}}} - l \right)} + 2\bar{\varepsilon}.
    \end{split}
\end{equation}
Thus, choose a value of $l$ satisfying
\begin{equation}
  \begin{split}
      l \leq \Hmine[\bar{\varepsilon}](\vec{Z}|\vec{E}\vec{C} C_{EC}\vec{D})_{\rho_{|{\Omega_{\text{sift}}}}}
      - 2\log\left(\frac{1}{2\varepsilon_{\text{PA}}}\right)
  \end{split}			
\end{equation}
is enough to guarantee $( \varepsilon_\text{PA} + 2 \bar{\varepsilon})$-secrecy.
As usual, we split off the leakage \(C_{EC}\) during the error correction and verification steps by using \cite[Lemma 6.8]{Tomamichel2016}
\begin{equation}
  \begin{split}
      \Hmine[\bar{\varepsilon}](\vec{Z}|\vec{E}\vec{C} C_{EC}\vec{D})_{\rho_{|{\Omega_{\text{sift}}}}} \geq \Hmine[\bar{\varepsilon}](\vec{Z}|\vec{E}\vec{C}\vec{D})_{\rho_{|{\Omega_{\text{sift}}}}} \\ - \leak - \log(2/\epsEV).
  \end{split}			
\end{equation}
As the next step, we employ \cref{Lem:Bound Smooth Min-Entropy sifted signals} and we find
\begin{align}
    \Hmine[\bar{\varepsilon}](\vec{Z}|\vec{E}\vec{C}\vec{D})_{\rho_{|{\Omega_{\text{sift}}}}} \geq \Hmine[\bar{\varepsilon}](\vec{Z}_{\text{sift}}|\vec{E}_{\text{sift}}\vec{C}_{\text{sift}})_{\tau^{\text{sift}}_{|{\vec{D}=\vec{1}}}}.
\end{align}
Now, we invoke the asymptotic equipartition property \cite[Corollary 4.10]{Dupuis2020Commun.Math.Phys.}, which yields
\begin{equation}
    \begin{split}
        &\Hmine[\bar{\varepsilon}](\vec{Z}_{\text{sift}}|\vec{E}_{\text{sift}}\vec{C}_{\text{sift}} )_{\tau^{\text{sift}}_{|{\vec{D}=\vec{1}}}} \\ \geq &\lfloor n_{\text{sift}} - Nt_{\text{sift}} \rfloor H(Z_{\text{sift}}|E_{\text{sift}} C_{\text{sift}})_{\sigma_{|{D=1}}} \\ 
        &- 2\sqrt{n_{\text{sift}} -Nt_{\text{sift}}} \log(1+2\dim(Z))\sqrt{\log(\frac{2}{\bar{\varepsilon}^2})}.
    \end{split}
\end{equation}
Let us simplify the expression of the entropy even further. For the single round expression one can drop the subscript \(\text{sift}\) of the registers \(Z\) and \(E\), if we condition on \(D=1\). Furthermore, we only generate key if \(D=1\), otherwise \(H\left(Z|EC;D=0 \right) = 0 \). Hence, we find
\begin{equation}
    H\left(Z|EC;D=1 \right) 
    = \frac{1}{\Pr(D=1)}  H\left(Z|EC \right).
\end{equation}
The event \(D=1\) is equivalent to the state surviving sifting, thus,
\(\Pr(D=1) = \Pr(\text{sift})\).

Finally, Eve could have picked any of the states in \(S_{\vec{\nu}}\), thus we need to minimize over \(S_{\vec{\nu}}\). Then, combining all previous equations together the theorem statement follows.
\end{proof}

After seeing the proof of \cref{Thrm:PM QKD security}, one might wonder how to determine appropriate values for \(\bar{F}\) to define the acceptance set. One option would be a channel model, which determines the observations for the honest behavior of the channel. Additionally, one requires requires a characterization of both Alice's transmitter and Bob's receiver.

In the process of this channel characterization, it is important not to use the fluctuation parameters \(\vec{t}\) to compensate for poorly chosen acceptance sets that result from inadequate device characterization. Doing so would result in significantly reduced key rates.

We remedy this problem in \cref{sec:variable} where we adapt our work to variable length protocols which do not require a channel model anymore, however one still requires Alice's and Bob's devices to operate as specified by their POVMs.

\section{Decoy-State Methods in Finite-size regime}\label{sec:Decoy-State Methods in Finite-size regime}
Decoy-state protocols still fall into the generic framework of a P\&M protocol stated in \cref{Prot:PM Protocol} but have slightly different steps which we highlight below.

\begin{prot}[Decoy-State Prepare-and-Measure Protocol]\label{Prot:Decoy Protocol}
	All parameters are chosen as in \cref{Tab:Protocol parameters}. The intensity \(\mu_s\) stands for the signal intensity and is used primarily for key generation and \(\{\mu_2, \mu_3, \dots\}\) are the decoy intensities. For a decoy-state version of \cref{Prot:PM Protocol}, only step (1.) is replaced with step (1$^\prime$.) from below. Furthermore, Alice additionally announces the particular intensity choice used for each signal sent, which is still contained in register \(C_A\).	
	\begin{enumerate}[1$^\prime$.]
		\item \textbf{State preparation and transmission:} In each round Alice decides with probability \(p(\text{gen})\) whether she sends a test or generation round. In the case of a generation round Alice prepares one of \(d_A\) states \(\{\ket{s_i}\}_{i=1\dots d_A}\) with intensity \(\mu_s\). For test rounds Alice selects the intensity \(\mu_i\) with probability \(p(\mu_i | \text{test})\). She stores a label for her signal state in register \(X_i\). Depending on her choices Alice computes her announcement and stores it in register \(C_A\). Finally, Alice sends the signal state to Bob via a quantum channel.
	\end{enumerate}	
\end{prot}
\begin{rem}
    Results from Refs.~ \cite{Tupkary2024,Lim2014Phys.Rev.A,Rusca2018Appl.Phys.Lett.} require all intensities to be sent in key generation rounds, potentially resulting in suboptimal rates. 
    In contrast, our proof techniques allow us to consider protocols in which \emph{only} the signal intensity is used in key generation rounds. 
    The examples presented later utilize this structure. In particular, we can use an arbitrary distribution over the basis choices, signal/decoy intensities as shown in \cref{sec:Example Decoy 4-6}.
\end{rem}

\subsection{Source Replacement Scheme}
Before, we start with the finite-size analysis of decoy-state protocols we recapitulate the relevant notation. First, including a shield system \cite{Horodecki2009IEEETrans.Inf.Theorya}, and applying the source replacement scheme \cite{Bennett1992Phys.Rev.Lett., Ferenczi2012Phys.Rev.A} as before, Alice prepares the following state each round
\begin{equation}\label{eq:Decoy source replacement}
    \ket{\xi}_{AA_SA'} = \sum_{n=0}^{\infty} \sum_{a=1}^{d_A} \sqrt{\Pr(a,n)} \ket{a}_A \ket{n}_{A_S} \ket{s_n^{a}}_{A'},
\end{equation}
where \( \Pr(n) = \sum_{a,\mu} \Pr(a, \mu ,n) = \sum_{a,\mu} \Pr(a) \Pr(\mu|a) \Pr(n|\mu, a) \). Note, that for fully phase randomized weak coherent pulses, the signal states \(\ket{s_n^{a}}\) only depend on \(a\), thus no explicit \(\mu\) dependence is stated.

Since Eve could always perform a QND measurement we can assume \(\rho_{AA_SB}\) to be block-diagonal in the photon number \(n\) (sent by Alice) as shown in \cite{Li2020Phys.Rev.Research}. Under a collective IID attack, Eve's attack can be represented as the same channel \(\mathcal{E}_{A'\rightarrow B}\) acting on system \(A'\) of \(\ket{\psi}_{AA_SA'}\) each round. Furthermore, we still assume that Eve has no access to Alice's lab, which is represented as constraints of Alice's marginal state, see \cref{eq:rhoA Decoy} below.

Then, the shared state between Alice and Bob is
	\begin{equation}\label{eq:Decoy shared state}
		\rho_{AA_SB} = \sum_{n=0}^{\infty} \Pr(n) \dyad{n}_{A_S} \otimes \rho_{AB}^{(n)},
	\end{equation}
where
\begin{equation}\label{eq:rho conditioned on n}
    \begin{split}
    \rho_{AB}^{(n)} = \sum_{\substack{a, a'}}  &\sqrt{\Pr(a|n)\Pr(a'|n)}\dyad{a}{a'}_A \\ &\otimes \mathcal{E}_n\left( \dyad{s_n^{a}}{s_n^{a'}}_{A'} \right)
    \end{split}
\end{equation}
as shown in \cite{Li2020Phys.Rev.Research} or \cite{Kamin2024Phys.Rev.Res.} for intensities being different for each signal state. Again, due to the source replacement scheme, we only have to consider states which satisfy 
	\begin{equation}\label{eq:rhoA Decoy}
		\Tr_{A'}[\ketbra{\xi}] = \sum_n \Pr(n) \dyad{n}{n}_{A_S}\otimes\rho_{A}^{(n)}.
	\end{equation}
Furthermore, Bob receives the following state conditioned on Alice choosing signal \(a\) and sending \(n\) photons
\begin{equation}
    \sigma_B^{(a,n)} = \mathcal{E}_{A'\rightarrow B}\left( \ketbra{s^{a}_n}_{A'} \right).
\end{equation}
Since \(\ket{s^{a}_n}_{A'}\) is independent of the choice of the intensity \(\mu\), \(\sigma_B^{(a,n)} \) is also independent of \(\mu\).

\subsection{Asymptotic Scenario for Decoy State Methods}
In this section we will quickly summarize asymptotic decoy-state methods and introduce a common notation before turning our attention to the finite-size case. As in the asymptotic case \cite{Wang2022Phys.Rev.Res.}, let us define the yields as
\begin{equation}
    Y_n^{a,b} := \Pr(b|a,n,\mu) = \Tr[M_b^{B} \sigma_B^{(a,n,\mu)}],
\end{equation}
the probability of Bob detecting outcome \(b\) given that Alice sent state \(a\) with \(n\) photons. For fully phase randomized coherent states, the \(n\)-photon yields \(Y_n^{a,b}\) are independent of the intensity \(\mu \), i.e. \(\Pr(b|a,n) = \Pr(b|a,n,\mu) = Y_n^{a,b}\) for all \(\mu\). Therefore, using simple chain rules for probabilities,
\begin{equation}\label{eq:yield relation asymptotic}
    \begin{aligned}
        \Pr(b|a,\mu) &= \sum_{n=0}^{\infty} \Pr(n|a,\mu) \Pr(b|a,n,\mu) \\
         &= \sum_{n=0}^{\infty} \Pr(n|a,\mu) Y_n^{a,b},
    \end{aligned}
\end{equation}
which holds for all intensities \(\mu\).

As described in \cref{Prot:Decoy Protocol}, Alice decides the choice between a test and a generation round. Thus, if we knew the \(n\)-photon yields \(Y_n^{a,b}\), we would find the following constraints in the test rounds
\begin{equation}\label{eq:Decoy relation trace yield}
    \begin{split}
        \Pr(\text{test}|a,n) \Tr[\Gamma_{a,b} \rho_{AB}^{(n)}] =\Pr(\text{test},a|n) Y_n^{a,b}.
    \end{split}
\end{equation}
Under an IID attack, Eve's attack (for each photon number) can still be described by probabilities, the yields, and the observations will give bounds on the probabilities.

Hence, the question is to find, or rather bound, the yields \(Y_n^{a,b}\) in the finite-size regime, and in the next section we will lay out how the methods from \cite{Wang2022Phys.Rev.Res.,Kamin2024Phys.Rev.Res.} need to be altered in order to be valid in the finite-size regime.

\subsection{Finite-Size Scenario for Decoy State Methods}
So far we assumed perfect knowledge of the observations \(\Pr(b|a,\mu)\) and found a relation to the yields \(Y_n^{a,b}\). However, in the finite-size regime, we do not have access to the probabilities (yields) \(Y_n^{a,b} = \Pr(b|a,\mu)\) directly.

Thus, we turn our attention to estimating the \(n\)-photon yields \(Y_n^{a,b}\) based on the \emph{finite} information generated in the test rounds. After we found a relation of the yields to the frequencies observed in test rounds, we will formulate the key-length expression. Again, since we assume an IID attack, the yields will play the same role as in the asymptotic case, representing the probabilities of Bob observing an event given that Alice sent \(n\) photons, thus encompassing Eve's attack. However, the estimation of the yields is now much looser due to finite-size effects.

Let us start by relating the probability \(\Pr(k) = \Tr[\Gamma_k \rho]\) of observing outcome \(k=(a,b,\mu,\text{test}) \in \Sigmat\) to the constraints imposed by the acceptance test. For any observed frequency \(F^{\text{obs}} \in \mathcal{Q}\) passing the acceptance test it holds for all \(k=(a,b,\mu,\text{test}) \in \Sigmat\),
\begin{equation}
    \abs{F^{\text{obs}}_{k} - \bar{F}_{k} } \leq t_{k},
\end{equation}
or equivalently
\begin{equation}\label{eq:Bounds Fbar}
   \bar{F}_{k} - t_{k} \leq F^{\text{obs}}_{k} \leq \bar{F}_{k} + t_{k}.
\end{equation}
Furthermore, for \(\rho_{AB} \in S_{\vec{\nu}}\) the observed frequency \(F^{\text{obs}} \in \mathcal{Q}\) satisfies,
\begin{equation}
     F^{\text{obs}}_{k} - \nu_k^L \leq \Tr[\Gamma_k \rho] \leq F^{\text{obs}}_{k} + \nu_k^U,
\end{equation}
for all \(k=(a,b,\mu,\text{test}) \in \Sigmat\). Therefore, inserting \cref{eq:Bounds Fbar} gives
\begin{equation}\label{eq:Intermediate bound tr rho decoy}
    \bar{F}_{k} - t_{k} - \nu_{k}^L \leq \Tr[\Gamma_k \rho] \leq \bar{F}_{k} + t_{k} + \nu_{k}^U,
\end{equation}
for all \(k=(a,b,\mu,\text{test}) \in \Sigmat\). Finally, we can identify \(\Tr[\Gamma_k \rho]\) with
\begin{equation}
\begin{split}
    &\Tr[\Gamma_k \rho] = \Pr(\mu,\text{test}|a,b) \Tr[\Gamma_{a,b} \rho] \\ &= \Pr(\mu,\text{test}|a,b) \Pr(a,b) = \Pr(a,b,\mu, \text{test})
\end{split}
\end{equation}
and inserting this into \cref{eq:Intermediate bound tr rho decoy} results in
\begin{equation}\label{eq:bounds pr(k) decoy}
    \begin{split}
        &\bar{F}_{k} - t_{k} - \nu_k^L \leq F^{\text{obs}}_{k} - \nu_k^L \\ &\leq \Pr(a,b,\mu, \text{test}) \\ &\leq F^{\text{obs}}_{k} + \nu_k^U \leq \bar{F}_{k} + t_{k} + \nu_k^U,
    \end{split}
\end{equation}
which again holds for all \(k=(a,b,\mu, \text{test}) \in \Sigmat \).

Now, note that during the acceptance test, we accept on the frequencies generated by the joint probability distribution \(\Pr(k) = \Pr(a,b,\mu,\text{test})\). However, in the asymptotic regime we formulated the decoy analysis \cref{eq:yield relation asymptotic} in terms of conditional probabilities. Therefore, applying the same conditioning on \((a,\mu, \text{test})\), we can write asymptotically
\begin{equation}
\begin{split}
    &\frac{\Pr(a,b,\mu, \text{test})}{\Pr(a,\mu,\text{test})} = \Pr(b|a,\mu,\text{test}) \\
    &= \sum_{n=0}^{\infty} \Pr(n|a,\mu,\text{test}) Y_n^{a,b},
    \end{split}
\end{equation}
where we remark that in the finite-size regime, we have the additional conditioning on the ``test'' event. 

Hence, inserting \cref{eq:bounds pr(k) decoy}, we find the following relation between the yields and our acceptance set characterized by \(\bar{F}\) and \(\vec{t}\)
\begin{equation}
\begin{aligned} \label{eq:decoy bounds yields}
    \frac{\bar{F}_{k} + t_{k} + \nu_{k}^U}{\Pr(a,\mu,\text{test})} &\geq \sum_{n=0}^{\infty} \Pr(n|a,\mu,\text{test}) Y_n^{a,b}, \\
    \frac{\bar{F}_{k} - t_{k} -\nu_{k}^L}{\Pr(a,\mu,\text{test})} &\leq \sum_{n=0}^{\infty} \Pr(n|a,\mu,\text{test}) Y_n^{a,b},
\end{aligned}
\end{equation}
which holds for all \(k=(a,b,\mu, \text{test})  \in \Sigmat \). Again, as in the asymptotic case, the yields \(Y_n^{a,b}\) satisfy
\begin{equation}\label{eq:yield relation trace finite}
    \begin{split}
        \Pr(\text{test}|a,n) \Tr[\Gamma_{a,b} \rho_{AB}^{(n)}] =\Pr(\text{test},a|n) Y_n^{a,b},
    \end{split}
\end{equation}
which allows us to constraint the \(n\)-photon states.

The expressions in \cref{eq:decoy bounds yields,eq:yield relation trace finite} would already be enough to formulate the upcoming security statement. However, \cref{eq:decoy bounds yields} contains an infinite number of unknown variables, the yields \(Y_n^{a,b}\), if one were to calculate them. Therefore, let us introduce the photon number cut-off \(N_{\mathrm{ph}}\) which introduces a small tolerance accounted by
\begin{equation}
    p_{\text{tot}}(\mu|a) := \sum_{n \leq N_{\mathrm{ph}}} \Pr(n|a,\mu,\text{test}),
\end{equation}
but allows us to reduce the infinite sum to a finite sum and formulate the finite linear program in \cref{eq:LP decoy} later. 

Furthermore, to simplify the notation, we note the following. Typically, the constraints on the \(n\)-photon states will be of the form of \cref{eq:yield relation trace finite} and are most naturally expressed in terms of the right-hand-side of \cref{eq:yield relation trace finite}, which we will call the joint yields. Therefore, let us define the joint yields as
\begin{equation}\label{eq:Defn gamma}
    \gamma_{n}^{ab} := \Pr(\text{test},a|n) Y_n^{a,b}.
\end{equation}

Whenever it will simplify the presentation, we will switch between the picture of the joint \(\gamma_{n}^{ab}\) and conditional yields \(Y_n^{a,b}\) but one could equally well express the decoy analysis in terms of either yield definitions.

Finally, before we state \cref{Cor:Decoy Security Single Opt}, a remark is in order. Again, the entry-wise bounding of \(\Pr(k) = \Pr(a,b,\mu, \text{test})\) is in contrast to previous numerical finite-size methods presented in \cite{George2020Phys.Rev.Res.}, where 1-norm constraints were used. With those constraints, one could still claim \(\abs{F^{\text{obs}}_{k} - \Pr(k)} \leq \nu\). However, this would be a "worst-case" upper bound because \(\nu\) satisfies
\begin{equation}
    \norm{\vec{F}^{\text{obs}} - \vec{p}}_1 \leq \nu,
\end{equation}
for \(p_k := \Pr(k)\). Hence, we would allow each element to vary in the worst possible way. On the other hand, entry-wise constraints are much more in line with previous numerical decoy state methods \cite{Wang2022Phys.Rev.Res.}.

After this discussion, we can state the following \cref{Cor:Decoy Security Single Opt}, which lets us evaluate the secure key length for decoy-state protocols.

\begin{cor}[Security for Decoy-state Protocols (Optimized Yields)]\label{Cor:Decoy Security Single Opt}
	Fix the parameters of \cref{Prot:Decoy Protocol}, and define \(\mathcal{Q}\) and \(S_{\vec{\nu}}\) as in \cref{Thrm:Secrecy outside Feasible Set}. Assuming a collective IID attack, the QKD protocol is \(\varepsilon_{\text{sec}} = \varepsilon_{\text{EV}} +  \max\{\varepsilon_{\text{AT}}, \varepsilon_{\text{PA}} + 2\bar{\varepsilon} \}\) secure, if the key length \(l\), in the case of not aborting, is chosen to be
	\begin{align}
		l \leq \; &\frac{\lfloor n_{\text{sift}} - Nt_{\text{sift}} \rfloor}{\bar{F}_\text{sift} + t_\text{sift} + \nu^U_\text{sift} }  \sum_{n=0}^{\infty} \Pr(n) \min_{\substack{\rho^{(n)} \in S_n\left(\gamma_n\right) \\ \{\gamma_n\}_n \in G}} H(Z|EC)_{\rho^{(n)}} \nonumber \\
		&- \sqrt{\lfloor n_{\text{sift}}-Nt_{\text{sift}} \rfloor} \Delta\left(\bar{\varepsilon} \right) - \leak - \log(2/\epsEV) \\
		& - 2\log(\frac{1}{2\varepsilon_{\text{PA}}}) \nonumber ,
	\end{align}
	where the sets \(S_n\) correspond to the feasible \(n\)-photon subspaces and are defined by
	\begin{align} \label{eq:decoy2}
		&S_n\left(\gamma_n\right) = \{\rho^{(n)}\in \mathcal{S}_{\circ}(\Hil_A\otimes \Hil_B) | \exists \{\gamma_n\} \in G \text{ s.t.} \: \nonumber \\
		&\hspace{50pt} \forall (a,b,\mu,\text{test}) \in \Sigmat, \\
		&\hspace{50pt} \Pr( \mathrm{test} |a,b,n)\Tr[\Gamma_{a,b} \rho^{(n)}] =\gamma_{n}^{ab}, \nonumber \\
		&\hspace{50pt} \rho_A^{(n)} = \Tr_{A_sA'}[\ketbra{n}_{A_s}\ketbra{\xi}] \nonumber,		
	\end{align}
    where
    \begin{align}\label{eq: G set}
        \begin{aligned} G:=
			\{ \{\gamma_m\} | &\gamma_{m}^{ab} = \Pr(\mathrm{test},a|m) Y_m^{a,b}, \\
            &\frac{	\bar{F}_{k} - t_{k} -\nu_{k}^L}{\Pr(a,\mu,\mathrm{test})} \leq \sum_{m \leq N_{\mathrm{ph}}} \Pr(m|a,\mu,\mathrm{test}) Y_m^{a,b} \\ + &\left(1- p_{\text{tot}}(\mu|a) \right), \\
			&\frac{	\bar{F}_{k} + t_{k} +\nu_{k}^U}{\Pr(a,\mu,\mathrm{test})} \geq \sum_{m \leq N_{\mathrm{ph}}} \Pr(m|a,\mu,\mathrm{test}) Y_m^{a,b}, \\
			&0 \leq Y_m^{a,b} \leq 1, \; \forall m\in \N_0, \\
			&\forall k=(a,b,\mu,\mathrm{test}) \in \Sigmat  \}.
		\end{aligned} 
    \end{align}
\end{cor}

\begin{proof}
    The correctness and secrecy for \(\rho \notin S_{\vec{\nu}}\) follow from \cref{Thrm:PM QKD security}. For \(\rho \in S_{\vec{\nu}}\) we continue from the key length expression in \cref{Thrm:PM QKD security} and only focus on the term containing the optimization over \(\rho \in S_{\vec{\nu}}\).

    First, we bound the fraction by the same lower bound on the denominator \(\Pr(\text{sift})\) as in \cref{eq:Lower bound psift} to find
    \begin{equation}\label{eq:lower bound psift decoy proof}
        \min_{\rho \in S_{\vec{\nu}} } \frac{ H(Z|EC)_{\rho}}{\Pr(\text{sift}) } \geq \frac{\min_{\rho \in S_{\vec{\nu}} } H(Z|EC)_{\rho}}{\bar{F}_\text{sift} + t_{\text{sift}} + \nu^U_{\text{sift}}}.
    \end{equation}
    
    Next, we will reformulate the entropy term in terms of contributions from different photon numbers. To do so, note that since we assume fully phase-randomized coherent states and as described in \cite[App. B]{Li2020Phys.Rev.Research}, Eve can perform a QND measurement on the photon number.  Hence, Eve knows the photon number of the states sent and we split Eve's register into \(E = E'M\), where \(M\) is classical and labels the photon number of Alice's sent states. Thus, 
    \begin{equation}
        \begin{split}
            &\min_{\rho \in S_{\vec{\nu}} } H(Z|EC)_{\rho} \\
            \geq & \min_{\rho \in S_{\vec{\nu}} } \sum_{n=0}^{\infty} \Pr(M=n)  H(Z|E'C)_{\rho{|{M=n}}}.
        \end{split}
    \end{equation}    
    All states \(\rho_{AB} \in S_{\vec{\nu}}\) must be of the form of \cref{eq:Decoy shared state} and furthermore all \(\rho_{AB{|{M=n}}} = \rho_{AB}^{(n)} \) satisfy \cref{eq:rho conditioned on n}. Additionally, \cref{eq:Decoy relation trace yield} establishes relations between the state \(\rho^{(n)}_{AB}\) and the yields. With the definition of the joint yields \(\gamma_n\) in \cref{eq:Defn gamma} and by applying the results of the acceptance test, we know that the yields need to satisfy the constraints summarized in \(G\), as laid out in the preceding section. Hence, we find
    \begin{align}
        &\min_{\rho \in S_{\vec{\nu}} } H(Z|EC)_{\rho} \\
        \geq &\sum_{n=0}^{\infty} \Pr(M=n) \min_{\rho \in S_{\vec{\nu}} } H(Z|E'C;M=n)_{\rho_{|_{M=n}}}\\
        \geq &\sum_{n=0}^{\infty} \Pr(M=n) \min_{\substack{\rho^{(n)} \in S_n(Y_n) \\ \{Y_n\}_n \in G}} H(Z|E'C;n)_{\rho^{(n)}},
    \end{align}
    where for the first inequality we swapped the minimization with the sum and for the second inequality we inserted the definitions of \(S_n\left(\gamma_n\right)\) and \(G\). Furthermore, we used that for \(\rho \in S_{\vec{\nu}}\) it holds \(\rho^{(n)} \in S_n(Y_n)\) and \(\{Y_n\} \in G\).

    Combining this with \cref{eq:lower bound psift decoy proof} the corollary follows.    
\end{proof}

The double optimization of \cref{Cor:Decoy Security Single Opt} can be numerically unstable and does not fit the framework for decoy-state protocols presented in \cite{Wang2022Phys.Rev.Res.} nor the general framework for QKD protocols shown in \cite{Winick2018Quantum}. In Ref.\cite{Wang2022Phys.Rev.Res.}, the decoy analysis was formulated as a linear program which outputs upper and lower bounds on the yields (or e.g. error rates) and then calculates the key rates based on these bounds.

Therefore, we will reformulate \cref{Cor:Decoy Security Single Opt}. However, the previous version will be tighter. As in the asymptotic case, we will formulate the key rate evaluation in a two-step process. First, find upper and lower bounds on the yields \(Y_n^{a,b}\) based on the observations and, second, calculate the key rate with those bounds.

We can solve for upper and lower bounds of the \(n\)-photon yields \(Y_n\) with a linear program given as
\begin{equation}\label{eq:LP decoy}
    \begin{aligned}
        Y_{n,L}^{ab} := &\min_{\mathbf{Y}_m}  Y^{a,b}_n\\
        \textrm{s.t.}\; & \frac{\bar{F}_{k} -t_k - \nu_k^L}{\Pr(a,\mu,\text{test})} \leq \sum_{m \leq N_{\mathrm{ph}}} \Pr(m|a,\mu,\mathrm{test}) \; Y_m^{a,b} \\&\hspace{80pt}  + \left(1- p_{\text{tot}}(\mu|a) \right), \\
        &\frac{\bar{F}_{k} +t_k + \nu_k^U}{\Pr(a,\mu,\text{test})} \geq \sum_{m\leq N_{\mathrm{ph}}} \Pr(m|a,\mu,\mathrm{test}) \; Y_m^{a,b}, \\
        &\forall  k = (a,b,\mu,\text{test}) \in \Sigmat,  \\
        &0 \leq Y^{a,b}_m \leq 1 \; \forall m \in \N_0,  
    \end{aligned}
\end{equation}
where the yields \(Y_m^{a,b}\) were rearranged into a vector \(\mathbf{Y}_m\) for each \(m \in \N_0\), and as before \(p_{\text{tot}}(\mu|a) = \sum_{m\leq N_{ph}}  \Pr(m|a,\mu,\mathrm{test})\). The same optimization problem stated with a maximization gives rise to the upper bounds.

As before, we define upper and lower bounds on the joint yields \(\gamma_{n,L}^{ab}\) and \(\gamma_{n,U}^{ab}\) by
\begin{equation}\label{eq:decoy3}
    \begin{aligned}
        \gamma_{n,L}^{ab} &= \Pr(a,\mathrm{test}|n) Y_{n,L}^{ab}, \\
        \gamma_{n,U}^{ab} &= \Pr(a,\mathrm{test}|n) Y_{n,U}^{ab}.
    \end{aligned}
\end{equation}
Thus, it follows
    \begin{equation}\label{eq:Upper and Lower bounds rhogen}
        \gamma_{n,L}^{ab} \leq \Pr( \mathrm{test} |a,b,n) \Tr[\Gamma_{a,b} \rho_{AB}^{(n)} ] \leq \gamma_{n,U}^{ab},
    \end{equation}
for all \(n \in \N_0\) and \((a,b) \in \Sigmat \), and we can state the following \cref{Cor:Decoy Security}.

\begin{cor}[Security for Decoy-state Protocols (Upper and Lower Bounds on Yields)]\label{Cor:Decoy Security}
    Fix the parameters of \cref{Prot:Decoy Protocol}, and define \(\mathcal{Q}\) and \(S_{\vec{\nu}}\) as in \cref{Thrm:Secrecy outside Feasible Set}. Assuming a collective IID attack, the QKD protocol is \(\varepsilon_{\text{sec}} = \varepsilon_{\text{EV}} +  \max\{\varepsilon_{\text{AT}}, \varepsilon_{\text{PA}} + 2\bar{\varepsilon} \}\) secure, if the key length \(l\), in the case of not aborting, is chosen to be
    \begin{align}
            l \leq \; &\frac{\lfloor n_{\text{sift}} - Nt_{\text{sift}} \rfloor}{\bar{F}_\text{sift} + t_\text{sift}+ \nu^U_\text{sift}}  \sum_{n=0}^{\infty} \Pr(n) \min_{\rho^{(n)} \in \tilde{S}_n } H(Z|EC)_{\rho^{(n)}} \nonumber \\
            &- \sqrt{\lfloor n_{\text{sift}}-Nt_{\text{sift}} \rfloor} \Delta\left(\bar{\varepsilon} \right) - \leak - \log(2/\epsEV)\\
            &- 2\log(1/2\varepsilon_{\text{PA}}), \nonumber
    \end{align}
    where the sets \(\tilde{S}_n\) correspond to the feasible \(n\)-photon subspaces and are defined by
    \begin{equation}
        \begin{aligned}
            \tilde{S}_n := \{&\rho^{(n)}\in \mathcal{S}_{\circ}(\Hil_A\otimes \Hil_B) | \forall (a,b, \mu, \mathrm{test}) \in \Sigmat \\ 
            &\gamma_{n,L}^{a,b} \leq \Tr[\Gamma_{a,b} \rho^{(n)}] \leq \gamma_{n,U}^{a,b}, \\
            & \rho_A^{(n)} = \Tr_{A_sA'}[\ketbra{n}_{A_s}\ketbra{\xi}] \}.
        \end{aligned}
    \end{equation}
\end{cor}
\begin{proof}
The correctness and secrecy for \(\rho \notin S_{\vec{\nu}}\) again follow directly from \cref{Thrm:PM QKD security}. For \(\rho \in S_{\vec{\nu}}\) we continue from the key length expression in \cref{Cor:Decoy Security Single Opt} and only focus on the entropy term.
For each \( \{Y_n\} \in G\), see \cref{eq: G set}, it holds \(S_n(Y_n) \subseteq \tilde{S}_n \), thus we can replace the lower bound on the conditional entropy from \cref{Cor:Decoy Security Single Opt} with the weaker one,
    \begin{align}
        &\min_{\rho \in S_{\vec{\nu}} } H(Z|EC)_{\rho} \\
        \geq &\sum_{n=0}^{\infty} \Pr(M=n) \min_{\rho^{(n)} \in \tilde{S}_n } H(Z|E'C;M=n)_{\rho^{(n)}},
    \end{align}
    which already concludes the proof.
\end{proof}

Instead of \cref{eq:LP decoy} one could also use improved decoy methods \cite{Kamin2024Phys.Rev.Res.} to achieve similar improvements in the finite-size regime, such as only requiring two total intensities. The appropriate optimization problem is stated in \cref{App:Improved Decoy}, which also allows the intensities to be different for each signal state.

This part concludes the fixed length security proofs and now we turn our attention to examples and optimizing finite-size key rates.

\section{Optimizing Security parameters}\label{sec:Optimizing Security parameters}

So far, in all security statements, \cref{Thrm:PM QKD security}, \cref{Cor:Decoy Security Single Opt} and \cref{Cor:Decoy Security}, one can arbitrarily choose the distribution of the individual \(\varepsilon\)-parameters to combine to the same final security parameter \(\varepsilon_{\text{sec}}\). However, some contributions to \(\varepsilon_{\text{sec}}\) result in a higher key rate reduction than others. Therefore, in this section, we aim to heuristically optimize the total key length in terms of the security parameters. 

The security parameter \(\varepsilon_{\text{sec}}\) in \cref{Thrm:PM QKD security}, \cref{Cor:Decoy Security Single Opt} and \cref{Cor:Decoy Security} is given by
\begin{equation}
    \varepsilon_{\text{sec}} = \max\{\varepsilon_{\text{AT}}, \varepsilon_{\text{PA}} + 2\bar{\varepsilon} \}.
\end{equation}

Heuristically, \(\varepsilon_{\text{AT}}\) has the greatest influence, as a smaller \(\varepsilon_{\text{AT}}\) results in a larger variational bound \(\nu\) and thus more states must be considered in minimization. To distribute \(\varepsilon_{\text{AT}}\) ideally under our assumptions, one should choose \(\varepsilon_{\text{AT}} = \varepsilon_{\text{PA}} + 2\bar{\varepsilon}\). Furthermore, if we made \(\varepsilon_{\text{AT}}\) bigger than \(\varepsilon_{\text{PA}} + 2\bar{\varepsilon}\), then one could increase the latter two parameters for free without any penalty in the security statement. Hence, \(\varepsilon_{\text{AT}} = \varepsilon_{\text{PA}} + 2\bar{\varepsilon}\) is the optimal choice.

Therefore, as a starting point, let us assume that the maximum in the security parameter is achieved by
\begin{equation}
    \max\{\varepsilon_{\text{AT}}, \varepsilon_{\text{PA}} + 2\bar{\varepsilon} \} = \varepsilon_{\text{PA}} + 2\bar{\varepsilon}.
\end{equation}
Thus, the total security parameter needs to satisfy
\begin{equation}
    \varepsilon_{\text{sec}} = \epsEV + \varepsilon_{\text{PA}} + 2\bar{\varepsilon}.
\end{equation}
Furthermore, the reduction of the key rate from the error verification parameter \mbox{\(\epsEV\)} is smaller than from privacy amplification. Hence, we make another choice, by assuming that their correction terms should be equal, i.e.
\begin{equation}
    \log(2/\epsEV) = 2\log(1/2\varepsilon_{\text{PA}}).
\end{equation}
This choice results in 
\begin{equation}\label{eq:Optimal epsilon EC}
    \varepsilon_{\text{EV}} = 8\varepsilon_{\text{PA}}^2.
\end{equation}
Then, for a fixed \(\varepsilon_{\text{sec}}\), the smoothing parameter is given by
\begin{equation}\label{eq:Optimal epsilon smooting}
    2\bar{\varepsilon} = \varepsilon_{\text{sec}} - \varepsilon_{\text{PA}} \left(8\varepsilon_{\text{PA}} +1 \right).
\end{equation}
Under these assumptions, the optimal choice for all \(\varepsilon\)-parameters is found by optimising the key length \(l\) in terms of \( \varepsilon_{\text{PA}}\). Fix \(n_{\text{sift}},N\), the security parameter \(\varepsilon_{\text{sec}}\) and define the remaining \(\varepsilon\) parameters by \cref{eq:Optimal epsilon EC} and \cref{eq:Optimal epsilon smooting}. Thus, differentiating \(l\) with respect to \(\varepsilon_{\text{PA}}\), yields
\begin{equation}
    \begin{split}
    \frac{\dif l}{\dif \varepsilon_{\text{PA}}} = \frac{1}{\ln(2)\varepsilon_{\text{PA}}} - &\sqrt{\lfloor n_{\text{sift}}-Nt_{\text{sift}} \rfloor} \\ 
    &\cdot \frac{\log(1+\dim(Z)) \left(8\varepsilon_{\text{PA}} +1 \right)}{\ln(2)\bar{\varepsilon} \sqrt{\log(\frac{2}{\bar{\varepsilon}^2})}}.
\end{split}
\end{equation}
The optimal value \(\varepsilon_{\text{PA}}^{*}\) is the solution to \(\frac{\dif l}{\dif \varepsilon_{\text{PA}}} = 0\), which can be done numerically. Since both \(\bar{\varepsilon}\) and \(\varepsilon_{\text{PA}}\) need to be positive and smaller than \(\varepsilon_{\text{sec}}\), the optimal value \(\varepsilon_{\text{PA}}^{*}\) needs to be in \( \left[0, -\frac{1}{16} + \frac{1}{\sqrt{8}} \sqrt{ \frac{1}{32} + \varepsilon_{\text{sec}}}\right]\).

In summary, the optimal \(\varepsilon_{\text{AT}}\) is given by \(\varepsilon_{\text{AT}}= \varepsilon_{\text{PA}}^{*} + 2\bar{\varepsilon}\). Thus, it makes up the majority of the available \(\varepsilon_{\text{sec}}\), since \(\varepsilon_{\text{EV}} = 8(\varepsilon_{\text{PA}}^*)^2 \ll \varepsilon_{\text{PA}}^*\).

\section{Optimising Expected Key Rates for Fixed Length Protocols}\label{sec:Optimising expected Key Rates}
The acceptance set \(\mathcal{Q}\) implicitly depends on the expected frequencies \(\bar{F}\) and the tolerated fluctuations characterised in terms of \(\vec{t}\) and \(t_{\text{sift}}\). So far, the tolerance parameters \(\vec{t}\) and \(t_{\text{sift}}\) in the acceptance set \(\mathcal{Q}\) are free and can be chosen arbitrarily. In terms of pure key rates, they should be chosen to be zero, however, then the protocol would almost always abort since observing a single point has vanishing probability.

Hence, one would want to choose them such that under the honest behavior we accept the observed frequencies with high probability while maintaining high secret key rates. However, both goals cannot be achieved simultaneously. A wider acceptance set will always decrease the key rate whereas a smaller acceptance set ultimately decreases the accept probability. Therefore, we introduce the expected key rate as a figure of merit because it captures the trade-off between both effects. In \cite{Kanitschar2023PRXQuantum,Tupkary2024Phys.Rev.Res.} expected key rates have been calculated by simulating the channel numerically. Here we will instead proceed by finding tight bounds on the accept probability.

At first, let us define the expected key rate as
\begin{equation}
    \begin{split}
        \bar{R}_{\mathcal{Q}(\vec{t})} = &\Pr[F^{\text{obs}} \in \mathcal{Q}(\vec{t})] \cdot R_{\text{acc}} \\ 
        &+ \Pr[F^{\text{obs}} \notin \mathcal{Q}(\vec{t})] \cdot R_{\text{abort}},
    \end{split}
\end{equation}
where \(R_{\text{acc}} \) is the usual finite-size key rate after passing the acceptance test. In the case of aborting \(R_{\text{abort}} = 0\) because then no key is generated.

In order to estimate the accept probability \(\Pr[\Omega_{\text{acc}}]\) for an acceptance set \(\mathcal{Q}\) characterized by \(\vec{t}\) and \(\bar{F}\), we assume a given channel model, which correctly describes the observations. This assumption means that the parameter \(\bar{F}\) in the security proof is equal to the honest behavior based on that channel model.

Additionally, we make certain assumptions about the accept probability, because it only influences the expected key rate, not the secret key rate upon acceptance. Specifically, \cref{Thrm:PM QKD security} proves security for any width of the acceptance set, although the resulting key length might be zero. 

Hence, we assume that the accept probability is mostly governed by \(\Pr[F^{\text{obs}} \in \mathcal{Q}]\), which is a reasonable assumption since under honest behavior, a well-designed error correction procedure will succeed with high probability. Thus, let us assume 
\begin{align}
    &\Pr[\Omega_{\text{acc}}] \approx \Pr[\Omega_{\text{AT}}] = \Pr[F^{\text{obs}} \in \mathcal{Q} ] \\
    &= \Pr[\abs{F^{\text{obs}}_k - \bar{F}_k} \leq t, \: \forall k \in \Sigma ].
\end{align}
In order to bound this last equation, let us first define the events
\begin{equation}
    A_k := \{F^{\text{obs}} \in \mathcal{P}(\Sigma) \text{  s.t.  } \abs{F^{\text{obs}}_k - \bar{F}_k} \leq t\}
\end{equation}
for all \( k \in \Sigmat \cup \{\text{sift} \}\). Then, we can write
\begin{equation}
    \Pr[F^{\text{obs}} \in \mathcal{Q} ] = \Pr[\bigcap_{k=1}^{|\Sigmat|+1} A_k],
\end{equation}
which allows us to apply the Frechet inequalities \cite{Frechet1935Fundam.Math.}. They state for events \(A_1, \dots, A_n\) (which can be dependent on each other), the probability of all events occurring simultaneously is bounded by
\begin{align}\label{eq:Frechet ineq}
    \Pr[\bigcap_{k=1}^n A_k] &\leq \min_{1\leq k \leq n} \Pr[A_k], \\
    \Pr[\bigcap_{k=1}^n A_k] &\geq\max\left\{0, \sum_{k=1}^{n} \Pr[A_k] -\left(n-1\right)  \right\}.
\end{align}
Thus, we can find upper and lower bounds on the accept probability given a channel model. Here it is important to note that we could not in general calculate the accept probability directly for a given channel model because the events \(A_k\) can arbitrarily depend on each other due to the underlying quantum state \(\rho_{AB}\), justifying the estimation via the Frechet inequalities.

Hence, we can evaluate the performance of our acceptance set choice in terms of both the maximal and minimal achievable expected key rate. Furthermore, we can estimate the optimal choice for allowed fluctuations \(\vec{t}\) and \(t_{\text{sift}}\) for a given channel model. 

\begin{figure}
	\centering
	\includegraphics[width=\linewidth]{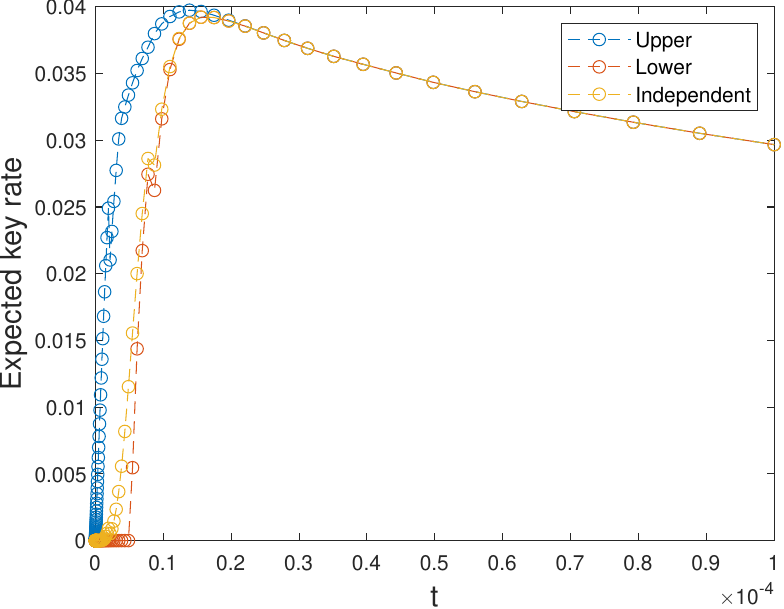}
	\caption{Plot of upper, lower and independent bounds on the expected key rate of the qubit BB84 protocol for \emph{IID attacks} with a channel loss of \(\unit[10]{dB}\) loss and a total number of signals of \(N=10^9\) over the tolerated fluctuation parameter with \(t_k=t_{\text{sift}} \equiv t\).}
	\label{fig:expectedqubitbb84}
\end{figure}

In \cref{fig:expectedqubitbb84} one can see the expected key rate of the qubit BB84 protocol plotted against the tolerated fluctuations \(t\) in the acceptance set for a loss-only channel model with \(\unit[10]{dB}\) loss and a total number of signals of \(N=10^9\). For each data point, we optimized the testing probability. In order to present the allowed fluctuations on a one-dimensional line, we chose \(t_k \equiv t\) for all \( k \in \Sigmat \cup \{\text{sift} \}\). The upper (blue circles) and lower (red circles) bounds use the Frechet inequalities of \cref{eq:Frechet ineq} and the independent bound assumes all \(A_k\) to be independent, i.e.
	\begin{equation}
		\Pr[\bigcap_{k=1}^n A_k] \approx \prod_{k=1}^{|\Sigmat|+1}  \Pr(A_k).
	\end{equation}
One can clearly see that there is an optimal choice of \(t\). The first increase in expected key rate originates from the accept probability increasing as we increase \(t\). Later, while increasing \(t\) further we allow more fluctuations and thus more states in \(S_{\vec{\nu}}\). This reduces the achievable key rate and therefore also reduces the expected key rate.

\section{Examples for Collective Attacks}\label{sec:Examples}
In this section, we only consider IID collective attacks and present key rates for fixed-length protocols with unique acceptance, i.e. \(\vec{t} = 0\). In general, the leakage depends on the actual error correction procedure. However, for all examples presented here we use the upper bound
\begin{equation}
	\begin{split}
		\leak = &N \bar{F}_\text{sift} f_{EC} H(X| Y; \text{sift} ), \\
	\end{split}
\end{equation}
which corresponds to the leaked number of bits during error-correction step for the expected observations. This expression for the leakage is not to be confused with a result from an optimization over the feasible set, it is just an upper bound on the error correction cost.

In each example we will optimize some free parameters such as the testing probability or the decoy intensities. Since this is different for each specific example we will discuss this in detail in the corresponding subsections.

Furthermore, for simplicity we define \(\nu_k := \max\{ \nu_k^U , \nu_k^L \}\) and set \(\nu_k^U = \nu_k, \nu_k^L = \nu_k\), which only has minuscule influence on the secret key rate. Moreover, the approximation becomes exact in the limit of infinite signals sent as the Binomial distribution converges to a Gaussian distribution. Even for the number of signals we are considering here, the difference between \( \nu_k^U \) and \(\nu_k^L\) is negligible.

\subsection{Qubit BB84 protocol with loss}\label{sec:Example Qubit BB84}
\begin{figure}
	\centering
	\includegraphics[width=\linewidth]{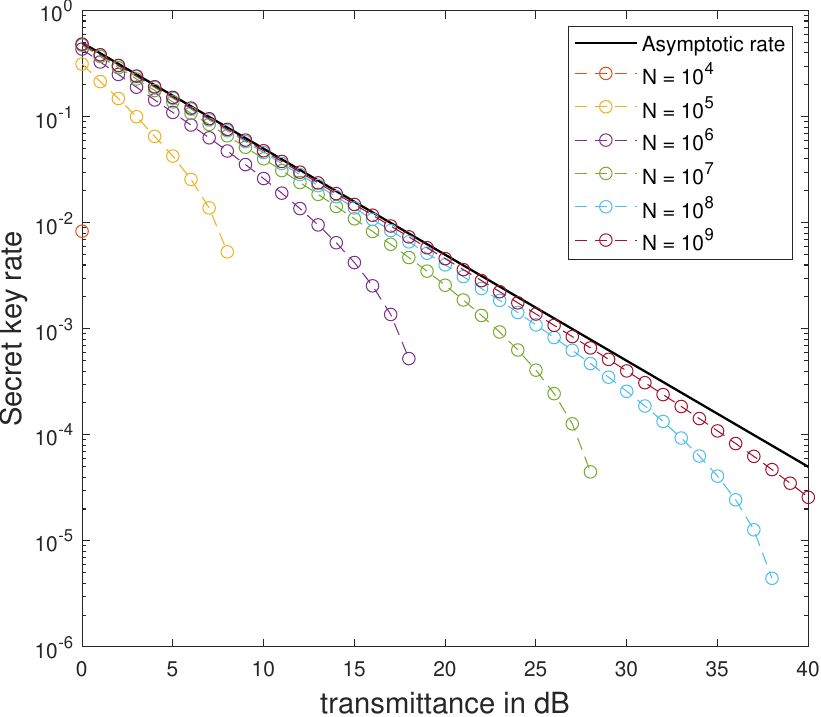}
	\caption{Secret key rate for \emph{IID attacks} of the qubit BB84 protocol against channel loss for varying number of total signals sent \(N=10^4, \dots, 10^{9} \).}
	\label{fig:qubit-bb84}
\end{figure}

As a simple example we present the BB84 protocol \cite{Bennett2014Theor.Comput.Sci.} with a perfect qubit source including loss in the channel. In this protocol, we assume that both the \(X\) and \(Z\) basis are used for key generation and with some probability each round is revealed as a test round. Thus, one acquires data from both bases to constraint the feasible set. We made the following parameter choices, \(Z\)-basis probability \(p_z = 1/2\) for both Alice and Bob, misalignment parameter \(\theta_{\text{misalign}} = 0.01\), error correction parameter \(f_{\text{EC}} = 1.16\) and a security parameter \(\varepsilon_{\text{sec}} = 10^{-8}\). We optimize the testing probability and all contributions to the final security parameter \(\varepsilon_{\text{sec}} \) for each data point.

To evaluate the secret key rate we used the methods from \cite{Winick2018Quantum}. The corresponding Kraus operators for this protocol can be found in \cref{App:Kraus ops BB84}. The resulting key rates can be seen in \cref{fig:qubit-bb84} for different numbers of total signals sent, \(N=10^4, \dots, 10^{9} \), over the total channel loss \(\eta\) in \(\unit{dB}\). One can clearly see the convergence to the asymptotic key rate (black solid line) as the number of signals increases.

\begin{figure}
	\centering
	\includegraphics[width=\linewidth]{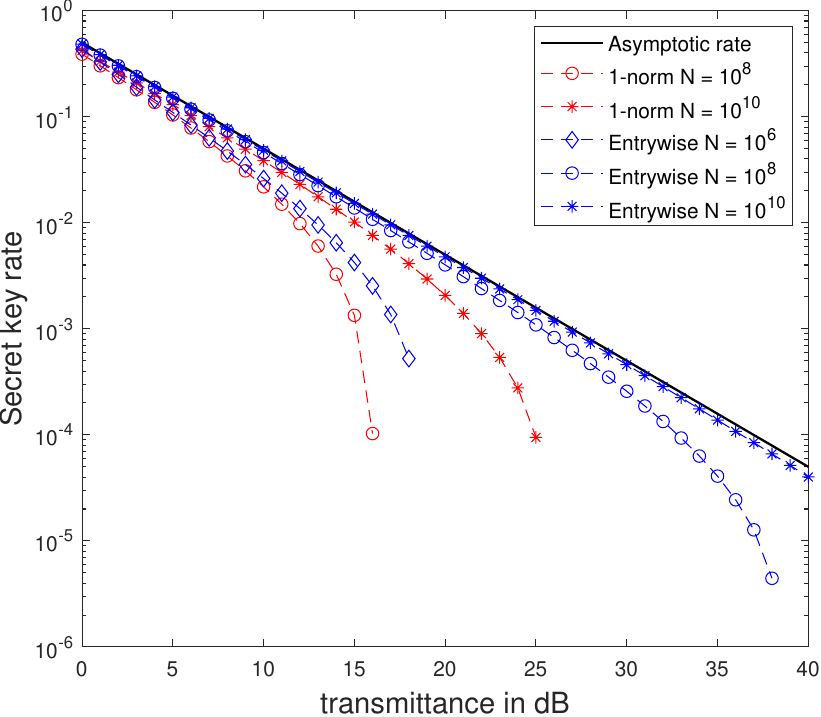}
	\caption{Comparison of the qubit BB84 secret key rates for \emph{IID attacks} between 1-norm constraints of \cite{George2020Phys.Rev.Res.} and our entry-wise constraints of \cref{Thrm:Secrecy outside Feasible Set}. The key rate is plotted against channel loss with a varying number of total signals sent \(N\).}
	\label{fig:Comparison Qubit BB84}
\end{figure}

Additionally, in \cref{fig:Comparison Qubit BB84} we compare our results for the qubit BB84 protocol with the ones from Ref.~\cite{George2020Phys.Rev.Res.}. The main difference in the techniques are the different constraint types. We use entry-wise constraints, while in Ref.~\cite{George2020Phys.Rev.Res.} 1-norm constraints were used.

Our entry-wise constraints give a clear improvement. This is due to the much tighter bounds derived in \cref{Thrm:Secrecy outside Feasible Set} and the improvements on applying the AEP due to \cref{Lem:Bound Smooth Min-Entropy sifted signals}. 

With the results presented in this work, we essentially achieve asymptotic rates with \(N=10^{10}\) signals whereas 1-norm constraints only achieve positive key rates until up approximately \(\unit[25]{dB}\).

As an aside, in \cite{George2020Phys.Rev.Res.} the authors derived an improvement on the variational bound \(\nu\). This improvement uses multiple coarse grainings to reduce the parameter \(\nu\). By incorporating all possible coarse grainings, they would effectively recover a vector \(\vec{\nu}\) just as we presented. However, their bounds originating from a 1-norm, are inherently worse than the ones in \cref{Thrm:Secrecy outside Feasible Set}. Therefore, even after including multiple coarse grainings, their results will remain worse than ours and our methods remain much simpler and easier to implement.

\subsection{Decoy BB84 protocol}\label{sec:Example Decoy BB84}
Next, we present a decoy-state version of the BB84 protocol. We assume a fully phase-randomized WCP source and a symmetric passive detection setup for Bob, i.e. Bob chooses the \(Z\)- and \(X\)-basis with probability \(p_z=1/2\). This allows us to use the squashing map from \cite{Beaudry2008Phys.Rev.Lett.,Gittsovich2014Phys.Rev.A}. We again use both bases for key generation, such that this protocol is equivalent to the qubit version apart from the additional overhead due to the decoy states. 

In each round Alice decides with probability \(p(\text{gen})\) if it is a test or generation round and always selects the signal intensity if it was a key generation round. In a test round Alice selects all intensities with equal probability, i.e. \(p(\mu|\text{test}) =1/3\). We use the same parameters as in \cref{sec:Example Qubit BB84} with the addition of two decoy intensities, which we set to be \(\mu_2 = 0.02, \mu_3 = 0.001\), and a signal intensity \(\mu_s\) as a free parameter.
Now, in addition to the testing fraction, we optimize the signal intensity \(\mu_s\) for each data point as well.

Since Eve could always perform a QND measurement to measure the photon number and since vacuum signals only contribute on the order of dark counts, we restrict the sum in \cref{Cor:Decoy Security} to single photons sent by Alice. Then, we can again use the methods from \cite{Winick2018Quantum} to evaluate the key rate. The required Kraus operators can be found in \cref{App:Kraus ops decoy BB84}.

\begin{figure}
    \centering
    \includegraphics[width=\linewidth]{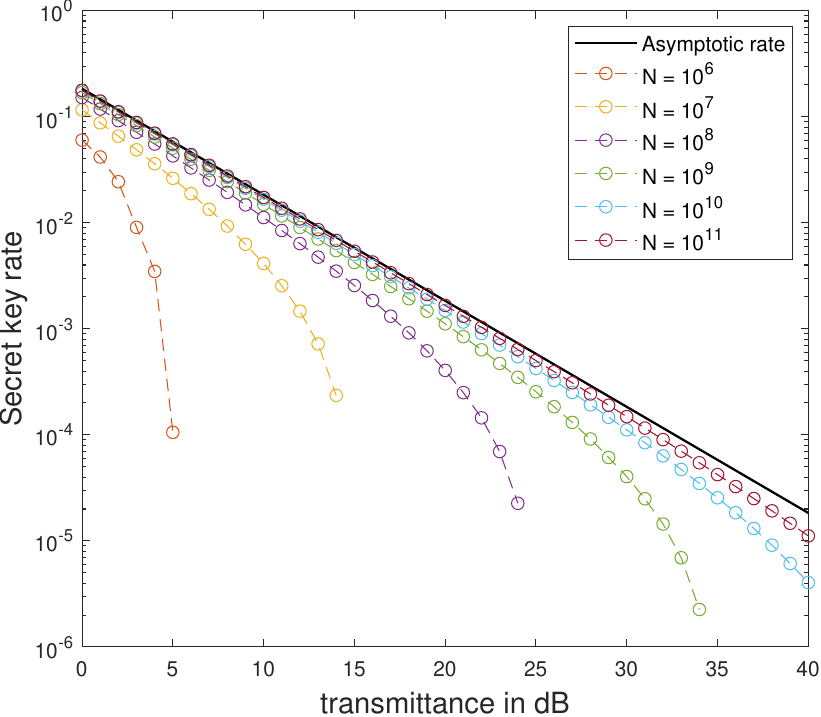}
    \caption{Secret key rate for \emph{IID attacks} of the decoy-state BB84 protocol using two decoy intensities against channel loss with a varying number of total signals sent \(N=10^6, \dots, 10^{11} \).}
    \label{fig:decoybb84}
\end{figure}

In \cref{fig:decoybb84} we show the resulting key rates over the total channel loss in \(\unit{dB}\) for a number of total signals ranging from \(N = 10^6\) to \(N=10^{11}\). With \(N =10^{11}\) we approximately recover the asymptotic limit up until \(\unit[40]{dB}\).

\section{Variable-length Decoy-state protocols} \label{sec:variable}

So far in this work we have studied \textit{fixed-length} QKD protocols, which either produce a key of fixed-length upon accepting or abort. Such protocols are in contrast with \textit{variable-length} protocols \cite{hayashiConcise2012,curras-lorenzoTight2021,kawakamiSecurity,Tupkary2024Phys.Rev.Res.} which allow Alice and Bob to determine the length of the key to be produced, and the number of bits to be used for error-correction, based upon their observed statistics $\Fobs$. 
Such protocols have the advantage that they do not suffer from the trade-off between the size of the acceptance test and the key rate upon acceptance. Furthermore, they do not require prior characterization of the honest behavior of the channel connecting the two users in order to determine a suitable acceptance test. Instead they can adapt to any value of $\Fobs$ that is observed. 

A generic security proof for variable-length protocols against IID collective attacks was recently obtained in \cite{Tupkary2024Phys.Rev.Res.}. In this section we will describe a variable-length security proof of decoy-state BB84. Similar to our earlier results, this can be lifted to coherent attacks using recent improvements \cite[Corollary 4.1] {nahar2024postselection} of
the postselection technique \cite{Christandl2009Phys.Rev.Lett.}.

 We will only present the final results here, and explain how to use them to implement a valid variable-length decoy-state BB84 protocol. The security proof combines techniques from \cite{Tupkary2024Phys.Rev.Res.} and this work, and is provided in \cref{App:Variable decoy,App:Renyi}.
 
\subsection{Variable-length protocol steps}
\begin{prot}[Variable-length Protocol]\label{Prot:Variable Protocol}
	In implementing a variable-length protocol, the following steps from \cref{Prot:PM Protocol} and \cref{Prot:Decoy Protocol} are modified.
	\begin{enumerate}[4$^\prime$.]
		\item \textbf{Variable length decision:} Alice and Bob compute the observed statistics $\Fobs$. They use this value to determine $\leak(\Fobs)$, the number of bits to be used for error-correction and $l(\Fobs)$, the length of the final key to be produced. $\leak(.)$ and $l(.)$ are fixed functions that are determined before the start of the QKD protocol.
        \item[6$^\prime$.] \textbf{Error Correction:} Alice and Bob use $\leak(\Fobs)$ bits for error-correction. The error-verification step is the same as before.
        \item[7$^\prime$.] \textbf{Privacy Amplification:} Alice and Bob apply a common two-universal hash function to produce a key of length $l(\Fobs)$. 
	\end{enumerate}	
\end{prot}

Note that in a QKD protocol, one starts with a fixed but unknown state $\rho_{AB}$ representing Eve's attack. This then gives rise to a random variable $\Fobs$. Our variable-length protocol will use $\Fobs$ to determine the parameters of the protocol.
In order to prove the security of a variable-length protocol, we  will first need to construct a set $V(\Fobs)$ such that
\begin{equation} \label{eq:VFobsrequirement}
    \Pr_{\Fobs}(\rho_{AB} \in V(\Fobs) ) \geq 1 - \epsAT.
\end{equation}
This set $V(\Fobs)$ is roughly analogous to the set $S_{\vec{\nu}}$ we constructed for fixed-length protocols.  

\subsection{Constructing $V(\Fobs)$}
A construction of $V(\Fobs)$ satisfying \cref{eq:VFobsrequirement} was provided in \cite{Tupkary2024Phys.Rev.Res.} for the qubit BB84 protocol, using the concentration inequalities utilized in \cite{George2020Phys.Rev.Res.}.
We will now provide a construction of $V(\Fobs)$ for \cref{Prot:Variable Protocol}. The following lemma constructs $V(\Fobs)$ by using the widely used ``Clopper-Pearson'' confidence interval \cite{Clopper1934Biometrikaa,Rao2000} on binomial random variables \footnote{These intervals can be easily constructed using preinstalled functions in MATLAB.}.

\begin{lem} \label{lemma:VFobsconstruction}
	For any state $\rho$ satisfying $\rho_A = \Tr_{A'}[\ketbra{\psi}]\}$, let $F^\text{obs} \in \mathcal{P}(\Sigma )$ where $\Sigma = \Sigmat\cup \{sift,\bot \}$ be the frequency vector obtained from measuring the state $N$ times. Define parameters
	\begin{equation} \label{eq:confinterval}
		\begin{aligned}
			\kappal{k} &:= \Fobs_k - B\left(\frac{\varepsilon_{\text{AT}}}{ 2| \Sigma|} ; NF^\text{obs}_k,  N-NF^\text{obs}_k+1\right) \\
			\kappau{k} &:= B\left(1-\frac{\varepsilon_{\text{AT}}}{2 | \Sigma|} ; NF^\text{obs}+1, N- NF^\text{obs}_k\right) - \Fobs_k \\
		\end{aligned}
	\end{equation}
	and the set
	\begin{equation}
 \begin{aligned} \label{eq:VFobs}
		V(F^\text{obs}) := \{  \rho \in  S_\circ(AB) \; | \;   \rho_A = \Tr_{A'}[\ketbra{\psi}]\} , \\
  \Fobs_k - \kappal{k} \leq \Tr(\Gamma_k \rho) \leq \Fobs_k + \kappau{k}, \forall k \in \Sigma  \}
  \end{aligned}
	\end{equation}
	where $B(p;x,y)$ is the $p$th quantile of the beta distribution with shape parameters $x,y$ \cite{Clopper1934Biometrikaa,Rao2000}, and $\Gamma_k$ is the POVM element corresponding to the $k$th measurement outcome.  
	Then, $\rho$ is contained in $V(F^\text{obs})$ with probability greater than $1-\varepsilon_{\text{AT}}$. That is,
	\begin{equation}
		\Pr_{F^\text{obs}} \left( \rho \in V(F^\text{obs}) \right) \geq 1-\varepsilon_{\text{AT}}.
	\end{equation}
\end{lem}

\begin{proof}
	Consider the frequency of the $k$th event, which happens in each round with probability $\Tr(\Gamma_k \rho) = p_k$.  This is a binomial random variable, since in each round it either happens or does not happen.  Given that one observed $F^\text{obs}_k N$ events after sampling $N$ times, we obtain the ``Clopper-Pearson confidence interval'' \cite{Clopper1934Biometrikaa,Rao2000} for $p_k$, as
		\begin{equation}
			\Pr_{F^\text{obs}} \left(  p_k \in [\Fobs_k - \kappal{k}, \Fobs_k+\kappau{k}]  \right) \geq 1-\frac{\varepsilon_{\text{AT}}}{|\Sigma|},
		\end{equation}
		where $\kappal{k},\kappau{k}$ are defined in \cref{eq:confinterval}. Combining the confidence intervals for each $k \in \Sigma$, we obtain
		\begin{equation}
			\begin{aligned}
				\Pr_{F^\text{obs}} \left( p_k \notin [\Fobs_k - \kappal{k}, \Fobs_k+\kappau{k}] \right) &\leq \frac{\varepsilon_{\text{AT}}}{|\Sigma|}, \\
				\Pr_{F^\text{obs}} \left( \bigcup_{ k \in \Sigma} p_k \notin [\Fobs_k - \kappal{k}, \Fobs_k + \kappau{k}]\right ) &\leq \sum_{k \in \Sigma}  \frac{\varepsilon_{\text{AT}}}{|\Sigma|} = \varepsilon_{\text{AT}}, \\
				\Pr_{F^\text{obs}} \left( \bigcap_{ k \in \Sigma} p_k \in [\Fobs_k - \kappal{k}, \Fobs_k+ \kappau{k}]  \right) &\geq 1-\varepsilon_{\text{AT}}, \\
				\Pr_{F^\text{obs}} \left( \rho \in V(F^\text{obs}) \right) &\geq 1-\varepsilon_{\text{AT}},
			\end{aligned}
		\end{equation}
		where we used the union bound for probabilities to obtain the second inequality.
  \end{proof}

\subsection{Variable-length key rates}

 Using the above construction of $V(\Fobs)$, we obtain the following statement concerning the variable-length security of QKD protocols. The theorem is proved in \cref{App:Variable decoy}.

\begin{restatable}{thrm}{varlengththeorem}\label{Thrm:Var length}
	Consider a variable-length protocol that, on obtaining $F^\text{obs}$ during the variable-length decision (acceptance test) and passing error-verification, produces a key of length $l(F^\text{obs})$ using $\leak{}(F^\text{obs})$ bits for error-correction. Let  $\lambda(F^\text{obs}) $ be any deterministic function of $F^\text{obs}$. Let $\Nsift$ denote the observed number of sifted rounds, and let $l(F^\text{obs})$ be given by
	\begin{equation} \label{eq:lfobsvalue}
		\begin{aligned}
		&l(F^\text{obs}):= \max \bigg( 0,  \Nsift  \min_{\rho \in V(F^\text{obs}) }  \frac{H(Z|CE)_\rho}{\Pr(\text{sift})} -  \leak(F^\text{obs})  \\
		&- \Nsift (\alpha-1)\log^2 (\dim(Z)+1) - \log(2/\epsEV) \\
  &-\frac{\alpha}{\alpha-1} \left( \log(\frac{1}{4\varepsilon_{\text{PA}}}) + \frac{2}{\alpha} \right)    \bigg)   
		\end{aligned}
	\end{equation}
	where $1 < \alpha \leq 1+ 1/\log(2 \dim(Z)+1)$.
	Then the variable-length protocol is $(\varepsilon_\text{EV}+\varepsilon_\text{AT} + \varepsilon_\text{PA})$-secure.
\label{Thrm:PM QKD variable security using Renyi Entropy}
\end{restatable}

Comparing with the fixed-length formula from \cref{Thrm:PM QKD security}, we notice two main differences. First, we optimize over a different set $V(\Fobs)$ in \cref{Thrm:PM QKD variable security using Renyi Entropy} as compared to  $S_{\vec{\nu}}$ in \cref{Thrm:PM QKD security}. (However, the optimization problem can be solved with the same methods). Second, the correction terms are different due to the use of R\'enyi entropies in the proof of \cref{Thrm:PM QKD variable security using Renyi Entropy}.
The above statement is valid for any value of $\alpha$ satisfying the requirements. However, if one expects to observe $n_\text{sift}$ sifted rounds (i.e $\Nsift \approx n_\text{sift}$), then it is optimal to choose
$\alpha=1+\frac{\sqrt{\log(1/\varepsilon_\text{PA})}  }{\log( \dim(Z)+1) \sqrt{n_\text{sift}}}$. 
The main difference between the \cref{Thrm:PM QKD variable security using Renyi Entropy} and Ref.~\cite[Theorem 2]{Tupkary2024Phys.Rev.Res.} is that the correction term in \cref{Thrm:PM QKD variable security using Renyi Entropy} depends on $\Nsift$ instead of $N$, and is therefore much smaller at high loss, leading to higher key rates. 
 In order to compute the minimization using the numerical framework of \cite{Winick2018Quantum}, we use 
\begin{equation}
\begin{aligned}
    \min_{\rho \in V(F^\text{obs}) }  \frac{H(Z|CE)_\rho}{\Pr(\text{sift})} &\geq \frac{ \min_{\rho \in V(F^\text{obs}) }   H(Z|CE)_\rho}{ \max_{\rho \in V(F^\text{obs}) }   \Pr(\text{sift})}  \\
    & =\frac{ \min_{\rho \in V(F^\text{obs}) }   H(Z|CE)_\rho}{   \Fobs_\text{sift} + \kappau{sift} }  
    \end{aligned}
\end{equation}

\subsection{Variable-length key rates for decoy states}

To obtain the statement for variable-length security of decoy-state protocols, we simply have to repeat the steps from \cref{sec:Decoy-State Methods in Finite-size regime}. These steps adapt the optimization problem used in key rate calculations to account for the specifics of the decoy state method.
This allows us to go from 
\cref{Thrm:PM QKD variable security using Renyi Entropy} to \cref{Cor:DecoyVarLength} for variable-length protocols in the same manner as we did from
\cref{Thrm:PM QKD security} to \cref{Cor:Decoy Security} for fixed-length protocols.

 The main difference is that we minimize $H(Z|CE)$  over the set $V(F^\text{obs})$ (\cref{eq:lfobsvalue}) instead of the set $S_{\vec{\nu}}$ (\cref{eq:lvalue}). This boils down to replacing the $\nu_k+t_k$ ($-\nu_k-t_k$) with $\kappau{k}$($-\kappal{k}$).
In doing so,  instead of \cref{eq:decoy bounds yields} we obtain

\begin{equation}
	\begin{aligned} \label{eq:decoy1varlength}
	\frac{F^\text{obs}_k + \kappau{k}}{\Pr(a,\mu,\text{test})} &\geq \sum_{n=0}^{\infty} \Pr(n|a,\mu,\mathrm{test}) \; Y_n^{a,b}, \\
	\frac{F^\text{obs}_k - \kappal{k}}{\Pr(a,\mu,\text{test})} &\leq \sum_{n=0}^{\infty} \Pr(n|a,\mu,\mathrm{test}) \; Y_n^{a,b}.
\end{aligned}
\end{equation}
Instead of \cref{eq:LP decoy}, we obtain
	\begin{equation}\label{eq:LP decoy Var}
	\begin{aligned}
		Y_{n,L}^{ab} := \min\;  & Y^{a,b}_n\\
		\textrm{s.t.}\; & \frac{F^\text{obs}_k - \kappal{k}}{\Pr(a,\mu,\text{test})} \leq \sum_{m \leq N_{\mathrm{ph}}} \Pr(m|a,\mu,\mathrm{test}) \; Y_m^{a,b} \\&\hspace{80pt}  + \left(1- p_{\text{tot}}(\mu|a) \right), \\
		&\frac{F^\text{obs}_k + \kappau{k}}{\Pr(a,\mu,\text{test})} \geq \sum_{m\leq N_{\mathrm{ph}}} \Pr(m|a,\mu,\mathrm{test}) \; Y_m^{a,b}, \\
		&\forall k = (a,b,\mu,\text{test}) \in \Sigma,  \\
		&0 \leq Y^{a,b}_m \leq 1 \; \forall m \in \N_0,  
	\end{aligned}
\end{equation}
Moreover, \cref{eq:decoy3} is unchanged, and we repeat it here:
\begin{equation} \label{eq:decoy3varlength}
	\begin{aligned}
		\gamma_{n,L}^{ab} &= \Pr(a,\mathrm{test}|n) Y_{n,L}^{ab}, \\
		\gamma_{n,U}^{ab} &= \Pr(a,\mathrm{test}|n) Y_{n,U}^{ab}.
	\end{aligned}
\end{equation}

 Thus, by repeating the analysis of the term $H(Z|CE)$ in the main text (from \cref{sec:Decoy-State Methods in Finite-size regime}), along with \cref{Thrm:PM QKD variable security using Renyi Entropy} we obtain the following Corollary for the variable-length BB84 decoy-state protocol, similar to \cref{Cor:Decoy Security}.

\begin{cor}[Variable-length Security for Decoy-state Protocols]\label{Cor:DecoyVarLength}
	Consider the variable-length decoy-state protocol that, on obtaining $F^\text{obs}$ during the variable-length decision (acceptance test) and passing error-verification, produces a key of length $l(F^\text{obs})$ using $\leak{}(F^\text{obs})$ bits for error-correction. Let $\leak(\Fobs)$ by any deterministic function of $\Fobs$.  Let $\Nsift$ denote the observed number of sifted rounds, and let $l(F^\text{obs})$ be given by
	\begin{equation}\label{eq:lfobsvaluefinal}				
		\begin{aligned}
		&l(F^\text{obs}):= \max \bigg( 0,    \\
        &\frac{\Nsift}{\Fobs_\text{sift} + \kappau{sift}} \sum_{n=0}^{\infty}  \Pr(n) \min_{\rho^{(n)} \in \tilde{V}_n(F^\text{obs})} H(Z|C E;n)_{\rho^{(n)}} \\
		& -  \leak(F^\text{obs})  - \Nsift (\alpha-1)\log^2 (d_Z+1) - \log(2/\epsEV) \\
		& - \frac{\alpha}{\alpha-1} \left( \log(\frac{1}{4\varepsilon_{\text{PA}}}) + \frac{2}{\alpha} \right)  \bigg)   
		\end{aligned}
		\end{equation}
	where $1 < \alpha \leq 1+ 1/\log(2 \dim(Z)+1)$. Moreover, the set $\tilde{V}_n(F^\text{obs})$ corresponds to  \(n\)-photon subspaces and are defined by
    \begin{equation}
        \begin{aligned}
        \tilde{V}_n(F^\text{obs}):= \{&\rho^{(n)}\in \mathcal{S}_{\circ}(\Hil_A\otimes \Hil_B) | \forall (a,b,\mu,\text{test}) \in \Sigmat \\ 
            &\gamma_{n,L}^{a,b} \leq \Tr[\Gamma_{a,b} \rho^{(n)}] \leq \gamma_{n,U}^{a,b}  \\
        &	\rho_A^{(n)} = \Tr_{A_sA'}[\ketbra{n}_{A_s}\ketbra{\xi}] \}.
        \end{aligned}
    \end{equation}
	where $\gamma_{n,L}^{a,b},\gamma_{n,U}^{a,b} $ are computed using \cref{eq:decoy1varlength,eq:LP decoy Var,eq:decoy3varlength} and $\rho_A^{(n)} = \Tr_{A_sA'}[\ketbra{n}_{A_s}\ketbra{\xi}]$ is the same as in \cref{Cor:Decoy Security}. Then, the variable-length protocol is $(\varepsilon_\text{EV}+\varepsilon_{\text{AT}} + \varepsilon_{\text{PA}})$-secure.
\end{cor}

Thus, computing variable-length key rate only introduced minor modifications to the calculations for fixed-length key rates. In particular, the constraints in the optimization are modified, and the second-order correction terms in the key rate calculations are modified.

\section{Coherent Attacks}\label{sec:Coherent Attacks}
So far in this work, we have considered IID collective attacks. To lift our security statements (for both fixed-length and variable-length protocols) to  coherent attacks, we use the postselection technique \cite{Christandl2009Phys.Rev.Lett.,nahar2024postselection}. In this section we first explain how this lift can be applied to our protocols, and then apply it to a concrete example. We follow the recipe of applying the postselection technique outlined in \cite[Section V A]{nahar2024postselection}.

To use the postselection technique, one first needs to verify that the protocol in question is permutation invariant. For protocols implementing IID measurements, this requirement can be enforced by Alice and Bob performing a common random permutation of their measurement results \cite[Appendix B]{nahar2024postselection}. Then, one must compute the correct effective ``dimension" (denoted by $x$) for the protocol in question, which appears as a cost in using the postselection technique via the parameter 
\begin{equation}
	g_{n,x} = {n + x - 1 \choose n} \leq \left(\frac{e (n+x-1)}{x-1}\right)^{x-1} .
\end{equation}  Note that for optical protocols, Alice sends infinite-dimensional states and Bob measures using infinite-dimensional POVMs, and therefore the computation of $x$ is non-trivial. In \cite[Section V A]{nahar2024postselection}, a recipe for the computation of $x$ for optical protocols is outlined, which makes use of source maps and squashing maps, in additional to block-diagonal symmetry.  For decoy-state protocols, $x$ is computed via 
\begin{equation}
	 x = n_\text{int}^2  (N_{\mathrm{ph}}+2)d_A^2\left(\sum_{i=0}^{N_B } (i+1)^2 + n_\text{meas}\right). 
\end{equation}
where $ n_\text{int}$ is the number of decoy intensities used in the protocol, $d_A$ is the dimension of Alice's subsystem $A$, and $N_{\mathrm{ph}}$ is the photon number cutoff used for the decoy-analysis. Here $N_B$ denotes the photon number cutoff used in the weight-preserving flag-state squasher on Bob's detection setup, and $n_\text{meas}$ is the total number of flags in Bob's detection setup.

Once this $x$ is determined, the  lift to coherent attacks is then stated in the following Corollary

\begin{cor}(\cite[Corollary 4.1]{nahar2024postselection}) \label{cor:ps}
	Let $ \mathcal{E}$ be a permutation invariant QKD protocol such that the $\varepsilon_{\text{sct}}$-secret  property holds for all IID states $\rho_{ABE}$ satisfying $\Tr_{BE}(\rho_{ABE}) = \sigma_A$. We use $l_1 \dots l_M$ to denote the possible output lengths of the protocol.  Let $ \mathcal{E}^\prime$ be a protocol identical to $ \mathcal{E}$, except that it produces a key of length $l^\prime_i$ in place of $l_i$. Then, $ \mathcal{E}^\prime$  is $g_{n,x} \left( \sqrt{8 \varepsilon_{\text{sct}} }   + \tilde{\varepsilon}/2 \right)$-secret against coherent attack (for input states satisfying $ \Tr_{B^nE^n}(\rho_{A^nB^nE^n}) = \sigma^{\otimes n}_A $) if $l^\prime_i = l_ i -2\log(g_{n,x}) - 2 \log(1/\tilde{\varepsilon})$.
\end{cor}
	
In other words, we have to increase the security parameter and decrease the key lengths as specified in the above Corollary to obtain secrecy for coherent attacks. The restriction to states with a fixed marginal ($\sigma_A$) comes from the fact that we implement a prepare-and-measure protocol. Note that we only need the postselection lift for the $\varepsilon_{sct}$-secret property of QKD protocols. The $\varepsilon_{cor}$-correct property directly holds for coherent attacks as well. 
	
\subsection{Example: Variable-length Decoy 4-6 protocol against coherent attacks}\label{sec:Example Decoy 4-6}
As the final example we present the 4-6 protocol with one signal and one decoy intensity using the variable length framework and proving security against coherent attacks. We use polarization encoding in which Alice sends horizontal/vertical and diagonal/anti-diagonal light and Bob measures in the HV- (horizontal/vertical), DA- (diagonal/anti-diagonal) and RL-basis (right/left-circular).

This protocol is different from the previous ones in many ways. To illustrate the flexibility of our framework, we make the following choices. In key generation rounds Alice biases her signals towards the \(Z\)-basis and in test rounds she sends all states with equal probability, i.e.
\begin{align}
    p_{z|\text{gen}} &= \frac{3}{4} \quad p_{x|\text{gen}} = \frac{1}{4}, \\
    p_{z|\text{test}} &= \frac{1}{2} \quad p_{x|\text{test}} = \frac{1}{2}.
\end{align}
Furthermore, we assume that every signal state is sent with a different but time-constant intensity. This could occur in a real experimental setup if different laser diodes are used for each polarization. Again, to showcase the flexibility of our framework, we chose the following intensity sets (ordered H,V,D,A)
\begin{align}
    \mu_s = \{0.9,0.5,0.6,1.1\},\\
    \mu_2 = \{0.1,0.2,0.01,0.09\}
\end{align}
In key generation rounds only the signal intensity set is used and in test rounds all intensities are sent with equal probability, i.e. \(p(\mu|\text{test}) = 1/3\) for all \(\mu\).

On Bob's side, we selected symmetric passive 6-state receiver, thus \(p_z^B= p_x^B=p_y^B = 1/3\). We utilize the weight preserving flag-state squasher from \cite{nahar2024postselection} with the extensions to arbitrary passive linear optical setups of \cite{Kamin2024Phys.Rev.Res.}. This squashing map allows for deviations from the perfect symmetric setup in contrast to the previous squashing maps. It also requires a choice on the photon number cut-off for Bob, which we set to \(N_B=1\), as this proved to be sufficient in the asymptotic regime. For this example the same held true in the finite-size regime by our numerical testing. The Kraus operators for this protocol can be found in \cref{App:Kraus ops 4-6}.

For a given target secrecy parameter of $\varepsilon_{\text{sct}}$, using \cref{cor:ps}, we set $\tilde{\varepsilon} / 2  = \sqrt{8 \varepsilon_\text{sct}^{IID}}   = \varepsilon_{\text{sct}} / 2 g_{n,x}$. For the key rate calculation after using the postselection lift, we set $\epsAT =  \epsPA = \varepsilon_\text{sct}^{IID} / 2$. We chose a photon number cut-off of \(N_{\text{ph}} = 2\) and for simplicity kept Alice's system four dimensional.

Therefore, the computed value of $x$ of our protocol is given by
\begin{equation}
    x = 2^2 (4) 4^2 (5+7) = 3072.
\end{equation}
Finally, for error correction efficiency and security parameter we choose as before \(f_{\text{EC}} = 1.16\), a target security parameter \(\varepsilon_{\text{sec}} =  10^{-12}\) and also assumed a misalignment of \(\theta_{\text{misalign}}=0.01\). The only parameter we optimize in this example is the testing probability. We do not optimize the intensities because the optimum is always achieved for equal intensities and thus would contradict the goal of showing key rates of unequal intensities.

\begin{figure}[th]
	\centering
	\includegraphics[width=\linewidth]{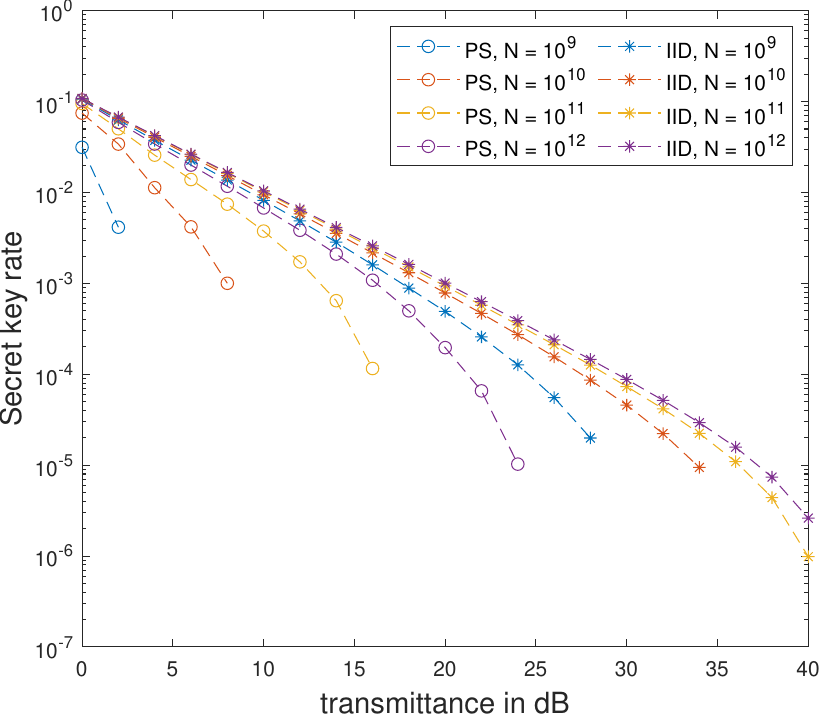}
	\caption{Comparison between secret key rates for \emph{coherent attacks} (PS) using \cref{cor:ps} and for \emph{IID attacks} (IID) using \cref{Cor:Decoy Security} of the decoy-state 4-6 protocol. Both protocols use one decoy intensity and the key rates are plotted against the channel loss with a varying number of total signals sent \(N=10^9, \dots, 10^{12} \).}
	\label{fig:decoy46}
\end{figure}

The resulting variable-length key rates for both IID (stars) and coherent attacks (circles) can be seen in \cref{fig:decoy46} for different numbers of total signals sent \(N=10^9, \dots ,10^{12}\). An equal number of signals sent is represented with the same color of the curve. We plot the secret key rate per signal when the observed frequency \(F^{\text{obs}}\) is equal to the frequency due to the honest behavior.

As expected, the key rates using the postselection technique are much smaller, since we allow for more general attacks. However, we are still able to achieve non-zero key rates for \(N= 10^{9}\) signals sent and can reach loss values of \(\unit[25]{dB}\) with \(N=10^{12}\) signals.

Finally, although not explicitly shown here, we also observed that due to finite-size effects two intensities actually perform better than three, which was as well previously seen in \cite{Rusca2018Appl.Phys.Lett.}.
	
\section{Conclusion}
First, we established a finite-size security proof against IID collective attacks for general P\&M protocols using probabilistic testing. Our approach improves upon previous methods by utilizing \cref{Lem:Bound Smooth Min-Entropy sifted signals} and \cref{Thrm:Secrecy outside Feasible Set}, leading to a significant increase in the finite-size scaling. These improvements stem from reducing the finite-size correction terms to scale with the detected signals and tightening the feasible set through the derivation of sharper concentration inequalities.

Next, we applied the IID security proofs from \cref{Thrm:PM QKD security} and \cref{Cor:Decoy Security} to qubit-based and decoy-state protocols, respectively. Notably, unlike methods based on the entropic uncertainty relation (EUR), such as those in \cite{Tomamichel2017Quantum, Lim2014Phys.Rev.A}, our approach is versatile and applicable to a broader range of setups, especially those with passive detection setups. As a concrete example, we presented a decoy-state BB84 protocol with a symmetric passive receiver.

Additionally, for collective attacks, we estimated the expected key rate, enabling a balance between achieving high acceptance probabilities and optimizing key rates.

Moreover, we extended our findings to the variable-length setting, as detailed in \cref{Thrm:Var length} for qubits and \cref{Cor:DecoyVarLength} for decoy-state protocols. These theorems build upon the results in Ref.~\cite{Tupkary2024Phys.Rev.Res.} by incorporating a tighter statistical analysis akin to the fixed-length approach, leading to improved performance.

Finally, we extended our results to coherent attacks using the postselection technique \cite{nahar2024postselection}, which notably avoids relying on specific assumptions about the detection setup. For coherent attacks, we presented the reference-frame independent 4-6 protocol as an example. Especially, we showcased the flexibility of our approach by incorporating intensity choices differing with each signal sent. Additionally, we extended the asymptotic results of requiring only one decoy intensity \cite{Kamin2024Phys.Rev.Res.} to the finite-size scenario.

\section*{Author Contributions}
The first author LK performed the derivations and proofs of the sections I - VII and section IX in the main text and App. A - C and App. F, the second author DT derived the results pertaining to variable-length key rates of Section VIII and App. D - E. NL provided support in the preparation of the manuscript and supervision.

\section*{Code Availability}
The code used to prepare the results in this paper will be available at \href{https://openqkdsecurity.wordpress.com/repositories-for-publications/}{https://openqkdsecurity.wordpress.com/repositories-for-publications/}.

\section*{Acknowledgments}
The research has been conducted at the Institute for Quantum Computing, at the University of Waterloo, which is supported by Innovation, Science, and Economic Development Canada and the NSERC Alliance. Support was provided by NSERC under the Discovery Grants Program, Grant No. 341495, Alliance Grant QUINT and Honeywell. Furthermore, this work was supported in part by funding from the Innovation for Defence Excellence and Security (IDEaS) program from the Department of National Defence (DND). DT is partially funded by the Mike and Ophelia Lazaridis Fellowship.

\bibliography{Finite_size_decoy}
	
\appendix
\section{Technical Definitions and Lemmas}\label{App:Technical Definitions and Lemmas}
\begin{defn}[Normalised and sub-normalised conditional states]
	Let \(\rho \in S_{\bullet}(DX)\) be a classical-quantum state, i.e. \(\rho_{DX} := \sum_{x \in \mathcal{X}} \alpha_x \sigma_x \otimes \ketbra{x}\) with \(\alpha_x >0\) and \(\sigma_x \in S_{\bullet}(D) \) for all \(x \in \mathcal{X}\). The \emph{sub-normalised} conditional state \(\rho_{DX \wedge \Omega}\) for an event \(\Omega:\mathcal{X} \rightarrow \{0,1\}\) is defined as
	\begin{equation}
		\rho_{DX \wedge \Omega} := \sum_{x \in \Omega} \alpha_x \sigma_x \otimes \ketbra{x}.
	\end{equation}
	The \emph{normalised} conditional state \(\rho_{DX_{|_{\Omega}}}\) is defined as
	\begin{equation}
		\rho_{DX|\Omega} := \frac{\Tr[\rho_{DX}]}{\Tr[\rho_{DX \wedge \Omega}]} \rho_{DX \wedge \Omega}.
	\end{equation}
\end{defn}

\begin{defn}[{Min-Entropy as in \cite[Definition 6.2]{Tomamichel2016}}]
	Let \(\rho_{AB} \in S_{\bullet}\). The min-entropy of \(A\) conditioned on \(B\) of the state \(\rho_{AB}\) is
	\begin{equation}
		\Hmin(A|B)_{\rho} := \sup_{\sigma_B \in S_{\bullet}(B)} \sup\{\lambda \in \R : \rho_{AB} \leq 2^{-\lambda} \1_A \otimes \sigma_B \}
	\end{equation}
\end{defn}
Note in \cite{Tomamichel2016} \(e^{-\lambda}\) was used instead of \(2^{-\lambda}\).

\begin{defn}[{Smoothing Ball as in \cite[Definition 6.4]{Tomamichel2016}}]
	Let \(\rho \in S_{\bullet}(A)\) and \(0 \leq \varepsilon < \sqrt{\Tr[\rho]}\). We define the \(\varepsilon\)-smoothing ball of states in \(S_{\bullet}(A)\) around \(\rho\) as
	\begin{equation}
		B_{\varepsilon}(\rho) := \{ \tau \in S_{\bullet}(A) : P(\tau,\rho) \leq \varepsilon \},
	\end{equation}
    where \(P(\cdot,\cdot)\) is the purified distance (see \cite[Definition 3.8]{Tomamichel2016}).
\end{defn}

\begin{defn}[{Smooth Min-Entropy as in \cite[Definition 6.5]{Tomamichel2016}}]
	Let \(\rho \in S_{\bullet}(A)\) and \(\varepsilon \geq 0\). We define the \(\varepsilon\)-smooth min-entropy of \(A\) conditioned on  \(S_{\bullet}(A)\) around \(\rho\) as
	\begin{equation}
		\Hmine(A|B)_{\rho} := \sup_{\tilde{\rho} \in B_{\varepsilon}(\rho_{AB})} \Hmin(A|B)_{\tilde{\rho}},
	\end{equation}
\end{defn}

\begin{thrm}[{Leftover Hashing Lemma as in \cite[Prop. 9]{Tomamichel2017Quantum}}]\label{Thrm:LHL}
	Let \(\sigma_{XD} \in \mathcal{S}_{\bullet}\left(XD\right)\) be classical on \(X\) and \(\varepsilon \in \left[ 0, \sqrt{\Tr[\sigma_{XD}]} \right)  \). Let \(\Hil \) be a \(\text{universal}_2\) family of hash functions from \(\mathcal{X} = \{0,1\}^n\) to \(\mathcal{K}=\{0,1\}^l\). Moreover, let \(\rho_{S^H} = \frac{1}{\abs{\Hil}}\sum_{h \in \Hil} \ketbra{h}_{S^H} \). Then,
	\begin{equation}
		\frac{1}{2} \norm{\omega_{KS^HD} - \chi_K \otimes \omega_{S^HD}} \leq \frac{1}{2} 2^{-\frac{1}{2} \left(\Hmine(X|D)_{\sigma} - l\right) } + 2\varepsilon,
	\end{equation}
	where \(\chi_K = \frac{1}{2^l} \id_K\) is the fully mixed state and \(\omega_{KS^HD} = \Tr_X\left[\mathcal{E}_f\left(\sigma_{XD} \otimes \rho_{S^H} \right) \right]\) for the function \(f: \left(x,h\right) \mapsto h(x) \) that acts on the registers \(X\) and \(S^H\).
\end{thrm}
\begin{proof}
	See \cite[Appendix B]{Tomamichel2017Quantum}.
\end{proof}

\begin{lem}[{Lemma 10 as in \cite{Tomamichel2017Quantum}}]\label{Lem:Lemma 10}
	Let \(\rho_{ABXY} \in \mathcal{S}_{\bullet}\left(ABXY\right)\) be classical on \(X\) and \(Y\) let \(\Omega: \mathcal{X} \cross \mathcal{Y} \rightarrow \{0,1\}\) be an event with \(\Pr[\Omega]_{\rho} > 0 \). Then for \(\varepsilon \in \left[0, \sqrt{\Pr[\Omega]_{\rho}} \right)\), we have
	\begin{align}
		\Hmine(AX | BY)_{\rho \wedge \Omega} &\geq \Hmine(AX|BY )_{\rho}, \\
		\Hmine(AX | B)_{\rho \wedge \Omega} &\geq \Hmine(AX|B)_{\rho}.
	\end{align}
\end{lem}
\begin{proof}
	See the proof in \cite[Lemma 10]{Tomamichel2017Quantum}.
\end{proof}

For the security proof of \cref{Thrm:PM QKD security} in the main text, we require a version of the leftover-hashing lemma as in \cite[Prop. 9]{Tomamichel2017Quantum} but with partial conditioning. Typically, the leftover-hashing lemma is applied to a subnormalized conditional state, which requires evaluating the smooth min-entropy of the subnormalized conditional state. By using chain rules, this subnormalized conditioning is often removed, allowing the smooth min-entropy of an unconditional and normalized state to be evaluated instead. 

However, for composite events such as \(\Omega := \Omega_1 \wedge \Omega_2\), it may be beneficial to retain some partial conditioning, particularly in the context of detected events. To address this, we present a version of the leftover-hashing lemma that allows partial conditioning to be preserved for composite events. The proof simply applies \cref{Thrm:LHL} and is stated for convenience of the reader.

\begin{cor}[Leftover-Hashing with Partial Conditioning]\label{Cor:Conditional LHL}
	Let \(\sigma_{XYD} \in \mathcal{S}_{\circ}\left(XYD\right)\) be classical on \(X\) and \(Y\) \(\varepsilon \in \left[ 0, \sqrt{\Tr[\sigma_{XYD}]} \right)  \). Let \(\Hil \) be a \(\text{universal}_2\) family of hash functions from \(\mathcal{X} = \{0,1\}^n\) to \(\mathcal{K}=\{0,1\}^l\). Moreover, let \(\rho_{S^H} = \frac{1}{\abs{\Hil}}\sum_{h \in \Hil} \ketbra{h}_{S^H} \). Similar to \cref{Thrm:LHL} let \(\omega_{KS^HYD_{|_{\Omega}}} = \Tr_X\left[\mathcal{E}_f\left(\sigma_{XYD_{|_{\Omega}}} \otimes \rho_{S^H} \right) \right]\) and \(\Omega := \Omega_1 \wedge \Omega_2: \mathcal{X} \cross \mathcal{Y} \rightarrow \{0,1\}\) be an event with \(\Pr[\Omega]_{\sigma} > 0 \). Then,
	\begin{equation}\label{eq:conditional LHL}
		\begin{split}
			\frac{1}{2} \Pr[\Omega] \; &\norm{\omega_{KS^HYD_{|_{\Omega}}} - \chi_K \otimes \omega_{S^HYD_{|_{\Omega}}} } \\ 
			&\leq \frac{1}{2} 2^{-\frac{1}{2} \left(\Hmine(X|YD)_{\sigma_{|_{\Omega_1}}} - l \right)} + 2\varepsilon,
		\end{split}
	\end{equation}
	where \(\chi_K = \frac{1}{2^l} \id_K\) is the fully mixed state.
\end{cor}
\begin{proof}
	First, observe that the probability of the event \(\Omega := \Omega_1 \wedge \Omega_2\) is bounded by
	\begin{equation}
		\Pr[\Omega] = \Pr[\Omega_2 | \Omega_1 ] \Pr[\Omega_1] \leq \Pr[\Omega_2 | \Omega_1 ].
	\end{equation}
	Furthermore, we can write the state \(\omega_{KS^HYD_{|_{\Omega}}}\) conditioned on the event \(\Omega\) as
	\begin{equation}
		\omega_{KS^HYD_{|_{\Omega}}} = \frac{\left(\omega_{KS^HYD_{|_{\Omega_1}}}\right)_{\wedge \Omega_2}}{\Pr[\Omega_2|\Omega_1]}.
	\end{equation}
	Hence, inserting both observations into the left-hand side of \eqref{eq:conditional LHL} gives
	\begin{align*}
		&\frac{1}{2} \Pr[\Omega] \; \norm{\omega_{KS^HYD_{|_{\Omega}}} - \chi_K \otimes \omega_{S^HYD_{|_{\Omega}}} } \\
		&\leq \frac{1}{2} \norm{\left(\omega_{KS^HYD_{|_{\Omega_1}}}\right)_{\wedge \Omega_2} - \chi_K \otimes \left(\omega_{S^HYD_{|_{\Omega_1}}}\right)_{\wedge \Omega_2} } \\
		&\leq \frac{1}{2} 2^{-\frac{1}{2} \left(\Hmine(X|DY)_{\left(\sigma_{|_{\Omega_1}}\right)_{\wedge \Omega_2}} - l\right) } + 2\varepsilon \\
		&\leq \frac{1}{2} 2^{-\frac{1}{2} \left(\Hmine(X|DY)_{\sigma_{|_{\Omega_1}}} - l\right) } + 2\varepsilon,
	\end{align*}
	where in the second step we used \cref{Thrm:LHL} and in the third step \cref{Lem:Lemma 10}.
\end{proof}

\begin{lem}[Conditioning on Classical Register]\label{Lem:Conditioning on Classical Register}
	Let $\rho_{ABZ} \in S_{\circ}(ABZ)$ be classical on \(Z\). Then,
	\begin{equation}
		\Hmine(A|BZ)_{\rho} \geq \min_{z \text{ s.t.} \Pr(z)\neq0} \Hmine(A|B)_{\rho^{(z)}}.
	\end{equation}
\end{lem}
\begin{proof}
    First, we will construct an explicit state in the \(\varepsilon\)-ball of \(\rho_{ABZ}\) and then show that it yields the claimed lower bound on the smooth min-entropy.
    
    By the definition of the smooth min-entropy, for each \(z \in \mathcal{Z}\), there exists an optimizer \(\sigma_z \in B_{\varepsilon}(\rho^{(z)})\) such that \(\Hmine(A|B)_{\rho^{(z)}} = \Hmin(A|B)_{\sigma_z}\). Next, let us define the state
	\begin{equation}
		\sigma := \sum_{z\in\mathcal{Z}} \Pr(z) \ketbra{z} \otimes \sigma_z \in S_{\bullet}(ABZ),
	\end{equation}
    which we will show to be contained in \(B_{\varepsilon}(\rho)\). For this purpose, consider the purified distance between \(\rho\) and \(\sigma\) 
	\begin{align}
		P(\rho,\sigma) &= P\left(\sum_{z\in\mathcal{Z}} \Pr(z) \ketbra{z} \otimes \rho^{(z)} , \sum_{z\in\mathcal{Z}} \Pr(z) \ketbra{z} \otimes \sigma_z \right) \\
		&\leq  \sup_{z\in \mathcal{Z}} P\left( \ketbra{z} \otimes \rho^{(z)} , \ketbra{z} \otimes \sigma_z \right) \\
		&= \sup_{z\in \mathcal{Z}} P\left( \rho^{(z)} , \sigma_z \right) \leq \varepsilon,
	\end{align}
	where for the first inequality \cite[eq. (3.59)]{Tomamichel2016} was used and for the third line we used the multiplicativity of the fidelity, which still applies to the generalized fidelity. The last inequality results from \(\sigma_z \in B_{\varepsilon}(\rho^{(z)})\).
    Hence, we found a state \(\sigma \in B_{\varepsilon}(\rho)\). 
    
    Next, we will use \(\sigma \in B_{\varepsilon}(\rho)\) to bound the smooth min-entropy of \(\rho\) by
	\begin{align}
		&\Hmine(A|BZ)_{\rho} \geq \Hmin(A|BZ)_{\sigma} \\
		&\geq \min_{z \text{ s.t.} \Pr(z)\neq0} \Hmin(A|B)_{\sigma_z} \\ &= \min_{z \text{ s.t.} \Pr(z)\neq0} \Hmine(A|B)_{\rho^{(z)}},
	\end{align}
	where for the second inequality the property of the min-entropy \cite[eq. (6.25)]{Tomamichel2016} was applied and the last equality follows from each \(\sigma_z\) being the optimizer such that \(\Hmine(A|B)_{\rho^{(z)}} = \Hmin(A|B)_{\sigma_z}\). 
\end{proof}

\section{Improved Decoy Methods}\label{App:Improved Decoy}
The improved decoy methods of \cite{arx_KL24} can also be adapted to the finite-size regime. The equivalent SDP results in
\begin{equation}\label{eq:SDP decoy}
    \begin{aligned}
        Y_{n,L}^{ab} := &\min_{\mathbf{Y}_m, J_m}  Y^{a,b}_n\\
        \textrm{s.t.}\; & \frac{\bar{F}_{k} -t_k - \nu_k^L}{\Pr(a,\mu,\text{test})} \leq \sum_{m \leq N_{\mathrm{ph}}} \Pr(m|a,\mu,\mathrm{test}) \; Y_m^{a,b} \\&\hspace{80pt}  + \left(1- p_{\text{tot}}(\mu|a) \right), \\
        &\frac{\bar{F}_{k} +t_k + \nu_k^U}{\Pr(a,\mu,\text{test})} \geq \sum_{m\leq N_{\mathrm{ph}}} \Pr(m|a,\mu,\mathrm{test}) \; Y_m^{a,b}, \\
		&Y_m^{a,b} = \Tr\left[J_m \left(M_b^B \otimes \left(\rho_a^{(m)}\right)^T \right)\right], \\
		&0 \leq Y^{a,b}_m \leq 1, \\
		&J_m \succeq 0, \; \Tr_B\left[J_m \right] = \1_{A'}, \\ 
		&\forall \mu \in \{\mu_1, \mu_2, \dots\}, \; \forall a,b, \;\forall m \in N_0, \\
        &\forall  k = (a,b,\mu,\text{test}) \in \Sigmat.
	\end{aligned}    
\end{equation}

\section{Kraus Operators for Example Protocols}\label{App: Kraus Ops}
Again, we are using the framework of \cite{Winick2018Quantum} together with the simplifications of \cite[App. A]{Lin2019Phys.Rev.X}.

\subsection{Qubit BB84}\label{App:Kraus ops BB84}
In the main text, we set \(p_z^B = p_x^B = 1/2\), but for generality, we provide the Kraus operators for arbitrary basis choices. Given a perfect qubit protocol, Bob's qubit POVM elements are
\begin{equation}
	\begin{aligned}
		M^B_{(Z,0)} &= p_z^B \begin{pmatrix} 0 & 0 &0 \\ 0& 1 & 0\\ 0&0 & 0 \end{pmatrix}, \;
		M^B_{(Z,1)} = p_z^B \begin{pmatrix} 0 & 0& 0\\ 0& 0 & 0\\ 0&0 & 1 \end{pmatrix}, \\
		M^B_{(X,0)} &= \frac{p_x^B}{2} \begin{pmatrix} 0&0 & 0 \\  0&1 & 1 \\ 0& 1 & 1 \end{pmatrix},
		\;
		M^B_{(X,1)} = \frac{p_x^B}{2} \begin{pmatrix} 0&0 &0   \\  0&1 & -1  \\ 0 & -1 & 1  \end{pmatrix}, \\
		M^B_{\bot} &=\begin{pmatrix} 1 & 0 & 0 \\ 0 & 0 & 0 \\ 0 & 0 & 0 \end{pmatrix}.
	\end{aligned}
\end{equation}
For Alice's side, the POVM elements are determined by the source replacement scheme \cite{Bennett1992Phys.Rev.Lett., Ferenczi2012Phys.Rev.A} and given by 
\begin{equation}
	\begin{aligned}
		M^A_{(Z,0)} &= \ketbra{0}, M^A_{(Z,1)} = \ketbra{1}, \\
		M^A_{(X,0)} &= \ketbra{2}, M^A_{(X,1)} = \ketbra{3}.
	\end{aligned}
\end{equation}
We only generate secret key from the key generation rounds. This can be viewed as a post-processing where a generation round happens for each of Alice's choices with probability \(p(\text{gen}|a)\). However, the generation round decision is independent of Alice's choices in this particular protocol (see main text for details \cref{sec:Example Qubit BB84}), hence \(p(\text{gen}|a) = p(\text{gen})\). The resulting Kraus operators are
\begin{equation}
	\begin{aligned}
		K_Z &= \left[
		\begin{pmatrix} 1 \\ 0 \end{pmatrix}
		_R  \otimes  \begin{pmatrix} 1 &  & &  \\ & 0 & & \\
			& & 0 & \\
			& & & 0 \end{pmatrix}_A +
		\begin{pmatrix} 0 \\ 1 \end{pmatrix}
		_R \otimes  \begin{pmatrix} 0 &  & &  \\ & 1 & & \\
			& & 0 & \\
			& & & 0 \end{pmatrix}_A \right] \\
		&\otimes \sqrt{p_z^B}
		\begin{pmatrix} 0 & & \\ & 1 & \\& & 1 \end{pmatrix}
		_B \otimes
		\begin{pmatrix} 1 \\ 0 \end{pmatrix}
		_{C}, \sqrt{p(\text{gen})}\\
		K_X &= \left[
		\begin{pmatrix} 1 \\ 0 \end{pmatrix}
		_R  \otimes  \begin{pmatrix} 0 &  & &  \\ & 0 & & \\
			& & 1 & \\
			& & & 0 \end{pmatrix}_A +
		\begin{pmatrix} 0 \\ 1 \end{pmatrix}
		_R \otimes  \begin{pmatrix} 0 &  & &  \\ & 0 & & \\
			& & 0 & \\
			& & & 1 \end{pmatrix}_A \right] \\
		&\otimes \sqrt{p_x^B}
		\begin{pmatrix} 0 & & \\ & 1 & \\& & 1 \end{pmatrix}
		_B  \otimes
		\begin{pmatrix} 0 \\ 1 \end{pmatrix}
		_{C} \sqrt{p(\text{gen})},
	\end{aligned}
\end{equation}
and
\begin{equation}
	\begin{aligned}
		Z_1 &=
		\begin{pmatrix} 1 & \\ & 0 \end{pmatrix}
		\otimes \1_{\dim_A \times\dim_B \times 2}, \\
		Z_2 &=
		\begin{pmatrix} 0 & \\ & 1 \end{pmatrix}
		\otimes \1_{\dim_A \times\dim_B \times 2}.
	\end{aligned}
\end{equation}

\subsection{Decoy BB84}\label{App:Kraus ops decoy BB84}
After applying the squashing map from \cite{Gittsovich2014Phys.Rev.A,Beaudry2008Phys.Rev.Lett.}, Bob's measurements act on a qubit again. In the main text, we set \(p_z^B = p_x^B = 1/2\), but for generality, we provide the Kraus operators for arbitrary basis choices. The resulting POVM elements coincide with the ones for a perfect qubit protocol, which again are
\begin{equation}
	\begin{aligned}
		M^B_{(Z,0)} &= p_z^B \begin{pmatrix} 0 & 0 &0 \\ 0& 1 & 0\\ 0&0 & 0 \end{pmatrix}, \;
		M^B_{(Z,1)} = p_z^B \begin{pmatrix} 0 & 0& 0\\ 0& 0 & 0\\ 0&0 & 1 \end{pmatrix}, \\
		M^B_{(X,0)} &= \frac{p_x^B}{2} \begin{pmatrix} 0&0 & 0 \\  0&1 & 1 \\ 0& 1 & 1 \end{pmatrix},
		\;
		M^B_{(X,1)} = \frac{p_x^B}{2} \begin{pmatrix} 0&0 &0   \\  0&1 & -1  \\ 0 & -1 & 1  \end{pmatrix}, \\
		M^B_{\bot} &=\begin{pmatrix} 1 & 0 & 0 \\ 0 & 0 & 0 \\ 0 & 0 & 0 \end{pmatrix}.
	\end{aligned}
\end{equation}
For Alice's side, the POVM elements are determined by the source replacement scheme \cite{Bennett1992Phys.Rev.Lett., Ferenczi2012Phys.Rev.A} and given by 
\begin{equation}
	\begin{aligned}
		M^A_{(Z,0)} &= \ketbra{0}, M^A_{(Z,1)} = \ketbra{1}, \\
		M^A_{(X,0)} &= \ketbra{2}, M^A_{(X,1)} = \ketbra{3}.
	\end{aligned}
\end{equation}
Still, as in the qubit case we only generate secret key from the key generation rounds. This can again be viewed as a post-processing, but now we additionally need to condition on single photons. Thus, a generation round happens for each of Alice's choices with probability \(p(\text{gen}|a,1)\). The generation round decision remains independent of Alice's choices in this particular protocol (see main text for details \cref{sec:Example Decoy BB84}), hence \(p(\text{gen}|a,1) = p(\text{gen}|1)\). The resulting Kraus operators are
\begin{equation}
	\begin{aligned}
		K_Z &= \left[
		\begin{pmatrix} 1 \\ 0 \end{pmatrix}
		_R  \otimes  \begin{pmatrix} 1 &  & &  \\ & 0 & & \\
			& & 0 & \\
			& & & 0 \end{pmatrix}_A +
		\begin{pmatrix} 0 \\ 1 \end{pmatrix}
		_R \otimes  \begin{pmatrix} 0 &  & &  \\ & 1 & & \\
			& & 0 & \\
			& & & 0 \end{pmatrix}_A \right] \\
		&\otimes \sqrt{p_z^B}
		\begin{pmatrix} 0 & & \\ & 1 & \\& & 1 \end{pmatrix}
		_B \otimes
		\begin{pmatrix} 1 \\ 0 \end{pmatrix}
		_{C} \cdot \sqrt{p(\text{gen}|1)}, \\
		K_X &= \left[
		\begin{pmatrix} 1 \\ 0 \end{pmatrix}
		_R  \otimes  \begin{pmatrix} 0 &  & &  \\ & 0 & & \\
			& & 1 & \\
			& & & 0 \end{pmatrix}_A +
		\begin{pmatrix} 0 \\ 1 \end{pmatrix}
		_R \otimes  \begin{pmatrix} 0 &  & &  \\ & 0 & & \\
			& & 0 & \\
			& & & 1 \end{pmatrix}_A \right] \\
		&\otimes \sqrt{p_x^B}
		\begin{pmatrix} 0 & & \\ & 1 & \\& & 1 \end{pmatrix}
		_B  \otimes
		\begin{pmatrix} 0 \\ 1 \end{pmatrix}
		_{C} \cdot \sqrt{p(\text{gen}|1)},
	\end{aligned}
\end{equation}
and
\begin{equation}
	\begin{aligned}
		Z_1 &=
		\begin{pmatrix} 1 & \\ & 0 \end{pmatrix}
		\otimes \1_{\dim_A \times\dim_B \times 2}, \\
		Z_2 &=
		\begin{pmatrix} 0 & \\ & 1 \end{pmatrix}
		\otimes \1_{\dim_A \times\dim_B \times 2}.
	\end{aligned}
\end{equation}

\subsection{4-6 Protocol}\label{App:Kraus ops 4-6}
As described in the main text, see \cref{sec:Example Decoy 4-6}, we chose a photon number cut-off for Bob as \(N_B = 1\). After applying the flag-state squasher of \cite{Zhang2021Phys.Rev.Res.} with this choice, the POVM elements on Bob's \(\leq 1\)-photon subspace are
\begin{equation}
	\begin{aligned}
		\tilde{M}^{B}_{(Z,0)} &= p_z^B \begin{pmatrix} 0 & 0 &0 \\ 0& 1 & 0\\ 0&0 & 0 \end{pmatrix}, \;
		\tilde{M}^{B}_{(Z,1)} = p_z^B \begin{pmatrix} 0 & 0& 0\\ 0& 0 & 0\\ 0&0 & 1 \end{pmatrix}, \\
		\tilde{M}^{B}_{(X,0)} &= \frac{p_x^B}{2} \begin{pmatrix} 0&0 & 0 \\  0&1 & 1 \\ 0& 1 & 1 \end{pmatrix},
		\;
		\tilde{M}^{B}_{(X,1)} = \frac{p_x^B}{2} \begin{pmatrix} 0&0 &0   \\  0&1 & -1  \\ 0 & -1 & 1  \end{pmatrix}, \\
		\tilde{M}^{B}_{(Y,0)} &= \frac{p_y^B}{2} \begin{pmatrix} 0&0 & 0 \\  0&1 & -i \\ 0& i & 1 \end{pmatrix},
		\;
		\tilde{M}^{B}_{(Y,1)} = \frac{p_y^B}{2} \begin{pmatrix} 0&0 &0   \\  0&1 & i  \\ 0 & -i & 1  \end{pmatrix}, \\
		\tilde{M}^{B}_{\bot} &=\begin{pmatrix} 1 & 0 & 0 \\ 0 & 0 & 0 \\ 0 & 0 & 0 \end{pmatrix}.
	\end{aligned}
\end{equation}
These need to be padded with flags. Therefore, let \(E_i:=\diag(0,\dots,0,1,0, \dots ,0) \in \R^{7 \times 7}\), be a diagonal matrix with \(1\) at the \(i\)-th entry. Furthermore, one needs to include multi-clicks (mult). For \(N_B=1\), these are only part of the flags. Hence, Bob's full squashed POVM elements are
\begin{equation}
	\begin{aligned}
		M^{B}_{(Z,0)} &= \tilde{M}^{B}_{(Z,0)} \oplus E_1, \quad
		M^{B}_{(Z,1)} = \tilde{M}^{B}_{(Z,1)} \oplus E_2, \\
		M^{B}_{(X,0)} &= \tilde{M}^{B}_{(X,0)} \oplus E_3,
		\quad
		F^{B }_{(X,1)} = \tilde{M}^{B}_{(X,1)} \oplus E_4, \\
		M^{B}_{(Y,0)} &= \tilde{M}^{B}_{(Y,0)} \oplus E_5,
		\quad
		M^{B}_{(Y,1)} = \tilde{M}^{B}_{(Y,1)} \oplus E_6, \\
		M^{B}_{\text{mult}} &= \bar{0}_3 \oplus E_{7}, \\
		M^{B}_{\bot} &= \tilde{M}^{B}_{\bot} \oplus \bar{0}_{7},
	\end{aligned}
\end{equation}
where \(\bar{0}_m\) indicates a matrix with only zeros of dimension \(m\).

Regarding Alice's POVM elements, we apply the source replacement scheme without reductions because we consider different intensities for each signal state. Hence, Alice's POVM elements are as in the qubit case simply given by
\begin{equation}
	\begin{aligned}
		M^A_{(Z,0)} &= \ketbra{0}, M^A_{(Z,1)} = \ketbra{1}, \\
		M^A_{(X,0)} &= \ketbra{2}, M^A_{(X,1)} = \ketbra{3}.
	\end{aligned}
\end{equation}
Since we only consider the single photon contribution but incorporate different intensities for each signal state, we need to condition Alice's probabilities appropriately on single photons. To maintain a notation close to the one of Bob, let us define the probabilities
\begin{align}
\begin{aligned}
    &p_{z,i|1\wedge\text{gen}}^A := \\
        &\Pr\left(\text{Alice sends bit "i" in Z-basis}| \right. \\ 
        &\qquad \left. \text{1 photon and gen round} \right), \\
    &p_{x,i|1\wedge\text{gen}}^A := \\
        &\Pr\left(\text{Alice sends bit "i" in X-basis}| \right. \\
        &\qquad \left. \text{1 photon and gen round} \right).
    \end{aligned}
\end{align}
Finally, we apply one isometry to reduce the dimensions of the Kraus operators and arrive at
\begin{align}
    K_Z &=
	\begin{aligned}[t]
		 &\sqrt{p(\text{gen}|1)} \left[ \sqrt{p_{z,0|1\wedge\text{gen}}^A}
		\begin{pmatrix} 1 & 0 & 0 & 0\\ 0 & 0 & 0 & 0 \end{pmatrix} \right. \\
        &\left. + \sqrt{p_{z,1|1\wedge\text{gen}}^A}
		\begin{pmatrix} 0 & 0 & 0 & 0 \\ 0 & 1 & 0 & 0 \end{pmatrix} \right] \\
		&\otimes \left[0 \oplus \sqrt{p_z^B} \1_2 \oplus \left(E_1 + E_2 \right) \right]_B \otimes
		\begin{pmatrix} 1 \\ 0 \end{pmatrix}
		_{C}, \\
	\end{aligned}\\
    K_X &= 
    \begin{aligned}[t]
		 &\sqrt{p(\text{gen}|1)} \left[ \sqrt{p_{x,0|1\wedge\text{gen}}^A}
		\begin{pmatrix} 0 & 0 & 1 & 0\\ 0 & 0 & 0 & 0 \end{pmatrix} \right. \\
        &\left. + \sqrt{p_{x,1|1\wedge\text{gen}}^A}
		\begin{pmatrix} 0 & 0 & 0 & 0 \\ 0 & 0 & 0 & 1 \end{pmatrix} \right] \\
		&\otimes \left[0 \oplus \sqrt{p_z^B} \1_2 \oplus \left(E_1 + E_2 \right) \right]_B \otimes
		\begin{pmatrix} 0 \\ 1 \end{pmatrix}
		_{C},
	\end{aligned}
\end{align}
and key projectors
\begin{equation}
	\begin{aligned}
		Z_1 &=
		\begin{pmatrix} 1 & \\ & 0 \end{pmatrix}
		\otimes \1_{\dim_A \times\dim_B\times \dim_C}, \\
		Z_2 &=
		\begin{pmatrix} 0 & \\ & 1 \end{pmatrix}
		\otimes \1_{\dim_A \times\dim_B\times \dim_C}.
	\end{aligned}
\end{equation}

\section{Technical Definitions and Lemmas for R\'enyi entropy} \label{App:Renyi}
In this appendix, we include technical statements required for proving variable-length security of decoy-state BB84 from \cref{sec:variable}. These technical statements are obtained from \cite{Dupuis2023IEEETrans.Inf.Theory, Dupuis2020Commun.Math.Phys.,Tomamichel2016}. 

	\begin{defn}[R\'enyi Entropy] \label{def:renyientropy}
	For $\rho \in S_\circ(AB)$, and $\alpha \in (0,1) \cup (1,\infty)$, the sandwiched Renyi Entropy of $A$ given $B$ for a state $\rho_{AB}$ is given by
	\begin{equation}
		H_\alpha (A|B)_\rho := \max_{\sigma_B \in S_\circ(B)} H_\alpha(A|B)_{\rho | \sigma},
	\end{equation}
	where
	\begin{equation}
		\begin{aligned}
			& H_\alpha(A|B)_{\rho| \sigma} \\
			& := \begin{cases*}
				\frac{1}{1-\alpha} \log \Tr \left[   \left( \sigma_B^{\frac{1-\alpha}{2\alpha}} \rho_{AB}  \sigma_B^{\frac{1-\alpha}{2\alpha}} \right)^\alpha \right] & if $\rho \in A \otimes \text{supp}(\sigma)$ \\
				-\infty     & otherwise
			\end{cases*}
		\end{aligned}
	\end{equation}
\end{defn}
\begin{lem}[{Conditioning on Events, \cite[Lemma B.5]{Dupuis2020Commun.Math.Phys.}}] \label{lemma:renyiconditioning}
	Let $\rho_{AB} \in S_\circ(AB)$ be a state of the form $\rho_{AB}= \sum_x p_x \rho_{AB|x}$, where $p_x$ is a probability distribution. Then, for $\alpha>1$,
	\begin{equation}
		H_\alpha (A|B)_{\rho_{ | x}} \geq H_\alpha (A|B)_\rho - \frac{\alpha}{\alpha-1} \log \left( \frac{1}{p_x} \right).
	\end{equation}
\end{lem}

\begin{thrm}[{Leftover hashing Lemma using R\'enyi Entropy, \cite[Theorem 9]{Dupuis2023IEEETrans.Inf.Theory}}]\label{Thrm:LHLrenyi}
	Let \(\sigma_{XD} \in \mathcal{S}_{\circ}\left(XD\right)\) be classical on \(X\) and $\alpha \in (1,2]$. Let \(\Hil \) be a two-universal family of hash functions from \(\mathcal{X} = \{0,1\}^n\) to \(\mathcal{K}=\{0,1\}^l\). Moreover, let \(\rho_{S^H} = \frac{1}{\abs{\Hil}}\sum_{h \in \Hil} \ketbra{h}_{S^H} \). Then,
	\begin{equation}
		\frac{1}{2} \norm{\omega_{KS^HD} - \chi_K \otimes \omega_{S^HD}} \leq \frac{1}{2} 2^{-\frac{(\alpha-1)}{\alpha} \left(     H_\alpha(X|D)_\rho - l    \right) + \frac{2}{\alpha} - 1},
	\end{equation}
		where \(\chi_K = \frac{1}{2^l} \id_K\) is the fully mixed state and \(\omega_{KS^HD} = \Tr_X\left[\mathcal{E}_f\left(\sigma_{XD} \otimes \rho_{S^H} \right) \right]\) for the function \(f: \left(x,h\right) \mapsto h(x) \) that acts on the registers \(X\) and \(S^H\).
\end{thrm}

\begin{lem}[Conditioning on Classical Register using Renyi Entropy]\label{Lem:Conditioning on Classical Register using Renyi}
	Let $\rho_{ABZ} \in S_{\circ}(ABZ)$ be classical on \(Z\), and $\alpha>1$. Then,
	\begin{equation}
		H_\alpha(A|BZ)_{\rho} \geq \inf_{z \in Z} H_\alpha (A|B;Z=z)_{\rho}.
	\end{equation}
\end{lem}
\begin{proof}
	Follows trivially from \cite[Proposition 5.1]{Tomamichel2016}, which states
	\begin{equation}
		 \sum_z \Pr(z) 2^{-\frac{(\alpha-1)}{\alpha} H_\alpha(A|B;Z=z)_{\rho} } = 2^{-\frac{(\alpha-1)}{\alpha} H_\alpha(A|BZ)_\rho}
	\end{equation}
\end{proof}
	 
\begin{lem}[{Splitting off a classical register, \cite[Eq. 5.94]{Tomamichel2016}}]\label{lemma:renyisplitting}
 	Let $\rho_{ABC} \in S_\bullet(ABC)$ be classical on $C$. Then
 	\begin{equation}
 		\begin{aligned}
 		H_\alpha(A|BC) &\geq H_\alpha(AC|B)  - \log(\dim(C)) \\
 		&\geq  H_\alpha(A|B)  - \log(\dim(C)).
 		\end{aligned}
 	\end{equation}		
\end{lem}

\begin{lem}[{\cite[Lemma B.9 ]{Dupuis2020Commun.Math.Phys.}}]\label{lemma:contrenyi}
	For any $\rho_{AB}\in S_\circ (AB)$, and $1 < \alpha < 1+1/ \log(1+2\dim(A))$,
	\begin{equation}
		H_\alpha(A|B)_\rho >  H(A|B)_\rho - (\alpha-1)\log^2\left(1+2 \dim(A) \right).
	\end{equation}
\end{lem}

\begin{lem}[{Additivity of Renyi Entropy, \cite[Corollary 5.2]{Tomamichel2016}}]\label{lemma:additivity}
	For any two states $\rho_{AB} \in S_\circ(AB), \sigma_{CD} \in S_\circ(CD)$, and $\alpha \geq \frac{1}{2}$, we have
	\begin{equation}
		H_\alpha(AC | BD)_{\rho \otimes \sigma}  = H_\alpha(A|B)_\rho + H_\alpha(C|D)_\sigma.
	\end{equation}
\end{lem}

\begin{lem}[Bounds on the Renyi Entropy for States]\label{Lem:Bound Renyi Entropy sifted signals}
	Let \(\vec{Z},\vec{Y},\vec{C},\vec{E},\vec{D}\) be the registers corresponding to Alice's and Bob's raw keys, classical announcements, Eve's quantum side information and the locations of the sifted detections, respectively, and \(\Omega_{\Nsift}\) denote the event that $\Fobs$ such that $N\Fobs_\text{sift} = \Nsift$ is observed. We label \(D_i=1\) if a signal and passed sifting in round \(i\) and \(D_i=0\) otherwise. For \(\rho_{\vec{Z}\vec{Y} \vec{C} \vec{E}\vec{D}}\) such that
	\begin{equation}
		\rho_{\vec{Z}\vec{Y} \vec{C} \vec{E}\vec{D}} = \bigotimes_{i=1}^{N} \sigma_{ZYCED},
	\end{equation}
	it holds
	\begin{equation}
		H_\alpha(\vec{Z}| \vec{C} \vec{E}\vec{D})_{\rho {|{\Omega_{\Nsift}}}} \geq H_\alpha(\vec{Z}_{\text{sift}}| \vec{C}_\text{sift} \vec{E}_{\text{sift}})_{\tau^{\Nsift}_{| \vec{D} = 1} },
	\end{equation}
	where
	\begin{equation}
		\tau^{\Nsift}_{\vec{Z}_{\text{sift}} \vec{Y}_\text{sift} \vec{C}_\text{sift} \vec{E}_{\text{sift}}| \vec{D}=\vec{1}} := \bigotimes_{i=1}^{ \Nsift} \sigma_{ZYCE|D=1}.
	\end{equation}
\end{lem}
\begin{proof}
	The proof follows from same steps as the proof of \cref{Lem:Bound Smooth Min-Entropy sifted signals} after switching from smooth min-entropy to R\'enyi entropy. Instead of \cref{Lem:Conditioning on Classical Register} we use \cref{Lem:Conditioning on Classical Register using Renyi}, and instead of \cite[Lemma 6.7]{Tomamichel2016}, we use \cref{lemma:additivity} and the fact that R\'enyi Entropy on classical registers is always non-negative.
\end{proof}

The following lemma is the main reason for the use of R\'enyi entropy instead of smooth min entropy in the variable-length framework of \cite{Tupkary2024Phys.Rev.Res.}. 
	
\begin{lem}[{\cite[Lemma 10]{Tupkary2024Phys.Rev.Res.}}]  \label{lemma:renyiweightedaverage}
    Let $\rho_{ABCY} = \sum_{y \in \Lambda} p(y) \rho_{ABC| y} \otimes \ket{y} \bra{y} \in S_\circ(ABCY)$ be classical in $Y,C$, where $p(y)$ is a probability distribution over $\Lambda$, and $Y$ can be generated from $C$ (more precisely: $Y \leftrightarrow C \leftrightarrow AB$ forms a Markov chain).  Let $\Lambda^\prime \subseteq \Lambda$. Then, 
    \begin{equation}
          \sum_{y \in \Lambda^\prime}  p(y)	2^{-\frac{(\alpha-1)}{\alpha} H_\alpha( A| BC)_{\rho_{ |y} } } \leq 2^{-\frac{(\alpha-1)}{\alpha} H_\alpha( A| BC)_\rho } .
    \end{equation}
\end{lem} 

\section{Variable-Length Finite-size Key Rates} \label{App:Variable decoy}
In this appendix, we will prove \cref{Thrm:PM QKD variable security using Renyi Entropy} by combining the tools developed in this work, and \cite{Tupkary2024Phys.Rev.Res.}. Combining the improvements of both requires a few crucial (but mainly cumbersome) modifications to the arguments in these works at critical places. Where applicable, we will comment on the modifications and their necessity.

 Recall that the protocol steps of our variable-length protocols will be similar to \cref{Prot:PM Protocol} and \cref{Prot:Decoy Protocol}. However, in the variable-length protocol, we will use  $F^{\text{obs}}$ to determine the length $l(F^\text{obs})$ of the key to be produced, and number of bits $\leak(F^\text{obs})$ to be used for error-correction. This is in stark contrast to fixed-length protocols, where $F^{\text{obs}}$ is used to determine only whether the protocol accepts (in which case it produces a key of fixed-length $l$ using $\leak$ bits for error-correction) or aborts. Thus, we refer to this as the ``variable-length decision'', instead of ``acceptance test''

The variable-length framework in \cite{Tupkary2024Phys.Rev.Res.} critically relies upon the construction of the set $V(F^\text{obs})$ with the 
following property:
\begin{equation} \label{eq:VFobsproperty}
	\Pr_{F^\text{obs}} \left( \rho \in V(F^\text{obs}) \right) \geq 1-\varepsilon_{AT}
\end{equation}
Such a set was constructed in \cref{lemma:VFobsconstruction}. Using $V(\Fobs)$, we can construct a statistical estimator $\bstat(\Fobs)$ of the R\'enyi entropy of the underlying state in the protocol with the help of the following lemma.

\begin{lem}\label{lemma:bstat}
Let \(\vec{Z},\vec{Y},\vec{C},\vec{E},\vec{D}\) be the registers corresponding to Alice's and Bob's raw keys, classical announcements, Eve's quantum side information and the locations of the sifted detections, respectively, and \(\Omega_{\Nsift}\) denote the event that $\Fobs$ such that $N\Fobs_\text{sift} = \Nsift$ is observed. We label \(D_i=1\) if a signal was detected and sifted in round \(i\) and \(D_i=0\) otherwise. Let \(\vec{\rho}_{\vec{Z}\vec{Y} \vec{C} \vec{E}\vec{D}}\) be the state in the QKD protocol, such that 
	\begin{equation}
		\vec{\rho}_{\vec{Z}\vec{Y} \vec{C} \vec{E}\vec{D}} = \bigotimes_{i=1}^{N} \rho_{ZYCED},
	\end{equation}
where $\rho$ satisfies $ \rho_A = \Tr_{A'}[\ketbra{\psi}]$. Define			
    \begin{equation}
    \begin{aligned}
        \bstat(\Fobs) &\coloneqq  \Nsift \min_{\sigma \in V(\Fobs)  } \! \! \frac{H(Z | C E)_\sigma}{\Pr(\text{sift})} \\
        &-  \Nsift (\alpha-1) \log^2(d_Z+1).
        \end{aligned}
    \end{equation}
where $d_Z=\dim(Z)$ and $1 < \alpha < 1+ 1/\log(2d_Z+1)$.  Then, 
\begin{equation} \label{eq:bstatcondition}
    \Pr_{\Fobs} \left( \bstat(\Fobs ) \leq H_\alpha(\vec{Z} | \vec{C} \vec{E} )_{ \vec{\rho} | \Omega_{\Nsift}} \right) \geq 1-\epsAT.
\end{equation}
\end{lem}

\begin{proof}
We have \begin{equation} \label{eq:tempbstat}
    \begin{aligned}
    H_\alpha(\vec{Z}|\vec{C} \vec{E}\vec{D})_{ \vec{\rho} | \Omega_{\Nsift}} &\geq H_\alpha(\vec{Z}_{\text{sift}}|\vec{C} \vec{E}_{\text{sift}})_{\tau^{\Nsift}_{\vec{D}=1}} \\
    &\geq \Nsift H(Z | C E ; D=1)_{\rho} \\
    &- \Nsift (\alpha-1) \log^2(d_Z+1) \\
    &= \frac{\Nsift}{\Pr(\text{sift})}_\rho H(Z|CE)_\rho \\
    &- \Nsift (\alpha-1) \log^2(d_Z+1) 
	\end{aligned}
\end{equation}
where we used \cref{Lem:Bound Renyi Entropy sifted signals} for the first inequality (where we also define $\tau^{\Nsift}_{\vec{D}=1} $), and \cref{lemma:contrenyi} and \cref{lemma:additivity} for the second.  Since $V(\Fobs)$ contains $\rho$ with high probability ($\geq 1 - \epsAT$),the required claim follows by minimizing the RHS of \cref{eq:tempbstat} over $V(\Fobs)$.
\end{proof}

This $\bstat(\Fobs)$ is used the variable-length decision of the QKD protocol as follows.

Let $\mathcal{F}$ be the (possibly infinite) set of all possible observations $\Fobs$ in the variable-length decision step of the protocol. Then, 
\begin{enumerate}
    \item From public announcements $\vec{C}$, Alice and Bob compute $\Fobs$ and $\bstat(\Fobs)$ .
    \item They compute $\leak{}(\Fobs)$, the number of bits to be used for error-correction information, where  $\leak{}(\cdot) : \mathcal{F} \rightarrow \{0,1,\dots,\lambda_{\text{max}}\}$ is some predetermined function. 
    \item They compute $l(\Fobs)$, the length of the final key to be produced, where  $l(\cdot) : \mathcal{F} \rightarrow \{0,1,\dots,l_{\text{max}}\}$ is a function defined as 
     \begin{equation}  \label{eq:livalue}
        \begin{aligned}
            l(\Fobs)& \coloneqq\max \bigg( 0,  \\
            &\left \lfloor \bstat(\Fobs) - \leak{}(\Fobs)  -\PAcost \right \rfloor \bigg), \\
            \PAcost &\coloneqq  \frac{\alpha}{\alpha-1} \left( \log(\frac{1}{4\epsPA} ) +  \frac{2}{\alpha}  \right) + \log(\frac{2}{\epsEV}).
        \end{aligned}
    \end{equation}
 \end{enumerate}
 We setup a partition of $\mathcal{F}$ based on the observed number of sifted events, and the length of the final key and error-correction amount as follows. We let
     \begin{equation} \label{eq:setsandevents}
     \begin{aligned}
         \accsetadapt{\nsift}{i} = \{ \Fobs \in \mathcal{F} | \Nsift = \nsift, \\ \! l(\Fobs) = l_i, \leak{}(\Fobs) = \lambda_{i} \} 
         \end{aligned}
     \end{equation}
Furthermore, we order these sets such that $l_i + \lambda_{i}$ forms a non-increasing sequence in $i$. We emphasize that this is purely for convenient notation, and such an ordering can always be enforced without any loss of generality. We let $\Omega_{\nsift,i}$ denote the event that $\Fobs \in \accsetadapt{\nsift}{i}$. Note that number of possible events $\Omega_{\nsift,i}$ is always finite, and we denote it by $M$.
This ordering allows us to state the following lemma.
 
\begin{lem} \label{lemma:ordering}
	Let $\mathcal{T}_p = \{ i | l_i > 0\}$ be the set of values of $i$ that lead to non-trivial length of the key. Then,
	\begin{equation}
		\begin{aligned}
		&\sum_{\substack{i=1 \\ i\in \mathcal{T}_p}}^j \Pr( \Omega_{\nsift,i}) \leq \\&\Pr(\bstat(\Fobs) \geq l_j + \lambda_{j} + \PAcost \wedge  \Nsift = \nsift),
			\end{aligned}
	\end{equation}
	 for any $j \in \{1,2,\dots,M\}$.
\end{lem}

\begin{proof}
	For any $i \in \mathcal{T}_p$, using \cref{eq:livalue} we have
	\begin{equation}\label{eq:temp} 
 \begin{aligned}
 \Omega_{\nsift,i} \implies	\bstat(\Fobs) &\geq l_i + \lambda_{i} +\PAcost \\
 & \wedge \Nsift = \nsift
 \end{aligned}
 \end{equation}
	Therefore
		\begin{equation}
			\begin{aligned}
	&\sum_{\substack{i=1 \\ i \in \mathcal{T}_p} }^{j}  \Pr(\Omega_{\nsift,i})  =    \Pr(\bigcup_{\substack{i=1 \\ i \in \mathcal{T}_p} }^{j} \Omega_{\nsift,i} ) \\
		&\leq \Pr(\bigcup_{\substack{i=1 \\ i \in \mathcal{T}_p}}^j \bstat(\Fobs) \geq l_i + \lambda_{i}  +\PAcost \wedge \Nsift = \nsift) \\
		&\leq \Pr( \bstat(\Fobs) \geq l_j + \lambda_{j}  +\PAcost \wedge \Nsift = \nsift)
		\end{aligned}
	\end{equation}
	where we used the fact that $\Omega_{m,i}$ are disjoint events in the first equality, \cref{eq:temp} for the second inequality, and the ordering on $l_i + \lambda_{i}$ in the final inequality.
\end{proof}

We now have all the tools necessary to prove the variable-length security of the QKD protocol , analogous to \cite[Theorem 2]{Tupkary2024Phys.Rev.Res.}.

\varlengththeorem*
Notice that with the definition of $\bstat$ in \cref{lemma:bstat},  $l(\Fobs)$ above corresponds to \begin{equation}
	\begin{aligned}
		&l(F^\text{obs}):= \max \bigg( 0,  \bstat(\Fobs) -  \leak(F^\text{obs})  - \PAcost \bigg)\\
		&\PAcost \coloneqq  \frac{\alpha}{\alpha-1} \left( \log(\frac{1}{4\epsPA} ) +  \frac{2}{\alpha}  \right) + \log(\frac{2}{\epsEV}).
	\end{aligned}
\end{equation}  
We will use the above formulation of $l(\Fobs)$ in our proof. Moreover, in this work, we have provided one construction of $\bstat$ satisfying \cref{eq:bstatcondition}. 
\begin{proof}
The fact that the variable-length protocol is $\epsEV$-correct follows from the properties of error-verification process. We do not repeat the proof here, and it can be found in Ref.~\cite{Tupkary2024Phys.Rev.Res.} for variable-length protocols.
Thus, we focus on only proving the  $(\epsAT + \epsPA)$-secrecy of the protocol. The protocol is then $(\epsAT+\epsPA+\epsEV)$-secure (again, we refer the reader to \cite{Tupkary2024Phys.Rev.Res.} for a proof). 
From the proof of \cite[Theorem 2]{Tupkary2024Phys.Rev.Res.}, proving $(\epsPA + \epsAT)$-secrecy of the protocol is equivalent to proving the following statement \footnote{Instead of considering a sum over events that only correspond to different output key lengths, we are allowed to sum over events with different output key lengths \textit{or error-correction lengths or number of sifted signals}. This can be shown using basic properties of the trace norm, and by noting that the additional events are known to Eve via classical announcements, and hence live on orthogonal supports.}:

\begin{equation}
    \begin{aligned}
        \sum_{i,\nsift} &\frac{\Pr(\Omega_{\nsift,i} \wedge \Omega_\text{EV} ) }{2} \norm{\rho^{(l_i,\lambda_{i})}_{K_A  \tilde{C}  \vec{E} | \Omega_{m,i} \wedge \Omega_{\text{EV}}} - \rho^{(l_i,\lambda_{i},\text{ideal})}_{K_A \tilde{C}  \vec{E} | \Omega_{m,i} \wedge \Omega_{\text{EV}} }}_1 \\
        &\leq \epsPA + \epsEV,
    \end{aligned}
\end{equation}
where $\tilde{C} = \vec{C} \CEV C_P$ denotes all the classical communication during the protocol, for public announcements ($\vec{C}$), error-correction and verification ($\CEV$), and the choice of hash function ($C_P$). Thus, $\rho^{(l_i,\lambda_{i})}_{K_A \tilde{C}  \vec{E} | \Omega_{m,i} \wedge \Omega_{\text{EV}}}$ denotes the output of the QKD protocol (with Bob's registers omitted), conditioned on the event $\Omega_{m,i} \wedge \Omega_{\text{EV}}$ (\cref{eq:setsandevents}), which leads to a key length of $l_i$ bits and error-correction using $\lambda_i$ bits. We use $\rho^{(l_i,\lambda_{i}, \text{ideal})}_{K_A \tilde{C}  \vec{E} | \Omega_{m,i} \wedge \Omega_{\text{EV}}}$ to denote the output of the ideal QKD protocol, where the final key register is replaced with the perfect keys.

We use $\Omega_{m}$ to denote the event that $\Nsift=m$. Recall that the values $l_i+ \lambda_{i}$ are ordered such that they form a non-increasing sequence.  Thus, for \textit{any} $\rho_{AB}$ that the protocol can start with, the R\'enyi entropy $H_\alpha(\vec{Z} | \vec{C} \vec{E})_{\rho | \Omega_{m}}$ has to fall under at least one of the following three cases: 
\begin{enumerate}
    \item $H_\alpha(\vec{Z} | \vec{C} \vec{E} )_{\rho | \Omega_{m}} \geq l_1+\lambda_{1}+\PAcost$.
    \item $ l_j + \lambda_{j} + \PAcost \geq H_\alpha(\vec{Z} | \vec{C} \vec{E})_{\rho | \Omega_{m}} \geq l_{j+1} + \lambda_{j+1} + \PAcost$ for some $j \in \{1,...,M-1\}$.
    \item $ l_{M} + \lambda_{M} + \PAcost\geq  H_\alpha(\vec{Z} | \vec{C} \vec{E})_{\rho | \Omega_{m}}$.
\end{enumerate} 
We will prove the secrecy claim separately for each case. Suppose that for every value of $\nsift$, $\rho$ is such that it satisfies case 2, for some value  $j^*_{\nsift}$. In this case, the secrecy bound can be obtained by splitting up the sum into two convenient parts. The first part groups the set of events that happen with low probability, and is given by,

\begin{widetext}
\begin{equation} \label{eq:boundingfirsthalf}
    \begin{aligned}
        & \sum_{\nsift} \sum_{i=1}^{j^*_{\nsift}}\frac{\Pr(\Omega_{\nsift,i}  \wedge \Omega_{\text{EV}}) }{2} \norm{\rho^{(l_i,\lambda_{i})}_{K_A  \tilde{C}  \vec{E} | \Omega_{m,i} \wedge \Omega_{\text{EV}}} - \rho^{(l_i,\lambda_{i},\text{ideal})}_{K_A \tilde{C}  \vec{E} | \Omega_{m,i} \wedge \Omega_{\text{EV}} }}_1 \\
        &=    \sum_{\nsift} \sum_{\substack{i=1 \\ i\in \mathcal{T}_p}}^{j^*_{\nsift}} \frac{\Pr(\Omega_{\nsift,i} \wedge \Omega_{\text{EV}} ) }{2} \norm{\rho^{(l_i,\lambda_{i})}_{K_A  \tilde{C}  \vec{E} | \Omega_{m,i} \wedge \Omega_{\text{EV}} } - \rho^{(l_i,\lambda_{i},\text{ideal})}_{K_A \tilde{C}  \vec{E} | \Omega_{m,i} \wedge \Omega_{\text{EV}}}}_1\\
        &\leq    \sum_{\nsift} \sum_{\substack{i=1 \\ i\in \mathcal{T}_p}}^{j^*_{\nsift}} \Pr(\Omega_{\nsift,i} \wedge \Omega_{\text{EV}}) \\
        &\leq    \sum_{\nsift} \sum_{\substack{i=1 \\ i\in \mathcal{T}_p}}^{j^*_{\nsift}} \Pr(\Omega_{\nsift,i} ) \\ 
        &\leq   \sum_{\nsift} \Pr(\bstat(\Fobs) \geq l_{j^*_{\nsift}}+ \lambda_{j^*_{\nsift}} + \PAcost \wedge \Nsift = \nsift) \\
        &\leq  \sum_{\nsift} \Pr(\bstat(\Fobs) \geq H_\alpha(\vec{Z} | \vec{C} \vec{E})_{\rho | \Omega_{m}}\wedge \Nsift = \nsift) \\
      &= \Pr(\bstat(\Fobs) \geq H_\alpha( \vec{Z} | \vec{C}  \vec{E})_{\rho | \Omega_{\Nsift}} ) \\
        &\leq \epsPA
    \end{aligned}
\end{equation}
\end{widetext}
where the first inequality follows from the fact that when the key length is zero, the real and the ideal outputs of the protocol are identical. The second inequality follows from the properties of the trace norm, and the third inequality follows from basic properties of probabilities. We used \cref{lemma:ordering} for the fourth inequality, and the definition of $j^*_{\nsift}$ for the fifth inequality. The sixth equality follows from the definition of these events, and the final inequality follows from the \cref{lemma:bstat}. For the second part, we obtain the following chain of inequalities (explained below):
    
\begin{widetext}
\begin{equation} \label{eq:boundingsecondhalf}
    \begin{aligned}
         &\sum_{\nsift} \sum_{i=j^*_{\nsift}+1}^{M}\frac{\Pr(\Omega_{\nsift,i} \wedge \Omega_\text{EV}  ) }{2} \norm{\rho^{(l_i,\lambda_{i})}_{K_A \tilde{C} \vec{E} | \Omega_{m,i} \wedge \Omega_\text{EV}} - \rho^{(l_i,\lambda_{i},\text{ideal})}_{K_A\tilde{C} \vec{E} | \Omega_{m,i} \wedge \Omega_\text{EV}} }_1 \\
         &=\sum_{\nsift} \sum_{i=j^*_{\nsift}+1, i\in \mathcal{T}_p}^{M}\frac{\Pr(\Omega_{\nsift,i} \wedge \Omega_\text{EV} ) }{2} \norm{\rho^{(l_i,\lambda_{i})}_{K_A \tilde{C} \vec{E} | \Omega_{m,i} \wedge \Omega_\text{EV} } - \rho^{(l_i,\lambda_{i},\text{ideal})}_{K_A\tilde{C} \vec{E} | \Omega_{m,i} \wedge \Omega_\text{EV} } }_1 \\
         &\leq   \sum_{\nsift} \sum_{i=j^*_{\nsift}+1, i \in \mathcal{T}_p}^{M}\frac{\Pr(\Omega_{\nsift,i} \wedge \Omega_\text{EV} ) }{2} 2^{-\frac{\alpha-1}{\alpha} \left( H_\alpha(\vec{Z} | \vec{C} C_E C_V \vec{E})_{\rho | \Omega_{\nsift,i} \wedge \Omega_\text{EV} } - l_i  \right) + \frac{2}{\alpha} -1 } \\
         &\leq \sum_{\nsift} \sum_{i=j^*_{\nsift}+1, i \in \mathcal{T}_p}^{M}\frac{\Pr(\Omega_{\nsift,i} \wedge \Omega_\text{EV} ) }{2} 2^{-\frac{\alpha-1}{\alpha} \left( H_\alpha(\vec{Z} | \vec{C} C_V \vec{E})_{\rho | \Omega_{m,i}\wedge \Omega_\text{EV}} - l_i -\lambda_{i} \right) + \frac{2}{\alpha} -1 } \\
       &\leq   \sum_{\nsift} \sum_{i=j^*_{\nsift}+1, i \in \mathcal{T}_p}^{M}\frac{\Pr(\Omega_{\nsift,i} \wedge \Omega_\text{EV} ) }{2} 2^{-\frac{\alpha-1}{\alpha} \left( H_\alpha(\vec{Z} | \vec{C} C \vec{E})_{\rho | \Omega_{\nsift,i} \wedge \Omega_\text{EV}} - l_{(j^*_{\nsift}+1)} -\lambda_{({j^*_{\nsift}+1})} \right) + \frac{2}{\alpha} -1 } \\
       &\leq \sum_{\nsift} \frac{\Pr(\Omega_{\nsift} ) }{2} 2^{-\frac{\alpha-1}{\alpha} \left( H_\alpha(\vec{Z} | \vec{C} C_V  \vec{E})_{\rho | \Omega_{\nsift}} - l_{(j^*_{\nsift}+1)} -\lambda_{({j^*_{\nsift}+1})} \right) + \frac{2}{\alpha} -1 } \\
       &\leq \sum_{\nsift} \frac{\Pr(\Omega_{\nsift} ) }{2} 2^{-\frac{\alpha-1}{\alpha} \left( H_\alpha(\vec{Z} | \vec{C}  \vec{E})_{\rho | \Omega_{\nsift}} - l_{(j^*_{\nsift}+1)} -\lambda_{({j^*_{\nsift}+1})} - \log(2/\epsEV) \right) + \frac{2}{\alpha} -1 } \\
         &\leq \sum_{\nsift} \frac{\Pr(\Omega_{\nsift} ) }{2} 2^{-\frac{\alpha-1}{\alpha} \left( \PAcost - \log(2/\epsEV) \right) + \frac{2}{\alpha} -1 } \\
         &\leq \epsPA,
    \end{aligned}
\end{equation}
\end{widetext}
where the first equality follows from the fact that real and ideal output are the same when key length is zero. We used leftover-hashing-lemma for R\'enyi entropy for the second inequality, and \cref{lemma:renyisplitting} to split off the error-correction information in the third inequality. We used the ordering on $l_i + \lambda_i$ for the fourth inequality, \cref{lemma:renyiweightedaverage} for the fifth inequality, and \cref{lemma:renyisplitting} to split off the error-verification information in the sixth inequality. We used the fact that $H_\alpha(\vec{Z} | \vec{C} \vec{E})_{\rho | \Omega_{\nsift}} \geq l_{j^*_{\nsift}+1} + \lambda_{j^*_{\nsift}+1} + \PAcost$ for the seventh inequality, and the definition of $\PAcost$ for the final inequality. 

The required claim follows from combining \cref{eq:boundingfirsthalf,eq:boundingsecondhalf}. Furthermore, it is easy to see that the same bound holds even if case (1) or case (3) is satisfied for some values of $\nsift$, instead of case (2). In this case, only one term out of \cref{eq:boundingfirsthalf,eq:boundingsecondhalf} appears, and it can be bounded in the same way as above.
\end{proof} 

Note that we had to split off the $C_E,C_V$ registers separately in the above proof, because we need the register $C_V$ to be known to Eve in order to use \cref{lemma:renyiweightedaverage}. Furthermore, the above proof is quite similar to the proof of \cite[Theorem 2]{Tupkary2024Phys.Rev.Res.}. However, we deal with different events $\Omega_m$ that correspond to various numbers of observed sifted rounds, in a manner that leads to the finite-size correction costs only depending on $\Nsift$ instead of $N$. Another minor point of difference, is that \cite{Tupkary2024Phys.Rev.Res.} uses a random subset of the rounds of fixed-length for testing, whereas in this work each round has some IID probability of being tested or used for key generation. 

\section{Confidence intervals and Variational bounds for small $\varepsilon$}
In order to apply the postselection technique to reduce the security analysis to the IID case, one requires very small security parameters for the IID calculations.

However, this will run into numerical issues as security parameters as small as \(\varepsilon \approx 10^{-3000}\) are needed. The usual methods for Clopper-Pearson intervals only estimate the interval numerically and hence are bounded by machine precision of \(\approx 10^{-16}\). Therefore, we use symbolic methods, which allow for much higher precision.

In essence, we will derive confidence intervals by applying bounds on the binomial distribution based on the normal distribution as shown in \cite{Zubkov2013TheoryProbab.Appl.}. This will require the calculation of the inverse of the cumulative distribution function of the normal distribution \(\Phi\). However, we note that the \texttt{MATLAB} function \texttt{erfcinv} supports symbolic expressions which can be used to recast the required inverse of \(\Phi\) as
\begin{equation}
    \Phi^{-1}(x) = -\sqrt{2} \mathrm{erfc}^{-1}(2x).
\end{equation}
Hence, one can use \texttt{erfcinv} together with symbolic expressions to achieve valid confidence intervals even for very small security parameters.

For fixed length protocols the problem is the same, here, the incomplete beta function only provides precision of \(\approx 10^{-16}\). We take a similar path and find bounds on the variational bounds \(\vec{\nu}^{U/L}\) by similarly bounding the binomial distribution by the normal distribution as in \cite{Zubkov2013TheoryProbab.Appl.}. We summarize the results in the following theorems.

\begin{lem}[Gaussian Bounds on Confidence intervals for the binomial distribution]\label{Lem:Gaussian Bounds on Confidence intervals for the binomial distribution}
    Let the conditions of \cref{lemma:VFobsconstruction} apply. One can bound \(\kappal{k}\) and \(\kappau{k}\) by
    \begin{align}
        \kappal{k} &\leq F^\text{obs}_k - \min_p \bigg\{ 1- C_{N,p}(\lfloor NF^\text{obs}_k \rfloor -1) \geq \frac{\varepsilon_{\text{AT}}}{ 2| \Sigma|} \bigg\},\\
        \kappau{k} &\leq \max_p \bigg\{ C_{N,p}( \lceil NF^\text{obs}_k \rceil +1 ) \geq \frac{\varepsilon_{\text{AT}}}{ 2| \Sigma|} \bigg\}  - F^\text{obs}_k.
    \end{align}
    Here the functions \(C_{N,p}\) are defined as in \cite{Zubkov2013TheoryProbab.Appl.} by
    \begin{align}
        C_{N,p}(k) = \Phi\left( \mathrm{sign}\left(\frac{k}{N} - p \right) \sqrt{2N H\left(\frac{k}{N},p \right)} \right),
    \end{align}
    where \(\Phi(x)\) is the cumulative-distribution function of the standard normal distribution and \(H(x,p)\) is given by
    \begin{equation}
        H(x,p) = x\log(\frac{x}{p}) + (1-x) \log(\frac{1-x}{1-p}).
    \end{equation}
\end{lem}
\begin{proof}
    For simplicity, we only consider the single event \(k\) in \(\Sigma\). Let \(X \sim \mathrm{Bin}_N(p)\) be binomially distributed, with parameter \(p\) which we aim to bound, \(N\) the number of trials and \(m\) the number of observed events. First, note that the bounds of the confidence interval with confidence level \( 1- \frac{\varepsilon_{\text{AT}}}{ |\Sigma| } \) can be written as \cite{Thulin_2014}
    \begin{align}
        p_{\max} &= \max_p\bigg\{ \Pr[X \leq m ] \geq \frac{\varepsilon_{\text{AT}}}{ 2| \Sigma|}\bigg\}, \\
        p_{\min} &= \min_p\bigg\{ \Pr[X \geq m ] \geq \frac{\varepsilon_{\text{AT}}}{ 2| \Sigma|} \bigg\}.
    \end{align}
    Next, we can apply the bounds from \cite{Zubkov2013TheoryProbab.Appl.} to find
    \begin{align}
        \Pr[X \leq m ] \leq C_{N,p}(m+1), \\
        \Pr[X \geq m ] \leq 1 - C_{N,p}(m-1).
    \end{align}
    If we observe the frequency \(F^\text{obs}_k\), then we can bound the number of observed \(m\) as 
    \begin{equation}
        \lfloor NF^\text{obs}_k \rfloor \leq m \leq \lceil NF^\text{obs}_k \rceil.
    \end{equation}
    The above step is unnecessary for implementations, where $F^\text{obs}_k B$ is guaranteed to be an integer. However, it may be necessary for theoretical simulations, if $F^\text{obs}_k$ is obtained from some model of the channel. Applying the previous step, we can bound \(p_{\min}\) and \(p_{\max}\) by
    \begin{align}
        p_{\max} &\leq \max_p\bigg\{ C_{N,p}( \lceil NF^\text{obs}_k \rceil +1 ) \geq \frac{\varepsilon_{\text{AT}}}{ 2| \Sigma|}\bigg\}, \\
        p_{\min} &\geq \min_p\bigg\{ C_{N,p}(\lfloor NF^\text{obs}_k \rfloor -1) \leq 1 - \frac{\varepsilon_{\text{AT}}}{ 2| \Sigma|} \bigg\}.
    \end{align}
    Thus, \(\kappal{k}\) and \(\kappau{k}\) are given as in the theorem statement, by noting that
    \begin{align}
        \kappal{k} &\leq F^\text{obs}_k - p_{\min},\\
        \kappau{k} &\leq p_{\max} - F^\text{obs}_k.
    \end{align}
\end{proof}

\begin{lem}[Gaussian Bounds on Variational bounds]\label{Lem:Gaussian Bounds on Variational bounds}
    Let the acceptance set \(\mathcal{Q}\) and the feasible set \(S_{\vec{\nu}}\) be defined as in \cref{Thrm:Secrecy outside Feasible Set}. Furthermore, choose the variational bounds \(\nu_k^{L/U}\) as the solutions 
    \begin{align}
         \nu_k^{L} &=
        \begin{aligned}[t] \argmin_{\nu'} \{ &\nu' \in [0,1] \; | \\ &1 - C_{N,F^L_k-\nu'}(\lceil N F^L_k \rceil - 1) = \varepsilon_{AT} \},
        \end{aligned} \\      
        \nu_k^{U} &=
        \begin{aligned}[t]
         \argmin_{\nu'} \{&\nu' \in [0,1] \; | \\ &C_{N,F_k^U+\nu'}(\lfloor N F_k^U \rfloor + 1) = \varepsilon_{AT} \},
        \end{aligned}
    \end{align}
    where \(C_{N,p}(k)\) are defined as in \cref{Lem:Gaussian Bounds on Confidence intervals for the binomial distribution} and \(F^{L/U}_k = \bar{F}_k \mp t_k\) as before. 
    Then, it holds for all \(\rho_{\vec{A}\vec{B}\vec{E}} = \bigotimes_{i=1}^{N} \sigma_{ABE}\) and \(\sigma \notin S_{\vec{\nu}}\) that,
    \begin{equation}
        \Pr[\Omega_\text{acc}] \leq \varepsilon_{\text{AT}}.
    \end{equation}
\end{lem}
\begin{proof}
    Following the same steps as in \cref{Thrm:Secrecy outside Feasible Set} and again defining \(p_k:= \Tr\left[\Gamma_k \sigma \right]\) one can show that for \(p_k > \bar{F}_k +t_k+\nu\)
    \begin{equation}
        \begin{aligned}
            &\Pr[\abs{F^{\text{obs}}_k - \bar{F}_k} \leq t_k]\Big|_{p_k > F^{U}_k+\nu} \\
    			&\leq \Pr[X_k \leq \lfloor N F_k^U\rfloor]\Big|_{p_k > F^{U}_k +\nu},
        \end{aligned}
	\end{equation}
    where we again note that \(X_k\) is binomial distributed. Now, one can apply the bounds from \cite{Zubkov2013TheoryProbab.Appl.} and note that it is monotonically decreasing in \(p\) to find 
    \begin{equation}
        \begin{aligned}
            &\Pr[\abs{F^{\text{obs}}_k - \bar{F}_k} \leq t_k]\Big|_{p_k > F^{U}_k+\nu} \\
    			&\leq C_{N,F_k^U+\nu'}(\lfloor N F_k^U \rfloor + 1).
        \end{aligned}
	\end{equation}
    Similarly, one finds 
    \begin{equation}
        \begin{aligned}
            &\Pr[\abs{F^{\text{obs}}_k - \bar{F}_k} \leq t_k]\Big|_{p_k < F^{L}_k-\nu} \\
    			&\leq 1 - C_{N,F_k^L-\nu'}(\lceil N F^L_k \rceil - 1).
        \end{aligned}
	\end{equation}
    Thus, if \(\nu^{U/L}_k\) are chosen as in the theorem statement we find 
    \begin{align}
        &\Pr[\abs{F^{\text{obs}}_k - \bar{F}_k} \leq t_k] \\ 
		&\leq \begin{aligned}[t]
		      \max \bigg\{ &\Pr[\abs{F^{\text{obs}}_k - \bar{F}_k} \leq t_k]\Big|_{p_k > F^{U}_k+\nu_k^U} ,  \\
            &\Pr[\abs{F^{\text{obs}}_k - \bar{F}_k} \leq t_k]\Big|_{p_k < F^{L}_k+\nu_k^L} \bigg\}
		\end{aligned}\\
        &\leq \varepsilon_{\text{AT}},
    \end{align}
    for all \(\sigma \notin S_{\nu}\).
\end{proof}

\end{document}